\documentclass[12pt,a4paper]{article} 

\usepackage[left=2.2cm,right=2.2cm,top=3cm,bottom=4cm,]{geometry}

\usepackage{graphicx}
\usepackage[textwidth=8em,textsize=small]{todonotes}
\usepackage{amsmath}
\usepackage{natbib}

\usepackage{url}

\usepackage{color}

\usepackage{amssymb}
\usepackage{booktabs}
\usepackage{bm}
\usepackage[affil-it]{authblk}

\usepackage{rotating}
\newcommand{\nc}{\newcommand}
\nc{\tb}[1]{\textcolor[rgb]{0.00,0.00,0.00}{#1}}
\nc{\trd}[1]{\textcolor[rgb]{1.00,0.00,0.00}{#1}}
\nc{\tp}[1]{\textcolor[rgb]{1.00,0.00,1.00}{#1}}

\nc{\tg}[1]{\todo[inline, color=green!40]{#1}}

\newcommand{\trans}{^{\text T}}
\def\bSig\bm{\Sigma}

\newcommand{\Ex}{{\rm E} }

\definecolor{airforceblue}{rgb}{0.773, 0.859, 0.914}
\definecolor{lightpink}{rgb}{0.992, 0.894, 0.863}

\newcommand{\solidblueline}{\raisebox{2pt}{\tikz{\draw[-,airforceblue,solid,line width = 1pt](0,0) -- (10mm,0);}}}

\newcommand{\dashblueline}{\raisebox{2pt}{\tikz{\draw[-,airforceblue,dashed,line width = 1pt](0,0) -- (10mm,0);}}}

\newcommand{\solidredline}{\raisebox{2pt}{\tikz{\draw[-,lightpink,solid,line width = 1pt](0,0) -- (10mm,0);}}}

\newcommand{\dashredline}{\raisebox{2pt}{\tikz{\draw[-,lightpink,dashed,line width = 1pt](0,0) -- (10mm,0);}}}

\usepackage{color}
\usepackage{tikz,graphicx}
\usetikzlibrary{shapes}

\begin{document}

\title{Accommodating  informative visit times for analysing irregular longitudinal data: a sensitivity analysis approach with balancing weights estimators}
\author{Sean Yiu~  and Li Su\thanks{E-mail address: \texttt{li.su@mrc-bsu.cam.ac.uk}; corresponding author}}
\affil{MRC Biostatistics Unit, School of Clinical Medicine, University of Cambridge,  Cambridge, CB2 0SR, UK}
\date{}
\maketitle

\baselineskip=24pt
\setlength{\parindent}{.25in}

\begin{abstract}

Irregular longitudinal data with informative visit times arise when patients' visits are partly driven by concurrent disease outcomes. \tb{However, existing methods such as inverse intensity weighting (IIW),  often overlook or have not adequately assess the influence of informative visit times on estimation and inference.}
Based on novel balancing weights estimators, we propose a new sensitivity analysis approach to addressing informative visit times \tb{within the IIW framework}. The balancing weights are obtained by balancing observed history variable distributions over time and \tb{including a selection function with specified sensitivity parameters to characterise the additional influence of the concurrent outcome on the visit process. A calibration procedure is proposed to anchor the range of the sensitivity parameters to the amount of variation in the visit process that could be additionally explained by the concurrent outcome given the observed history and time.} Simulations demonstrate that our balancing weights estimators outperform existing weighted estimators for robustness and efficiency. We provide an \texttt{R Markdown} tutorial of the proposed methods and apply them to analyse data from a clinic-based cohort of psoriatic arthritis.

\end{abstract}

\section{Introduction}
\label{sec1}

\subsection{Informative visit times}
Irregular longitudinal data with informative visit times arise when patients' follow-up visits are partly driven by concurrent longitudinal outcomes (e.g., ongoing disease activities or symptoms), which induces selection bias because the visit and outcome processes are associated. However, \tb{existing methods for analysing irregular longitudinal data, such as \emph{inverse intensity weighting} (IIW) and other methods reviewed in \cite{Pullenayegum2016a}, often overlook or do not adequately assess the influence of informative visit times on estimation and inference.}

First proposed by \cite{Lin2004}, IIW is a useful approach to analysing irregular longitudinal data when the \emph{visiting at random} assumption holds, that is, when visiting at time $t$ is independent of the longitudinal outcome at $t$, given the observed covariate and outcome histories up to $t$. The aim of IIW is to remove the selection bias from irregular visit times by creating a pseudo-population after weighting that is representative of the target population, i.e., the  study population that could be continuously observed from baseline until the study end.  Specifically, the observed longitudinal data are weighted by the inverse of the visit intensities estimated from a semi-parametric Cox model given the observed history.
Provided that the visiting at random assumption is satisfied and the Cox model is correctly specified, the inverse intensity weighted estimators (IIWEs)  based on weighted generalised estimating equations (GEEs) are consistent estimators of the parameters in a marginal regression model of the longitudinal outcome \citep{Lin2004, Buzkova2008, Pullenayegum2013}. 

Unfortunately, the visiting at random assumption is unverifiable from the observed data and is clearly violated in the presence of informative visit times, where patients' visits are partly urged by concurrent longitudinal outcomes. 
Therefore, it is necessary to
assess the sensitivity of conclusions drawn from the IIWEs to the violations of the visiting at random assumption by accommodating informative visit times. 
However, there is very limited research on this topic in the IIW literature \tb{and more broadly in the literature of handling irregular and potentially informative visit or assessment times \citep{Pullenayegum2022}.}
Using an augmented IIWE based on the efficient influence function, \cite{Smith2022} recently proposed a sensitivity analysis approach to addressing informative assessment times in clinical trials. 
However, their methods are restricted to estimating treatment-arm-specific outcome means over time in clinical trial settings, and thus are not applicable to typical marginal regression analyses of irregular longitudinal data for examining associations with time-invariant and time-varying covariates.

Outside the IIW framework, \cite{Wang2020}  also developed sensitivity analyses for estimating treatment-arm-specific outcome means over time to address informative assessment times in clinical trials. \tb{Their approach was based on discretising the time scale (in days) and applying G-computation (standardisation) using fully parametric models for estimation \citep{robins1986}}, unlike the semi-parametric estimation used in IIWEs. 
At the end of Section~\ref{proposal}, we will elaborate on the distinctions between the proposed methods and the methods by  \cite{Wang2020} and \cite{Smith2022}.

\tb{Other approaches for analysing irregular longitudinal data include joint modelling of the visit and outcome processes by incorporating random effects to induce the correlations between the two processes \cite[]{Liang2009,Sun2011,Cai2012}, which provides a useful framework for characterising the underlying data generating mechanism.  By introducing specific dependence structures of the outcome and visit processes, these joint models can allow for particular types of informative visit times, therefore they have been used to develop diagnostic tests to check the dependence between the outcome and visit processes conditional on observed covariates \citep{McCulloch2018}. Focusing on multi-state models for life history data,  \cite{Lange2015} and \cite{Cook2019} proposed joint models of the life history process and the visit process, assuming that the visit intensity depends on the current partially unobserved life history state to account for informative visit times. \cite{Lange2015} and \cite{Cook2019} recommended collecting auxiliary information, such as the reasons why an individual paid a visit, to check the plausibility of the assumption that the visit process and the life history process are conditionally independent given the observed history (Definition 2.1 in \citealp{Cook2019}). To the best of our knowledge, there is no published research on sensitivity analysis for unverifiable assumptions made in the joint modelling approach based on random effects for irregular longitudinal data. One possible reason is that it is difficult to integrate random effects and derive the extrapolation distribution of unobserved outcomes given observed data in these models (Section 8.5 in \citealp[]{Daniels2008}). For simpler settings with informative dropout, \cite{Su2019} developed a sensitivity analysis approach for joint models of a longitudinal outcome and dropout by deriving a closed-form extrapolation distribution of the unobserved outcomes after dropout given observed information. Sensitivity analysis is an open area of research for the joint modelling approaches for irregular longitudinal data.   }

\subsection{Overview of contributions}\label{proposal}

In this paper, to adequately accommodate informative visit times within the IIW framework, we propose  a new sensitivity analysis approach
for marginal regression analyses of irregular longitudinal data, based on novel balancing weights estimators. 
Specifically, motivated by improving the robustness and efficiency of existing IIWEs, we develop a new class of balancing weights estimators of marginal regression coefficients, \tb{where the balancing weights are estimated to satisfy the conditions that, after weighting,  the observed history variable distributions influencing the visit and outcome processes are representative of the observed history variable distributions in the target population.}

Balancing weights \tb{focusing on optimising the balance of covariate distributions} were shown to be more stable than weights obtained by maximum likelihood estimation (MLE), \tb{which considerably improved the performance of inverse probability weighted estimators for average treatment effect estimation  (e.g., see \citealp{ Hainmueller2012, Imai2014, Chan2016,Yiu2017, Tan2017, Chattopadhyay2020}) }and for handling missing data in cross-sectional settings and dropout in longitudinal studies  (e.g., see \citealp{Graham2012,Zubizarreta2015, Han2016,  Yiu2022}). The simulations study in Section~\ref{simulation} also demonstrates that,  under both correct and incorrect model specifications for weight estimation, our balancing weights estimators outperform the existing estimators that use inverse intensity weights estimated by maximum partial likelihood in terms of reducing finite-sample bias and mean squared error. \tb{ We provide  details about the rationale behind the proposed balancing weight estimators in  Section~\ref{prop_set_up}.  }

More importantly, anchored at the visiting at random assumption, we introduce a selection function into the model for estimating the balancing weights, similar to the sensitivity analysis approaches for handling missing longitudinal data \cite[]{Rotnitzky1998,Scharfstein1999,Vansteelandt2007, Wen2018}. 
This selection function characterises the residual dependence of the visit intensity at time $t$ on the concurrent outcome at $t$, given the observed history up to $t$. A sensitivity parameter (or a vector of sensitivity parameters) is used to describe the strength of this
residual dependence.  When the sensitivity parameter is set to zero, visiting at random is assumed.  By varying this sensitivity parameter over a range of plausible values and re-estimating the balancing weights, the
sensitivity of the substantive conclusions to the deviations from the visiting at random assumption can be
assessed. As a result, the influence of a wide range of \tb{informative visit times} on the regression coefficient estimation is accommodated in our sensitivity analysis approach with the balancing weights estimators. 

\tb{We would like to emphasise that similar to the semiparametric weighted estimating equation approaches used for handling missing data in longitudinal studies \citep{Rotnitzky1998,Scharfstein1999,Vansteelandt2007, Wen2018}, the sensitivity parameters specified in our selection function cannot be identified from observed data. This feature enables us to develop a  transparent sensitivity analysis strategy as advocated by \cite{Daniels2008}  for handling missing data in general in longitudinal studies, where in principle sensitivity parameters should characterise the extrapolation distribution of the unobserved outcomes given observed data and should not be identifiable from observed data (see Definition 8.1 in \citealp{Daniels2008}).}

\tb{In addition, inspired by the recent literature on sensitivity analysis for unmeasured confounding \citep{Franks2019}, we  propose a calibration procedure to anchor the range of the sensitivity parameter to the amount of variation in the visit process that could be \emph{additionally} explained by the concurrent outcome given the observed history and time. This can help practitioners to gauge the sensitivity analysis to the observed information and facilitate the interpretation of the sensitivity parameter in their study settings.  }




 {Our sensitivity analysis approach can also be applied to the existing IIWEs; see details in Section~\ref{IIW}. Moreover, coupling the enhanced performance of the balancing weights estimators with our sensitivity analysis approach will hopefully promote more widespread utilisation of the IIW approach for analysing irregular longitudinal data in practice.  
Compared with the methods by \cite{Wang2020} and  \cite{Smith2022}, our sensitivity analysis approach has the following features and advantages.}
\begin{enumerate}
    \item[(a)] {The methods by \cite{Wang2020} and  \cite{Smith2022} can only estimate the treatment-arm-specific outcome means over time, not the covariate effects in marginal regression models, whereas our approach can achieve both. }

    \item[(b)]  {\cite{Wang2020}'s methods treat the visit times as discrete, whereas our methods and the methods by \cite{Smith2022} are within the IIW framework where the visit times are continuous. }
    \item[(c)] {\cite{Wang2020}'s methods were based on G-computation instead of weighting, therefore they required specifying two fully parametric models for the visit process and the distribution function of the observed outcome.  \tb{Specifically, \cite{Wang2020} applied penalised logistic regression to model the visit probability in discrete time and zero-inflated negative binomial regression to model the observed outcomes, both conditional on the observed history. The G-computation started by sampling from the empirical distribution of the observed history variables. Given the sample of the observed history variables,  the observed outcomes given a visit being made were sampled from the fitted zero-inflated negative binomial model. These samples were then weighted by the estimated visit probabilities and an exponential tilting function for the unobserved outcome distribution to account for the extra influence of the concurrent outcome on visit probabilities. Finally, the weighted samples were averaged to obtain the marginal mean of the outcome at time $t$. } In contrast,  our estimators are semi-parametric like the existing IIWEs and do not involve fully parametric modelling of the observed outcome. }  

          \item[(d)] {\tb{\cite{Smith2022}  proposed augmented IIWEs based on the efficient influence function of the parameters in a spline model for treat-arm-specific outcome means, where the extra influence of the concurrent outcome was also  characterised by an exponential tilting function for the unobserved outcome distribution when estimating the nuisance  parameters in their estimators. } As a result, their estimator 
          also requires specifying a model for the distribution function of the observed outcome, whereas our approach does not. Moreover, the validity of their variance estimator and bootstrap confidence intervals requires that both the visit process model given the observed history and the model for the observed outcome be correctly specified (\citealp[Theorem 2]{Smith2022}). In contrast, our estimators only require the visit process model to be correctly specified for valid inference based on bootstrap.}

     \item[(e)] {The estimator by \cite{Smith2022} was developed for estimating the outcome mean with the \emph{identity}  link only, thus cannot be applied to other types of outcomes, such as the count outcome in our data example in Section~\ref{Application}. In contrast, similar to the existing IIWEs, our balancing weights estimators can be applied to different types of outcomes as they are in the framework of weighted GEEs.  }

 \item[(f) ]{\tb{\cite{Smith2022} selected the range of their sensitivity parameter based on domain experts' opinions on the minimum and maximum of the outcome mean at each visit for each treatment arm. Our calibration procedure for determining the magnitude of the sensitivity parameter is based on the additional variation of the visit process explained by the concurrent outcome, above and beyond what has been accounted for by the observed history and time. By linking  the target value of this additional variation towards the variation explained by  the observed history variables,  
  the range of our sensitivity parameter values can be anchored to the observed information.}}

        \item[(g) ]{\tb{\cite{Smith2022} specified their sensitivity parameter in the exponential tilting function for the unobserved outcome distribution, which is equivalent to a special case of our selection function specification that includes the concurrent outcome at the original scale. In contrast, our selection function specification could include transformations of the concurrent outcome, which could facilitate the proposed calibration procedure (see details in Section~\ref{IIW}).} }
        

    \item[(h)] {Our sensitivity analysis approach based on the existing IIWEs can be implemented using standard software such as the \texttt{coxph} function of the  \texttt{R} \citep{RCoreTeam2014} package  \texttt{Survival} \citep{Therneau2023}. The balancing weights estimators are implemented using the  \texttt{R} package  \texttt{nleqslv} \citep{Hasselman2016}.  We provide an \texttt{R Markdown} tutorial to demonstrate the implementation of our methods (see \url{https://github.com/lisu-stats/IIW_SA}). \tb{\cite{Wang2020}  did not provide code/programs, while \cite{Smith2022}  added the \texttt{R} code for implementing their methods in the latest version of their manuscript. }} 

    \end{enumerate}


\subsection{Motivating data}\label{datainfo}
This research is motivated by data from
the University of Toronto Psoriatic Arthritis  (PsA)  Clinic cohort.
PsA is inflammatory arthritis associated with the skin disease psoriasis. Manifestations of PsA include the development of joint activity, which is characterised by the occurrence of pain and/or swelling in the joints (i.e., active joints).  The University of Toronto PsA clinic is one of the largest
cohorts of PsA patients in the world. More than 1000 patients are assessed in the clinic about every 6-12 months, where data on their demographics,  clinical factors, and treatment information are recorded. 
 Despite the regular scheduling of patients' visits to the clinic (6-12 months apart), the actual visit times varied considerably from patient to patient. \cite{Zhu2017} noted that the gap times between a patient's clinic visits appeared to be related to his/her prior disease history, past visit history and various factors associated with these processes.  They addressed this problem of the PsA cohort data by applying IIW in a parametric failure time model to examine the association of biologics use with PsA disease progression. The variables considered in the visit process model of  \cite{Zhu2017} included demographics, PsA disease duration, erythrocyte sedimentation rate, histories of various disease activity and progression variables, treatment use history as well as the median of past visit gap times. However, as the PsA cohort is clinic-based, patients could visit the clinic due to a PsA flare at non-scheduled times. In addition, important socioeconomic factors (e.g., the ability to pay for treatments) that are indirectly associated with the PsA disease activity and the clinic visits are not recorded in the PsA cohort. Therefore, there is likely residual dependence of the visit intensity on the ongoing PsA disease activity. In this paper, we aim to accommodate the likely informative visit times in the PsA clinic cohort using the proposed methods in order to examine the association between biologics use and  PsA active joint counts over time in a marginal regression analysis.

\section{Methods}
\label{sec2}

\subsection{Notation,  setting and assumptions}
\label{notation}
We consider a longitudinal study where  $n$ patients are enrolled at baseline  (time $0$) and each of the $i=1, \ldots, n$ patients are followed to a maximum time of $\tau$. Patients' data were assumed independent and identically distributed in the study and thus we suppress the patient-specific subscript $i$ for now and reintroduce it when we describe our sensitivity analysis approach in Section~\ref{IIW}.
Let $Y(t)$ be a longitudinal outcome measurable at time $t$ for $0\le t \le \tau$. Let $\bm{X}(t)$ denote a vector of exogenous (possibly time-varying) covariates, which is assumed to be known prior to $t$.
We are interested in a   marginal  regression model for $Y(t)$, 
\begin{equation}\label{marginalmodel}
   g[\Ex\{Y(t)\mid \bm{X}(t)\}]= g[\mu\{t,\bm{X}(t);\bm{\beta}_0\}]=\bm{\beta}_0\trans\bm{X}(t)
\end{equation}
 for $t \in [0, \tau]$, where $g(\cdot)$ is a known link function and $\bm{\beta}_0$ is a $p$-dimensional   vector of regression coefficients.
Note that by specifying $\bm{X}(t)$  to include indicators of treatment arms, time $t$ and their interactions, Model~\eqref{marginalmodel} can be used to estimate the treatment-arm-specific outcome means over time, which are the focus of the clinical trial settings described in \cite{Wang2020} and \cite{Smith2022}. But clearly, Model~\eqref{marginalmodel} is more general for examining the associations of the covariates with the outcome means.

 Model~\eqref{marginalmodel} is formulated for the outcome process  in $0\le t \le \tau$. However,  the outcome measurements are not available continuously over time but only at visit times. 
 Let $N(t)$  be the counting process for the number of visits made by time $t$ and $N(t)= \int_0^t d N(s)$, where  $d N(s)= N(s) -N(s-)$ is the indicator of a visit made at time $s$.  Let $C$ be the follow-up time measured from the baseline of the study.   Then  $\xi(t)=\mbox{I}\{t < \min(C, \tau)\} $ is a left-continuous at-risk process, indicating whether a patient is still under follow-up. We denote the uncensored visit process by $N^*(t)= \xi(t) N(t)$. In addition, let $\bm{Z}(t)$ be a vector of auxiliary covariates that
 are associated with the visit and outcome processes but are excluded from Model~\eqref{marginalmodel} because their associations with the outcome means are not of scientific interest. We assume that $\bm{Z}(t)$ is known at all times \citep{Buzkova2008}.

Following \cite{Buzkova2008} and \cite{Pullenayegum2013}, we make the \emph{non-informative censoring} assumption for the marginal mean of the outcome 
\[
\Ex\{Y(t)\mid \bm{X}(t), t \le C \le \tau \}= \Ex\{\xi(t)Y(t)\mid \bm{X}(t)\}=\Ex\{Y(t)\mid \bm{X}(t)\}, 
\] such that the marginal mean $\Ex\{Y(t)\mid \bm{X}(t)\}$ is the same for those patients whose follow-up is censored at time $t$ and for those who remain in the study at  $t$.  

Let $H(t)= \{Y(s): s < t;   \bm{X}(s), \bm{Z} (s), N^*(s): s \le t\}$ include the outcome process history prior to $t$,  the covariate and visit process histories  up to  $t$.  Let  \tb{ $O(t)= \{dN^*(s)Y(s): s < t; dN^*(s)\bm{X}(s), \bm{Z} (s), N^*(t): s \le t\}$ }include the \emph{observed} outcome history prior to $t$, the \emph{observed} covariate history and the uncensored visit process history up to  $t$.  We assume that the uncensored visit process follows 
\begin{equation}\label{visitassumption}
    \Ex\{d N^*(t) \mid H(t-), Y(t), \underline{Y}(t)\}=\Ex\{d N^*(t) \mid H(t-), Y(t)\}=\Ex\{ d N^*(t) \mid O(t-), Y(t)\}, 
\end{equation}
 where $\underline{Y}(t)=\{Y(s): s > t\}$ is the future of the outcome process beyond $t$. The first equality in \eqref{visitassumption}  is the \emph{non-future dependence} assumption, which means that conditional on the process history  $H(t-)$, and the (possibly unobserved) concurrent outcome $Y(t)$, the uncensored visit process is independent of future outcomes beyond time $t$. \tb{The non-future dependence assumption is plausible because of the temporal order of $d N^*(t)$ and $\underline{Y}(t)$. } The second equality requires that the uncensored visit process be independent of past unobserved values of the outcome and covariates given observed history $O(t-)$ and the concurrent outcome  $Y(t)$. \tb{This assumption is unverifiable from observed data but} it is \emph{less restrictive} compared to the visiting at random assumption which does not allow the visit process to depend on the concurrent outcome $Y(t)$ \tb{as well as the past \textit{unobserved }values of the outcome and covariates given observed history $O(t-)$.}
 By allowing the uncensored visit process to additionally depend on the concurrent outcome $Y(t)$, we are able to accommodate informative visit times in clinic-based studies or electronic health record databases where patients' visits could be partly driven by ongoing disease activities.  
 
Let $\Ex\{ d N^*(t) \mid O(t-), Y(t)\}=\xi(t)\lambda\{t, O(t-), Y(t)\}$, where $\lambda\{t, O(t-), Y(t)\} $ is the visit intensity at time $t$ that depends on the observed history and the concurrent outcome. We further make the positivity assumption for the visit intensity that $\lambda\{t, O(t-), Y(t)\}>0$ when $\xi(t)=1$. 

\subsection{Sensitivity analysis for  inverse intensity weighted estimators under informative visit times}\label{IIW}
To facilitate understanding,   \tb{ we first explain the main idea behind IIW, and describe our sensitivity analysis approach based on the existing IIWEs along with the calibration procedure} before introducing the proposed balancing weights estimators in Section~\ref{prop_set_up}.

Under the assumptions made in Section~\ref{notation} and assuming that the marginal regression model in \eqref{marginalmodel} is correctly specified, $\bm{\beta}_0$ can be consistently estimated by solving weighted GEEs  with an independence working correlation structure of the following form 
\begin{equation}\label{IIW_es}
\sum_{i=1}^n \int_0^{\tau} U\left\{Y_i(t),\bm{X}_i(t);\bm{\beta}\right\}w_i(t)  dN^*_i(t)=0,
\end{equation} where $U\{Y_i(t),\bm{X}_i(t);\bm{\beta}\}=D\{t,\bm{X}_i(t);\bm{\beta}\}\left[Y_i(t)-\mu\{t,\bm{X}_i(t);\bm{\beta}\}\right]$, \\ $D\{t,\bm{X}_i(t);\bm{\beta}\}=\left[\partial\mu\{t,\bm{X}_i(t);\bm{\beta}\}/\partial\bm{\beta}\right]\mbox{Var}^{-1}\left[t,\mu\{t,\bm{X}_i(t);\bm{\beta}\}\right]$, \tb{$\mbox{Var}\left[t,\mu\{t,\bm{X}_i(t);\bm{\beta}\}\right]$}  is a conditional variance function, $w_i(t)=s(t)/\lambda\{t, O_i(t-), Y_i(t)\}$ and \tb{$s(t)$ is a user-specified function of $t$  that stabilise the weights $w_i(t)$}  \cite[]{Lin2004}. 
This is justified because 
\begin{equation}\label{IIW_id}
\begin{split}
&\Ex\left[\int_0^{\tau}U\{Y(t),\bm{X}(t);\bm{\beta}_0\}\frac{s(t)}{\lambda\{t, O(t-), Y(t)\}}dN^*(t)\right]\\
=&\Ex\left(\int_0^{\tau} \Ex \left[U\{Y(t),\bm{X}(t);\bm{\beta}_0\}\frac{s(t)}{\lambda\{t, O(t-), Y(t)\}}dN^*(t) \,\middle\vert\, O(t-), Y(t)\right]\right)\\
=&\Ex\left[\int_0^{\tau}U\{Y(t),\bm{X}(t);\bm{\beta}_0\}\frac{s(t)}{\lambda\{t, O(t-), Y(t)\}}\,\Ex\left\{dN^*(t) \mid O(t-), Y(t)\right\}\right]\\
=&\Ex\left[\int_0^{\tau}U\{Y(t),\bm{X}(t);\bm{\beta}_0\}s(t) \xi(t)\,dt\right]=0.
\end{split} 
\end{equation}
Intuitively, the purpose of IIW is to create a  pseudo-population that is representative of the potentially  observable population from baseline until time $\tau$ (i.e., the target population), but with a visit process with an intensity function $s(t)$ that no longer depends on the observed history and the concurrent outcome.
However, the true visit intensity ${\lambda}\{t, O(t-), Y(t)\}$ is unknown. Even if we make the visiting at random assumption such that  ${\lambda}\{t, O(t-), Y(t)\}={\lambda}\{t, O(t-)\}$, the visit intensity given the observed history  ${\lambda}\{t, O(t-)\}$ still needs to be estimated in practice. 

In the IIW literature, ${\lambda}\{t, O(t-)\}$ is typically estimated by specifying a semi-parametric Cox model  
 ${\lambda}\{t,O(t-);\bm{\bm{\gamma}}_0\}={\lambda}_0(t)\exp\{\bm{\bm{\gamma}}^{\text T}_0 \widetilde{\bm{Z}}(t)\}$,
where ${\lambda}_0(t)$ is an unspecified baseline intensity function of $t$,  $\widetilde{\bm{Z}}(t)$ are functions of $O(t-)$ that may contain interactions and transformations of the variables \cite[]{Buzkova2008,Pullenayegum2013}. Then, the regression parameter vector $\bm{\bm{\gamma}}_0$ is estimated by solving the score equations of the  Cox partial likelihood
\begin{equation}\label{Cox_score}
\sum_{i=1}^n\int_{0}^{\tau}\left\{\widetilde{\bm{Z}}_i(t)-\frac{\sum_{l=1}^n\xi_l(t)\widetilde{\bm{Z}}_l(t)\exp\{\bm{\bm{\gamma}}^{\text T} \widetilde{\bm{Z}}_l(t)\}}{\sum_{l=1}^n\xi_l(t)\exp\{\bm{\bm{\gamma}}^{\text T} \widetilde{\bm{Z}}_l(t)\}}\right\}dN^*_i(t)=0.
\end{equation}
${{\lambda}}_0(t)$ can be estimated  by smoothing Breslow's estimate of the cumulative baseline intensity ${\Lambda}_0(t)$ \cite[]{Lin2004, Smith2022}, 
\begin{equation}\label{Breslow}
\hat{{\Lambda}}_0(t;\hat{\bm{{\gamma}}})=\int_{0}^t\frac{\sum_{i=1}^ndN^*_i(s)}{\sum_{i=1}^n\xi_i(s)\exp\{\hat{\bm{\gamma}}^{\text T} \widetilde{\bm{Z}}_i(s)\}},
\end{equation}
where $\hat{\bm{\bm{\gamma}}}$ is obtained from solving \eqref{Cox_score}.  
Alternatively, \tb{if we set  the numerator term  $s(t)={\lambda}_0(t)$ for stabilisation}, then $w_i(t)$ can be estimated by $\exp\{-\hat{\bm{\bm{\gamma}}}\trans \widetilde{\bm{Z}}_i(t)\}$ \cite[]{Buzkova2008}. The estimator 
 in~\eqref{IIW_es} is consistent when the visiting at random assumption is satisfied and there is no model misspecification in the Cox model for ${\lambda}\{t, O(t-)\}$.

In our setting with informative visit times, however,   ${\lambda}\{t, O(t-), Y(t)\}$ cannot be identified and estimated from the observed data because $Y(t)$ is only observed when a visit is made. We instead propose a sensitivity analysis approach by assuming that
\begin{equation} \label{sfmodel}
{\lambda}\{t, O(t-), Y(t);\bm{\gamma}_{s0}, \bm{\phi}\}={\lambda}_0(t)\exp[\bm{\bm{\gamma}}_{s0}^{\text T} \widetilde{\bm{Z}}(t)+q\{O(t-), Y(t);\bm{\phi}\} ],
\end{equation} where $q\{O(t-), Y(t);\bm{\phi}\}$ is a \textit{known} selection function with a \textit{known} sensitivity parameter vector $\bm{\phi}$.  For example, in the PsA data analysis reported in Section~\ref{Application}, we specify $q\{O(t-), Y(t);\bm{\phi}\}= \phi \log\{Y(t)+1\}$ with a single sensitivity parameter $\phi$ that characterises the residual dependence of the visit intensity on the concurrent (log-transformed) PsA active joint count, after adjusting for the observed history variables $\widetilde{\bm{Z}}(t)$. \tb{ We will discuss the  form of the selection function in Section~\ref{calibration} as it is related to how we could calibrate the range of the sensitivity parameters against the observed information. }

Let $Q(t ;\bm{\phi})=\exp[-q\{O(t-), Y(t);\bm{\phi}\}]$. It is easy to see that  if the model in~\eqref{sfmodel} is correctly specified, 
\begin{equation}
  \Ex\left\{Q(t ;\bm{\phi})dN^*(t) \mid O(t-), Y(t) \right\}=\xi(t){\lambda}_{0}(t)\exp\{\bm{\bm{\gamma}}^{\text T}_{s0} \widetilde{\bm{Z}}(t)\},  
\end{equation} \tb{and 
$$
\mathcal{M}(t;\bm{\bm{\gamma}}_{s0}) = Q(t ;\bm{\phi})dN^*(t) -  \int_{0}^{t} \xi(s){\lambda}_{0}(s)\exp\{\bm{\bm{\gamma}}^{\text T}_{s0} \widetilde{\bm{Z}}(s)\} ds    $$ is a zero-mean random process described in \cite{Lin2000} and \cite{Buzkova2008}.} 

Thus, to estimate $\bm{\gamma}_{s0}$, we can replace   $dN^*_i(t)$  by $Q_i(t ;\bm{\phi})dN^*_i(t)$  in the estimating equation in~\eqref{Cox_score}, 
\begin{equation}\label{Cox_weighted}
\sum_{i=1}^n\int_{0}^{\tau}\left\{\widetilde{\bm{Z}}_i(t)-\frac{\sum_{l=1}^n\xi_l(t)\widetilde{\bm{Z}}_l(t)\exp\{\bm{\bm{\gamma}}_s^{\text T} \widetilde{\bm{Z}}_l(t)\}}{\sum_{l=1}^n\xi_l(t)\exp\{\bm{\bm{\gamma}}_s^{\text T} \widetilde{\bm{Z}}_l(t)\}}\right\}Q_i(t ;\bm{\phi})dN^*_i(t)=0.
\end{equation} 

Similarly, the Breslow estimator is modified as 
 \begin{equation}\label{Breslow2}
 \hat{{\Lambda}}_0(t;\hat{\bm{{\gamma}}}_s)=\int_{0}^t\frac{\sum_{i=1}^n Q_i(s ;\bm{\phi})dN^*_i(s)}{\sum_{i=1}^n\xi_i(s)\exp\{\hat{\bm{\gamma}}_s^{\text T} \widetilde{\bm{Z}}_i(s)\}},
\end{equation} where $\hat{\bm{\gamma}}_s$ is the solution to \eqref{Cox_weighted}. 
Note that  $Q_i(t ;\bm{\phi})$, or equivalently  the selection function $q\{O_i(t), Y_i(t);\bm{\phi}\}$,  only needs to be evaluated when $dN^*_i(t)=1$, i.e., when a patient makes a visit. \tb{As we have removed the impact of the selection function in the zero-mean random process   $\mathcal{M}(t;\bm{\bm{\gamma}}_{s0})$, the estimation of $\bm{\gamma}_{s0}$ is not directly influenced by $Y_i(t)$, thus 
$\widetilde{\bm{Z}}_i(t)$ can be correlated with  $Y_i(t)$.} 
When $q\{O_i(t), Y_i(t);\bm{\phi}\}=0$, \eqref{Cox_weighted}  reduce to the form in~\eqref{Cox_score} under the visiting at random assumption. We then update \eqref{IIW_es} by  setting $s(t)=\lambda_0(t)$ and replacing $w_i(t)$ with  $\hat{w}_i(t; \bm{\phi})=\exp[-\hat{\bm{\gamma}}_s^{\text T} \widetilde{\bm{Z}}_i(t)]Q_i(t ;\bm{\phi})$.  For each set of fixed values of $\bm{\phi}$, we can repeat the estimation of $\bm{\beta}_0$ using different sets of weights $\hat{w}_i(t; \bm{\phi})$ in~\eqref{IIW_es}  and evaluate the sensitivity of marginal regression analysis results under informative visit times. 

We provide  details about how to estimate $\bm{\gamma}_{s0}$ and obtain  $\hat{w}_i(t; \bm{\phi})$ using  \texttt{R}  in Section~\ref{implement}.   To distinguish $\hat{w}_i(t; \bm{\phi})$ from the covariate balancing weights proposed in the next section, we call 
 $\hat{w}_i(t; \bm{\phi})$ the  `MLE weights' for convenience, even though~\eqref{Cox_weighted} are not score equations for $\bm{\gamma}_{s0}$. \tb{Note that the MLE weights are stabilised as we set the numerator term as $s(t)=\lambda_0(t)$.  }

\subsection{Calibrating the range of the sensitivity parameters}\label{calibration}

\tb{Since the sensitivity parameters in~\eqref{sfmodel} are not identifiable from the observed data and the estimation of  $\bm{\gamma}_{s0}$ in~\eqref{Cox_weighted} and the baseline intensity function in~\eqref{Breslow2} is affected by the specification of the selection function including the magnitude of the sensitivity parameters, it is important to anchor the selection function specification to the observed information.  
Inspired by the calibration approach in sensitivity analysis for unmeasured confounding proposed by \cite{Franks2019}, we propose to calibrate our sensitivity parameters against the variation of the visit process explained by the concurrent outcome $Y(t)$, above and beyond what has been accounted for by the observed history variables $\widetilde{\bm{Z}}(t)$ and time $t$. In this section, first we will review the calibration approach proposed by \cite{Franks2019} for sensitivity analysis of unmeasured confounding. Then we describe the proposed calibration procedure assuming that the selection function does not depend on $\widetilde{\bm{Z}}(t)$,  which enables us to limit the number of sensitivity parameters. Afterwards, we will discuss the extension of our calibration procedure to allow dependence on covariates. }  

\subsubsection{Review of the calibration approach by \cite{Franks2019}}\label{Franksreview}

\tb{In their approach for calibrating sensitivity parameters for unmeasured confounding, \cite{Franks2019} assumed  a logistic model for treatment selection that depends on potential outcomes $Y(a)$ when receiving treatment $a$ ($a=0,1$) and observed covariates $\bm{W}$, 
\begin{equation}\label{treatmentmodel}
 P\{A=1 \mid Y(a), \bm{W}\} = \mbox{logit}^{-1} \{\alpha_a(\bm{W})+\gamma_a Y(a)\},    
\end{equation}
where $A$ is the treatment variable,   $\text{logit}^{-1}(x)= \{1+\exp(-x)\}^{-1}$, $\alpha_a(\bm{W})$ is  a linear predictor function of $\bm{W}$ and  $\gamma_a$ is the sensitivity parameter. In their setting, $Y(a)$ is only observable when $A=a$. Therefore, $\gamma_a$ cannot be identified and its magnitude needs to be specified.    }

\tb{ \cite{Franks2019} adopted the `implicit $R^2$' measure from \cite{Imbens2003} to calibrate the magnitude of $\gamma_a$ against the variation of the treatment selection model in~\eqref{treatmentmodel} additionally explained by $Y(a)$ than what has been explained by $\bm{W}$. Specifically, \cite{Franks2019} noticed that a logistic propensity score model for the treatment, e.g., $m(\bm{W})= \mbox{logit} \{P(A=1 \mid \bm{W})\}$, can be expressed using a latent variable formulation as
$$
L=m(\bm{W}) + \epsilon~~ \mbox{with}~~ \epsilon \sim \mbox{Logistic} (0,1), 
$$
$$
A= \left\{\begin{array}{c}
      0 ~~\mbox{if}~~ L <0  \\
      1 ~~\mbox{if}~~  L \ge 0.
\end{array}\right.
$$
Therefore, the variance of $L$ explained by   $\bm{W}$  can be defined as 
\begin{equation}\label{R2}
    \rho_{W}^2= \frac{\mbox{var}\{m(\bm{W})\}}{\mbox{var}\{m(\bm{W})\}+\pi^2/3}, 
\end{equation}
 where 
$\pi^2/3$ is the variance of the standard logistic distribution.  The partial variance explained by $Y(a)$  is defined as 
\begin{equation}\label{partialvar}
   \rho_{ Y(a) \mid \bm{W}}^2= \frac{\rho_{Y(a), \bm{W}}^2-\rho_{ \bm{W}}^2}{1-\rho_{ \bm{W}}^2}, 
\end{equation}  which represents the fraction of previously unexplained variance in $A$  that can now be explained by adding $Y(a)$ to the propensity score model. \cite{Franks2019} proposed to target the value of unidentified $ \rho_{ Y(a) \mid \bm{W}}^2$ towards the partial variance of an observed covariate or a group of observed covariates in $\bm{W}$.  }

\tb{Proposition 3 of \cite{Franks2019}   then established the one-to-one relationship between the magnitude of $\gamma_a$ and $ \rho_{ Y(a) \mid \bm{W}}^2$,
\begin{equation}\label{Franksgamma}
     |\gamma_a|=\frac{1}{\sigma_{ra}}\sqrt{\frac{\rho_{ Y(a) \mid \bm{W}}^2}{1-\rho_{ Y(a) \mid \bm{W}}^2}[\{\mbox{var}\{m(\bm{W})\}+\pi^2/3]},
\end{equation}
 where $\sigma_{ra}= \sqrt{ \Ex[\mbox{var}\{Y(a) \mid \bm{W}\}]} $. An estimate of  $\mbox{var}\{m(\bm{W})\}$ can be obtained from a logistic model for the propensity score. \cite{Franks2019} assumed that the observed outcome $Y(a) \mid A=a,\bm{W}$ follows a mixture of exponential family models. Proposition 2 of \cite{Franks2019} showed that, under the treatment selection model in~\eqref{treatmentmodel}, the missing potential outcome $Y(a) \mid A \ne a, \bm{W}$ also follows a mixture of exponential family models.
 If the conditional distribution of $Y(a)$ given $A$ and $\bm{W}$ is homoscedastic so that $\sigma_{ra}$ is independent of treatment and the observed covariates, $\sigma_{ra}$ can be estimated by the residual standard deviation from a regression model for the observed outcome given $A=a$ and $\bm{W}$.  Otherwise, $\sigma_{ra}$ also depends on $\gamma_a$ but is available in analytical form for the mixture of exponential family models. As a result,    $\gamma_a$ can be numerically solved by satisfying~\eqref{Franksgamma}. 
}

\subsubsection{Proposed calibration  procedure}\label{calibration_proc}
\tb{We notice the close resemblances of our setting with the setting in \cite{Franks2019} because both $Y(t)$ and $Y(a)$ are partially unobserved and their missingness depends on the visit indicator $dN^*(t)$ and the treatment variable $A$, respectively. Utilising the fact that a Cox model with time-varying covariates is approximately equivalent to a pooled logistic model when the probability of an event occurrence is small in a short interval \citep{dagostino1990}, we adapt the calibration approach in \cite{Franks2019} to our sensitivity analysis setting. We first assume that the selection function in ~\eqref{sfmodel} does not depend on $\widetilde{\bm{Z}}(t)$ so that
\begin{equation} \label{sfmodel2}
    {\lambda}\{t, O(t-), Y(t);\bm{\gamma}_{s0}, {\phi}\}={\lambda}_0(t)\exp[\bm{\bm{\gamma}}_{s0}^{\text T} \widetilde{\bm{Z}}(t)+\phi \cdot S\{Y(t)\} ] 
\end{equation}
where $S(\cdot)$ is a transformation function  and  $\phi$ is the sensitivity parameter. }
\tb{Consider a partition  $0= t_0 < t_1 < \cdots < t_R = \tau$ of $[0, \tau]$. Let $\Delta N( t_r)$ be the number of visits in $[t_r, t_{r+1}) $ and $\Delta t_r= t_{r+1}- t_r $. From the definition of the intensity function and the property that events cannot occur
simultaneously,   it follows that 
\begin{eqnarray}
    P\{\Delta N(t_r)=1 \mid O(t_r-), Y(t_r)\} &=&   {\lambda}\{t_r, O(t_r-), Y(t_r);\bm{\gamma}_{s0}, {\phi}\} \Delta t_r+ o(\Delta t_r), \nonumber \\
    &\approx& \exp\left[\log\{{\lambda}_0(t_r)\Delta t_r\} +\bm{\bm{\gamma}}_{s0}^{\text T} \widetilde{\bm{Z}}(t_r)+\phi \cdot S\{Y(t_r)\} \right]; \nonumber
\end{eqnarray}  see  Section 2.1 of \cite{cook2008}. 
We can also specify a logistic model for $P\{\Delta N(t_r)=1 \mid O(t_r-), Y(t_r)\}$ such that 
\begin{equation}\label{logistic}
 \mbox{logit}[P\{\Delta N(t_r)=1 \mid O(t_r-), Y(t_r)\} ] =\alpha_r+ \bm{\Gamma} \trans  \widetilde{\bm{Z}}(t_r)+ \Phi \cdot S\{Y(t_r)\},   
\end{equation}
 where $\bm{\Gamma}$ is the regression coefficient vector of $\widetilde{\bm{Z}}(t_r)$ and  $\Phi$ is the coefficient of $ S\{Y(t_r)\}$. As shown by 
\cite{dagostino1990}, under the condition that $P\{\Delta N(t_r)=1 \mid O(t_r-), Y(t_r)\}$ is small in the short interval  $[t_r, t_{r+1})$, 
\[
\alpha_r \approx \log\{{\lambda}_0(t_r)\Delta t_r\}, ~~~~~~\bm{\Gamma}\approx \bm{\gamma}_{s0}~~~\mbox{and}~~~\Phi \approx \phi.
\] Because of these approximate relationships between the parameters in the model in~\eqref{sfmodel2} and those in the logistic model in~\eqref{logistic},  we can now adapt the calibration approach of  \cite{Franks2019} to determine the range of the sensitivity parameter $\phi$.}

\tb{Following \cite{Franks2019}, we make parametric assumptions about $S\{Y(t)\}$ such that the conditional distribution of the observed $S\{Y(t)\}$ given $dN^*(t)=1$, $O(t-)$ and time $t$  follows an exponential family model. Together with the model in~\eqref{sfmodel2}, it can be shown that the conditional distribution of the \emph{unobserved} $S\{Y(t)\}$ given $dN^*(t)=0$, $O(t-)$ and time $t$ also follows an exponential family model but its density is multiplied by an exponential tilting function $\exp[-\phi \cdot S\{Y(t)\}]/\mathcal{C}\{O(t-), \phi\}$, where $\mathcal{C}\{O(t-), \phi\}$ is the normalising constant (see Proposition 2 in \citealp{Smith2022}). This result is consistent with Propositions 1 and 2 in \cite{Franks2019}, where a logistic model for treatment selection is assumed and the corresponding normalising constant  is analytically tractable when the observed outcome follows a mixture of exponential family models. } 

\tb{In the PsA clinic cohort data example, we apply the log transformation to the longitudinal active joint count so that  $S\{Y(t)\}= \log\{Y(t)+1\}$  and assume that  the log-transformed  outcome  $S\{Y(t)\} \mid O(t-), t $ follows a mixture of normal distributions.  Specifically,  we have $S\{Y(t)\} \mid dN^*(t)=1, O(t-),t \sim N\left( \mu\{O(t-)\}, \sigma^2\right)$, with mean $\mu\{O(t-)\}$,  and it follows that $S\{Y(t)\} \mid dN^*(t)=0, O(t-), t \sim N\left(\mu\{O(t-)\}+\phi, \sigma^2\right)$, with a shifted mean $\mu\{O(t-)\}+\phi$.  $S\{Y(t)\}$ has a constant residual standard deviation $\sigma$ that is independent of $O(t-)$, $t$ and $dN^*(t)$. 
\cite{Smith2022} focused on the specification with $S\{Y(t)\}= Y(t)$. Here we recommend applying a transformation $S(\cdot)$ that makes the mixture of exponential family models for $S\{Y(t)\}$ more plausible. 
}

\tb{In unmeasured confounding scenarios, domain expertise can often help to determine the target partial variance  $\rho_{ Y(a) \mid \bm{W}}^2$ by analogously using the partial variance explained by an important covariate or a group of important covariates \citep{Franks2019}. However, in informative visit time settings,  expert knowledge often lacks to choose an observed history variable or a group of observed history variables that could help to determine the target value of the partial variance, $\rho^2_{Y(t) \mid Z,  t}$, explained by $s\{Y(t)\}$, above and beyond what has been explained by all observed history variables and time $t$. As a conservative choice, we consider that   $\rho^2_{Y(t) \mid Z,  t}$ is no larger than $\rho^2_{Z \mid t}$, the partial variance explained by all observed history variables, above and beyond what has been explained by time $t$. Therefore, assuming that $\rho^2_{Y(t) \mid Z,  t} \le \rho^2_{Z \mid   t}$ and   $S\{Y(t)\} \mid O(t-), t $ follows a mixture of normal distributions,}
we summarise the proposed calibration procedure  in the following steps:
\tb{
\begin{enumerate}
    \item[(a)] Prepare the visit process data using the counting process format \texttt{[start, end)} based on the unique observed visit times across patients.  For all patients, calculate the lengths of the created intervals, e.g., the length of the $r^{\mbox{\scriptsize th}}$ interval is $\Delta t_r= t_{r+1}- t_r$.
    \item[(b)] Fit a Cox model to the visit process data under the visiting at random assumption with all observed history variables $\widetilde{\bm{Z}}(t_r)$ and obtain the corresponding intensity estimates for all patients during each interval,  $\hat{\lambda}_{Z}(t_r)$.
    \item[(c)] Calculate the sample variance of $\log \{\hat{\lambda}_{Z}(t_r)  \Delta t_r \}$, which is equivalent to the role of $\mbox{var}\{m(\bm{W})\}$ in~\eqref{Franksgamma}. Use the formula in~\eqref{R2} to calculate the implicit $R^2$ for the Cox model based on observed history variables $\widetilde{\bm{Z}}(t_r)$.
    \item[(d)] Obtain nonparametric estimates of the visit intensities when no observed history variables are included (i.e., the `null model'), denoted by  $\hat{\lambda}_{\mbox{\scriptsize{null}}}(t_r)$. 
    \item[(e)] Calculate the sample variance of $\log \{\hat{\lambda}_{\mbox{\scriptsize{null}}}(t_r)  \Delta t_r \}$. Use the formula in~\eqref{R2} to calculate the implicit $R^2$ for the null model.
    \item[(f)] Calculate the partial variance explained by all observed history variables, $\rho^2_{Z \mid t}$, in comparison with the null model,  using the formula in~\eqref{partialvar}.
    \item[(g)] Setting the target partial variance $\rho^2_{Y(t) \mid Z,  t}$ to be equal to $\rho^2_{Z \mid   t}$ obtained in Step (6).
    \item[(h)] Fit a linear model as flexible as possible for the mean of the observed $s\{Y(t_r)\}$ given observed history variables $\widetilde{\bm{Z}}(t_r)$ and time $t_r$, and obtain the residual standard deviation from this fitted model. 
    \item[(i)] Use the results obtained in Steps (c), (g) and (h) to estimate $|\phi|$ using the formula in~\eqref{Franksgamma}. 
\end{enumerate}
}
\tb{Following \cite{Franks2019}, other mixtures of exponential family models such as a mixture of Bernoulli distributions can be specified for $S\{Y(t)\} \mid O(t-), t $. In this case,  $\sqrt{ \Ex[\mbox{var}\{S\{Y(t)\} \mid O(t-), t\}]}$ depends on $\phi$ but has analytical form so that $|\phi|$ can be determined numerically based on~\eqref{Franksgamma}.  } 

\subsubsection{Extension to allow covariate dependence in the selection function}
\tb{We have focused on the selection function specification in~\eqref{sfmodel2} that is independent of the observed history variables $\widetilde{\bm{Z}}(t)$, which enabled us to limit the number of sensitivity parameters. In certain scenarios, it would be desirable to extend the selection function in~\eqref{sfmodel2} to allow dependence on the observed history variables. For example, instead of using $S\{Y(t)\}$, we may want to include the outcome change from the last visit in the selection function. It is likely that the impact of the outcome change  depends on the outcome  observed at the last visit and the time since the last visit, thus their interactions with the outcome change need to be accounted for in the selection function specification. Unfortunately, this will  increase the number of sensitivity parameters and create challenges for calibration and reporting of results.}

\tb{In a simpler setting with an important baseline categorical covariate, we could specify visit intensity models stratified by the covariate categories and then apply the calibration procedure in Section~\ref{calibration_proc} separately within the covariate categories.  
The sensitivity analysis can be performed by varying the sensitivity parameters over the calibrated ranges and assessing the impact of the combinations of sensitivity parameter values on marginal regression results.}  

\subsection{Balancing weights estimators}\label{prop_set_up}

\subsubsection{Limitations of the MLE weights}\label{limitation}
\tb{In this section, we discuss the limitations of the MLE weights estimated by maximum partial likelihood estimation, which motivated us to develop the proposed balancing weight estimators. For simplicity, we focus on the scenario where the visiting at random assumption is satisfied,  since, with informative visit times,  the selection function specifications are the same when estimating the MLE weights and the balancing weights. }

\tb{Recall that, under the visiting at random assumption,  the goal of IIW is to create a pseudo-population that is representative of the  target population with a visit process that no longer depends on the observed history variables.
To achieve this,  the distributions of the observed history variables $\widetilde{\bm{Z}}(t)$ from those who made a {\it{visit}} at $t$ after weighting need to be balanced with the distributions of $\widetilde{\bm{Z}}(t)$  from the {\it{risk set}}, i.e., those patients under follow-up at $t$ with $\xi(t)=1$. The main problem of using the MLE weights for IIW is that it often fails to achieve the goal of IIW and can result in large imbalances of $\widetilde{\bm{Z}}(t)$  in finite samples, especially if the sample size is small.
To see why this might be the case, recall that $\hat{\bm{\gamma}}$ satisfies the score equations of the Cox model partial likelihood in \eqref{Cox_score}. Thus, the MLE weights are estimated such that 
$\widetilde{\bm{Z}}(t)$   from those who visited at time $t$ are being predicted by the weighted sum of  $\widetilde{\bm{Z}}(t)$ from the risk set, where the $l$th patient in the risk set is weighted by his/her estimated conditional probability of visiting given a visit occurred at $t$,  $\exp\{\bm{\bm{\gamma}}^{\text T} \widetilde{\bm{Z}}_l(t)\}/\sum_{l=1}^n\xi_l(t)\exp\{\bm{\bm{\gamma}}^{\text T} \widetilde{\bm{Z}}_l(t)\}$. As this estimation approach does not align with the goal of IIW, it can lead to the MLE weights creating a highly non-representative sample of the target population. For instance, when the visit process is highly dependent on the observed history variables, extreme values of the MLE weights frequently arise because the estimation approach focuses on prediction, and not on creating a representative pseudo-population. Moreover, if $\widetilde{\bm{Z}}(t)$ is highly associated with the outcome $Y(t)$, the residual imbalances of $\widetilde{\bm{Z}}(t)$ after weighting by the MLE weights and thus the residual selection bias from irregular visit times could lead to large finite-sample estimation error of the marginal regression parameters.
Consequently,  IIWEs with the MLE weights can have large finite-sample biases (relative to their standard errors) and can be inefficient, even when the model for the visit process is correctly specified; see simulation results in \citealp{Pullenayegum2013} and in Section~\ref{simulation}. In Section 1 of the Supplementary Materials, 
we also take an asymptotic viewpoint to demonstrate that using the MLE weights for IIW can still be problematic in large samples when there is model misspecification. }


\subsubsection{Balancing weights estimators}
To improve the robustness and efficiency of the IIWEs, we consider an alternative weight estimation approach by covariate balancing and develop novel balancing weights estimators of $\bm{\beta}_{0}$. Specifically, we propose covariate balancing inverse intensity weights (`balancing weights' in short) of the form
\begin{equation}\label{weight_model_m}
W(t;\bm{\gamma}_\text{b0}, \bm{\phi})=\exp\left[\bm{\gamma}_\text{b0}^{\text T} h\{t, O(t-)\} \right]Q(t ;\bm{\phi}),
\end{equation} 
where  $\bm{\gamma}_\text{b0}$ is a vector of unknown parameters, $h\{t, O(t-)\}$ is a vector of functionals of the observed history $O(t-)$, which can contain functions of time $t$ and the interactions between  $t$ and the observed history variables.

For fixed $\bm{\phi}$,  we propose the following balancing conditions for estimating $\bm{\gamma}_\text{b0}$ and hence $W(t;\bm{\gamma}_\text{b0}, \bm{\phi})$, 
\begin{eqnarray}\label{bal_eq_m}
\sum_{i=1}^n\int_{0}^{\tau}h\{t, O_i(t-)\}\left\{W_i(t;\bm{\gamma}_\text{b}, \bm{\phi})dN^*_i(t) 
- \frac{\xi_i(t)\sum_{l=1}^n Q_l(t ;\bm{\phi})dN^*_l(t)}{\sum_{l=1}^n\xi_l(t)\exp\{\hat{\bm{\gamma}}_s^{\text T} \widetilde{\bm{Z}}_l(t)\}}\right\}=0,  
\end{eqnarray}
where $\frac{\sum_{l=1}^n Q_l(t ;\bm{\phi})dN^*_l(t)}{\sum_{l=1}^n\xi_l(t)\exp\{\hat{\bm{\gamma}}_s^{\text T} \widetilde{\bm{Z}}_l(t)\}}$ is the increment of the modified Breslow estimator in~\eqref{Breslow2}  at time $t$, which is used as an approximation of ${\lambda}_0(t) dt$.
Basically,  \eqref{bal_eq_m}  ensures that weighting $dN^*_i(t)$ with the balancing weights $W_i(t;\bm{\gamma}_\text{b}, \bm{\phi})$ creates a pseudo-population that is representative of a target \emph{at-risk} population at $t$ with a visit intensity equal to the uncensored baseline intensity  $\xi(t) {\lambda}_0(t)$, in terms of the functionals of observed history variables $h\{t, O_i(t-)\}$. In other words, the distributions of $h\{t, O_i(t-)\}$ over time are \emph{exactly} balanced between the weighted observed sample with $dN_i^*(t)=1$ and the target at-risk population with uncensored visit intensity $\xi(t) {\lambda}_0(t)$.  Thus the selection bias due to \tb{imbalances of  $h\{t, O_i(t-)\}$ between the patients who visited at $t$ and the target at-risk population} is removed after weighting. \tb{By enforcing balances of observed history variables in finite samples, the balancing weights estimators would typically have smaller mean squared errors than their counterparts with the MLE weights.}

Assuming that the true visit intensity follows the model in~\eqref{sfmodel}, \tb{in Section 2 of the Supplementary Materials,} we show that the expectations of the left-hand side of \eqref{bal_eq_m} are equal to zero if $W_i(t;\bm{\gamma}_\text{b}, \bm{\phi})$ is replaced by the true inverse intensity weight $\exp[-{\bm{\gamma}}_{s0}^{\text T} \widetilde{\bm{Z}}_i(t)]Q_i(t ;\bm{\phi})$  and $\hat{\bm{\gamma}}_s$ is replaced by  the true regression parameter vector $\bm{\gamma}_{s0}$. Thus the estimator of the balancing weights in \eqref{bal_eq_m} is consistent for the true inverse intensity weights if the weight model in \eqref{weight_model_m} includes the correct observed history variables $\widetilde{\bm{Z}}_i(t)$,  the selection function is correctly specified and the Cox model with estimates used in the modified Breslow estimator is correctly specified.  As a result,  the estimators in~\eqref{IIW_es} using the balancing weights are consistent for $\bm{\beta}_0$. 



For the choice of $h\{t, O_i(t-)\}$,  the same set of variables  $\widetilde{\bm{Z}}_i(t)$ as in~\eqref{sfmodel} can be included. In addition, to prevent extreme weights, we recommend that $h\{t, O_i(t-)\}$  contains $1$ such that   \eqref{bal_eq_m} imposes a constraint on the sum of the balancing weights. Moreover, we can add indicator variables that define a partition of the study follow-up period to $h\{t, O_i(t-)\}$ such that the sums of the balancing weights are bounded within the partitioned periods.  Ideally, for unbiased estimation of $\bm{\beta}_0$,  we would like to impose balancing conditions in~\eqref{bal_eq_m}  to ensure that the distributions of observed history variables are \emph{exactly} balanced for all potential visit times.  Clearly, this is not possible since $t$ is continuous and an infinite number of conditions are required. Instead, we recommend including  
 the interactions between time $t$ (either as a continuous variable or as indicators of partitioned time periods) and $\widetilde{\bm{Z}}_i(t)$ for balancing  in order to provide some parsimony for the conditions in~\eqref{bal_eq_m}.

\subsection{Implementation}\label{implement}
\tb{An \texttt{R} Markdown tutorial to demonstrate the implementation of the proposed methods can be found at \url{https://github.com/lisu-stats/IIW_SA}.} 
For fixed $\bm{\phi}$, the estimation of $\bm{\gamma}_{s0}$ using~\eqref{Cox_weighted} can be implemented by the \texttt{coxph} function in the \texttt{R} package \texttt{survival}, \tb{ where the counting process format \texttt{[start, end)} for the visit process data is used and each row of the data frame contains all observed history variables  that are potentially associated with the visit intensity in the interval specified in \texttt{[start, end)}. } We use a trick by 
defining $Q_i(t;\bm{\phi})$ as case weights if $dN_i^*(t)=1$, while the case weights are equal to $1$ if $dN_i^*(t)=0$.  We add an offset term in the model formula of the \texttt{coxph} function such that if $dN_i^*(t)=1$ the offset is $-\log\{Q_i(t ;\bm{\phi})\}$ and if $dN_i^*(t)=0$ the offset is $0$. Using these offsets prevents the \texttt{coxph} function to recalculate the weighted sum of $\widetilde{\bm{Z}}_i(t)$ in score functions of the Cox partial likelihood using the case weights we just defined. \b{As a result}, the \texttt{coxph} function can implement the estimation using the estimating equation in~\eqref{Cox_weighted}.
The balancing weights in~\eqref{bal_eq_m}  are estimated using the \texttt{R} package \texttt{nleqslv} \citep{Hasselman2016}. The estimation of $\bm{\beta}_0$ can be achieved using the \texttt{glm} or \texttt{lm} function in  \texttt{R} with the MLE weights or the balancing weights.  Finally, we use nonparametric bootstrap for inference, while for small samples jackknife can also be applied.  

\section{Simulation} \label{simulation}
In this section, we conducted a simulation study to assess the performance of the proposed methods in finite samples. Specifically, we compared the IIWEs with the MLE weights and the   balancing weights estimators under the combinations of the following scenarios: 
(1) The true visit process depended on \tb{baseline and time-varying covariates  only  (i.e., the visiting at random assumption is satisfied) or additionally on}  the concurrent outcome, but the selection function was omitted (i.e., assuming visiting at random) or included (i.e., accommodating informative visit times) when estimating the weights. (2) The visit process was highly or moderately dependent on time-varying covariates.  (3) Correct or incorrect (transformed) covariates were included in the visit process models for estimating the weights.

\tb{We simulated both longitudinal continuous outcomes and longitudinal count outcomes. For each scenario examined,  1000 data sets with different sample sizes ($n = 200, 500, 1000$) were generated. Full details of the simulation study can be found in Section 3 of the Supplementary Materials.  }

\tb{Overall, the simulation results showed that when the true visit process was highly dependent on time-varying covariates, the IIWEs with the MLE weights can have large finite-sample biases and mean squared errors (MSEs), even if the Cox model for the visit process was correctly specified. 
In contrast, the balancing weights estimators consistently had smaller MSEs than the IIWEs with the MLE weights under both correct and incorrect model specifications. Notably, the better performance of the balancing weights estimators was most prominent when the true visit process was highly dependent on time-varying covariates but not dependent or weakly dependent on the concurrent outcome  (i.e.,  when the visiting at random is  satisfied or nearly satisfied). This suggests that in these scenarios the balancing weights have the most potential for improving the robustness and efficiency of the IIWEs. }

\section{Analysis of the PsA clinic cohort data}\label{Application}

\subsection{The PsA Clinic cohort}\label{PsA_data}


Recall that we were interested in examining the associations between biologics use and PsA active joint counts in the PsA clinic cohort. Following  \cite{Zhu2017}, we created a sub-cohort by including a PsA patient who started his/her first course of biologics on a specific calendar day and another randomly selected patient who had never taken biologics as of the same calendar day and then assembling data from all such patients during the year 1981-2014.  This sub-cohort contained 414 patients, with the baseline defined at the calendar days when the patients started their first courses of biologics.  The cut-off date of this sub-cohort was 31 December 2014.
Overall, the patients with and without baseline biologics use had an average of 9.75 and 9.31 clinic visits, with mean inter-visit gap times of 7.29 and 7.44 months (standard deviation 3.07 and 3.45), respectively.


\subsection{Models and estimators}\label{psamodels}

Since substantial over-dispersion was observed for active joint counts due to a large proportion (about 50$\%$) of zero counts, we specified a negative binomial model with the log link for the marginal regression of the active joint count $Y(t)$, where the scale of time  $t$ was in calendar years. The positive active joint counts ranged from 1-59, with about 10$\%$ of the counts being above 12. For covariates, we included the indicators of the first year, second year and so on until after the fifth year since baseline, the biologic use at baseline (yes/no) and the interactions between baseline biologics use and the yearly indicators. In addition, we included indicators for calendar year periods $[1980, 2008)$, $[2008, 2012)$ and $[2012, 2015)$ to account for changing clinical practice, where roughly similar numbers of visits were made during these time periods. 
Finally, the following \textit{baseline covariates} were considered in the marginal regression model: gender, PsA disease duration, age, use of non-steroidal anti-inflammatory drugs (NSAIDs)  and disease-modifying anti-rheumatic drugs (DMARDs), erythrocyte sedimentation rate (ESR) with square root transformation,  active joint count and damaged joint count.  \tb{Details of the marginal regression model specification can be found in Section 4.1 of the Supplementary Materials.}

\tb{For the Cox model for the visit process, the scale of time  $t$ was also in calendar years.} We included the following variables in $\widetilde{\bm{Z}}(t)$:  gender, PsA disease duration at baseline, ESR at baseline, active joint count and damaged joint count at baseline,   PsA duration and age at $t$;   the interactions between calendar year periods and the following variables---ESR at the last visit before $t$ (with square root transformation), active joint count and damaged joint count at the last visit before $t$ (transformed using $\log(x+1)$), NSAID use, DMARD use,  biologics use, the median of previous inter-visit times at $t$ \tb{and the time since last visit}. Following \cite{Zhang2020}, we standardised all non-binary variables to have mean zero and standard deviation 0.5. 
We included the same set of variables for estimating the balancing weights, where additionally the indicators of calendar year periods were also included for balancing (i.e., we imposed constraints on the sums of the balancing weights within the calendar time periods $[1980, 2008)$, $[2008, 2012)$ and $[2012, 2015)$). 

We applied three estimators to estimate the marginal regression coefficients:
(1) the GEE estimators without weighting (the \tb{naive} estimator); 
(2) the IIWEs with the MLE weights;
(3) the balancing weights estimators. 
Standard errors of the parameter estimates were obtained by jackknife, as we had a large number of variables included in the analysis and this caused convergence problems when implementing the nonparametric bootstrap. We constructed 95\% Wald confidence intervals based on the jackknife standard errors.

\subsection{Calibrating the sensitivity parameter}

\tb{For sensitivity analysis, we assumed that the selection function followed the specification in~\eqref{sfmodel2} and   $q\{O(t-), Y(t);\bm{\phi}\}= \phi\log\{Y(t)+1\}$. The log transformation was chosen based on the empirical checks of the residuals (see Figure 13 of the Supplementary Materials for the histogram) of a linear regression model of the log-transformed observed active joint count $\log\{Y(t)+1\} \mid dN^*(t)=1$ given observed history variables $\widetilde{\bm{Z}}(t)$ and time $t$, where the time effect was modelled as natural cubic splines with five degrees of freedom. The histogram of these residuals suggested that the normality assumption for the log-transformed observed active joint count was plausible. As discussed in Section~\ref{calibration_proc},  the conditional distribution of the \textit{unobserved} log-transformed active joint count given all observed history variables would also follow a normal distribution with the same variance but a shifted mean. As a result, the complete log-transformed active joint count followed a mixture of normal distributions with equal variances.   }
 
\tb{We calibrated the range of the sensitivity parameter $\phi$ using the procedure outlined in Section~\ref{calibration_proc}. We set the target value of the additional variation explained by the concurrent active joint count to be equal to the partial variance explained by all observed history variables, with ${\rho}^2_{Z \mid t}=0.0315$. To map this target value of $\rho^2_{Y(t) \mid Z,  t}$ to the sensitivity parameter $\phi$, we used the estimated residual standard deviation ($0.349$) of the regression model of the log-transformed observed active joint count (i.e., $\log\{Y(t)+1\} \mid dN^*(t)=1$) given $\widetilde{\bm{Z}}(t)$ and $t$, which is equal to the residual standard deviation for the complete log-transformed active joint count (i.e., $\log\{Y(t)+1\}$) under our selection function specification. We then obtained that $|\phi| \approx 1.132$ using the formula in~\eqref{Franksgamma} and limited the values of the sensitivity parameter up to this magnitude. }

\tb{In the context of the PsA clinic cohort, we would expect that patients currently having a larger number of active joints were more likely to make a clinic visit. Therefore  $\phi>0$ and we set the sensitivity parameter at $\phi=0, 0.19, 0.38, 0.57, 0.76, 0.95,1.14$ and examined its impact on the marginal regression analysis results.} 


\subsection{Results}

\tb{Table 14 in the Supplementary Materials} presented the results of the fitted Cox model for the visit process in the PsA clinic sub-cohort, assuming the visiting at random assumption was satisfied (i.e., $\phi=0$).  \tb{Schoenfeld residual plots from this fitted model (Figure 14 in the Supplementary Materials) did not show patterns with time, thus suggesting that the proportionality assumption was not violated.}  It appears  that patients with more active joints at the last visit were more likely to visit, but those with more damaged joints at the last visit were less likely to visit. Patients with longer median previous inter-visit gap times were less likely to visit. \tb{However, patients were more likely to visit as the time since the last visit increased.}  Finally, visits occurred more frequently as time elapsed since baseline.



\begin{figure}[!p]
\centering\includegraphics[scale=0.5]{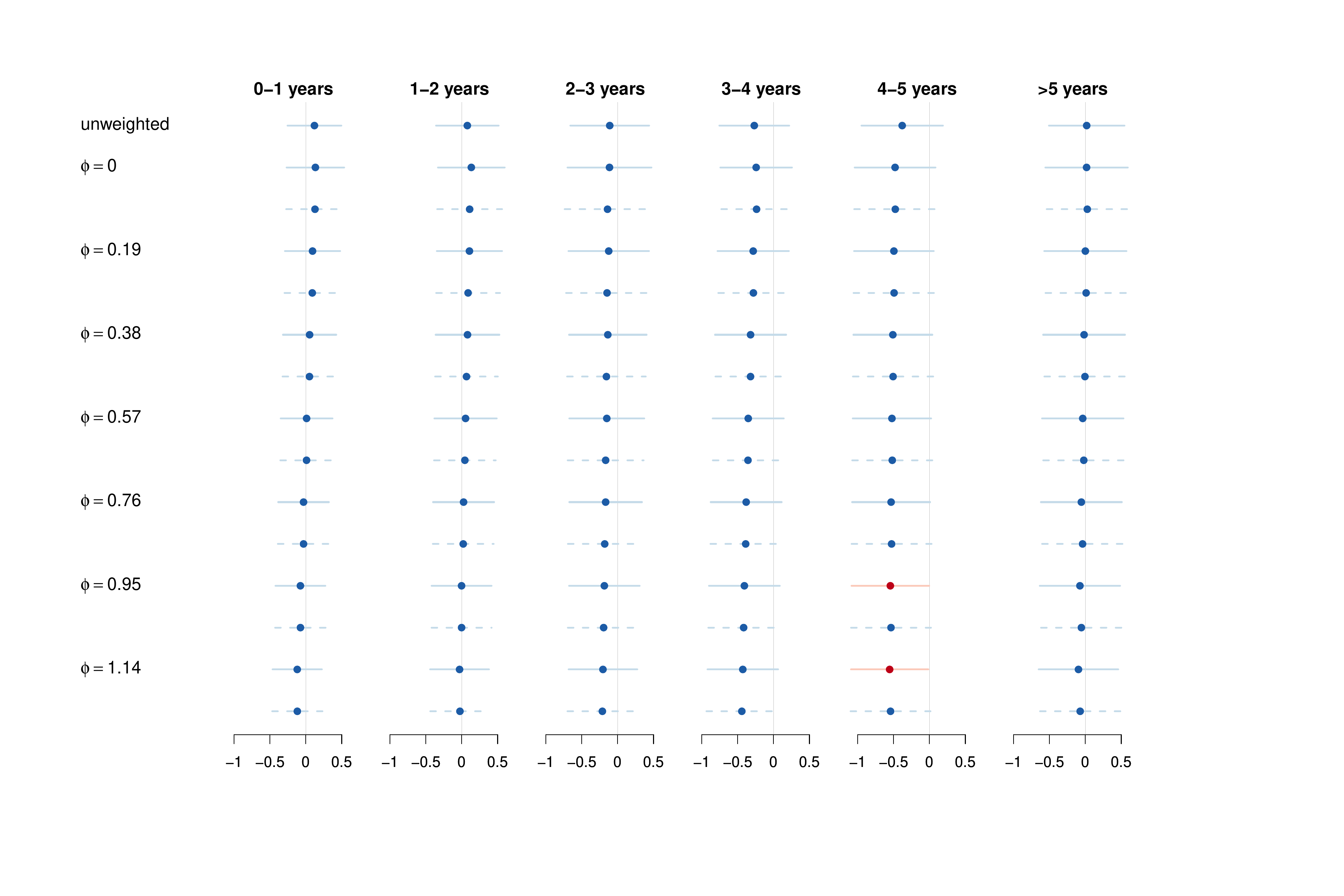}
\caption{Estimates and 95\% confidence intervals (based on jackknife standard errors) of the effect of baseline biologic use on active joint counts during the first year, the second year and so on until after the fifth year since baseline, in the PsA clinic sub-cohort.  Solid lines (\protect\solidblueline, \protect\solidredline): 95\% confidence intervals when using the  IIWEs with the MLE weights or the naive unweighted estimator; dashed lines (\protect\dashblueline,\protect\dashredline): 95\% confidence intervals when using the  balancing weights estimators.  The
estimated effects with 95\% confidence intervals covering zero and not covering zero are in light blue (\protect\solidblueline,\protect\dashblueline)  and pink  (\protect\solidredline,\protect\dashredline), respectively. 
}
 \label{Biologic}
\end{figure}

\tb{Since the sensitivity parameter was set at different values,  it was easier to graphically examine the impact of the sensitivity parameter on the estimates and 95\% confidence intervals (based on jackknife standard errors) of the marginal regression coefficients. 
Figure~\ref{Biologic} presented the estimates and 95\% confidence intervals  of the effect of baseline biologic use on active joint counts during the first year, the second year and so on until after the fifth year since baseline, 
using the naive (unweighted) estimator, the IIWEs with the MLE weights and the balancing weights estimators under the different values of the sensitivity parameter.}

 We noted that the results of the IIWEs using the MLE weights and the balancing weights estimators were almost identical, which might be because the estimated MLE weights and balancing weights were not drastically different. 
For example, when $\phi=0$,  the estimated MLE weights had a minimum of 0.116, a median of 0.287, and a maximum of 1.744, while the minimum, median and maximum of the balancing weights were 0.119, 0.288  and 1.418, respectively. Also, the estimated hazard ratios in the fitted Cox model from  \tb{Table 14 in the Supplementary Materials}  suggested that the visit process was not highly dependent on the observed history variables.  Therefore, the IIWEs with the MLE weights and the balancing weights estimators had similar results. 

\tb{In Figure~\ref{Biologic}, the results based on the naive estimator suggested that patients who took biologics at baseline had lower active joint counts around 3-5 years in follow-up, while the results based on the weighted estimators assuming $\phi=0$ were similar. When the magnitude of the sensitivity parameter was increased,  these negative associations between biologics use and active joint counts became more prominent. When $\phi=0.95$ and $\phi=1.14$, the negative association during the fifth year reached the  $5\%$ significance level or was close to the boundary }

\begin{figure}[!p]
\centering\includegraphics[scale=0.5]{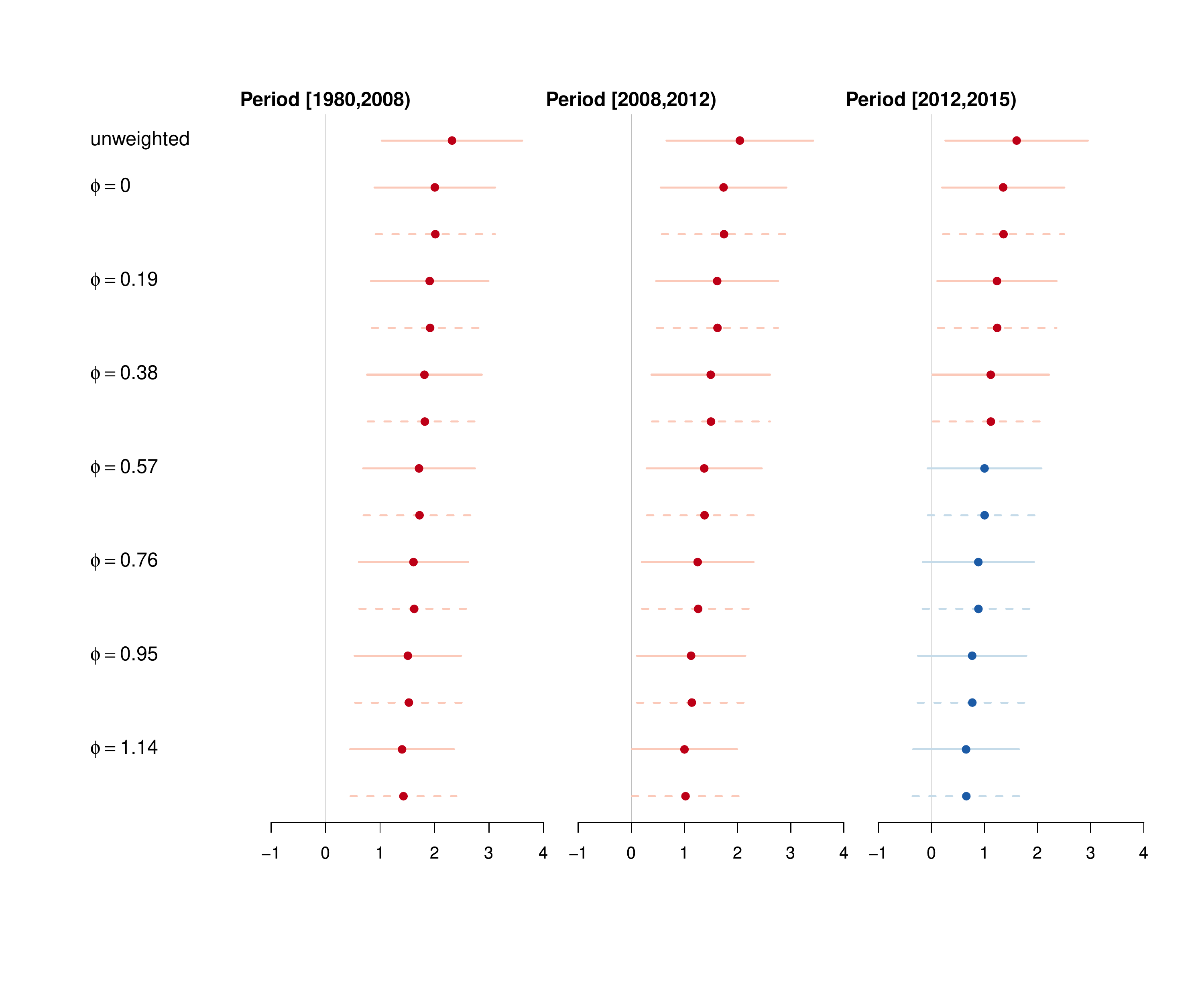}
\caption{Estimates and 95\% confidence intervals (based on jackknife standard errors) of the calendar time period effects on active joint counts in the PsA clinic sub-cohort.  Solid lines (\protect\solidblueline, \protect\solidredline): 95\% confidence intervals when using the  IIWEs with the MLE weights or the naive unweighted estimator; dashed lines (\protect\dashblueline,\protect\dashredline): 95\% confidence intervals when using the balancing weights estimators.  The
estimated effects with 95\% confidence intervals covering zero and not covering zero are in light blue (\protect\solidblueline,\protect\dashblueline)  and pink  (\protect\solidredline,\protect\dashredline), respectively. }
 \label{calendar}
\end{figure}

\tb{Figure~\ref{calendar} presented the estimates and 95\% confidence intervals of the calendar year effects on the active joint counts. It appears that when the sensitivity parameter was increased, the mean active joint count was reduced across all calendar periods.  This was not surprising since we assumed that patients with more active joint counts were more likely to visit, therefore after applying IIW to account for the patients who did not visit but with fewer active joints, the marginal mean of active joint counts given other covariates were adjusted downwards.}  

\tb{The results for the effects of other covariates such as the demographics variables and baseline clinical variables can be found in Figures 15--17 of the Supplementary Materials.  The effects of these variables were not as sensitive as the calendar time effects and the effects of biologics use over time.   }
Since the active joint count is an important clinical marker for PsA disease progression, our findings could inform further cost-benefit analyses of PsA treatment options \citep{DAngiolella2018} while accommodating the uncertainties due to informative visit times in the PsA clinic cohort.

\section{Conclusion and discussion}
In this paper, incorporating novel balancing weights estimators, we developed a new sensitivity analysis approach for accommodating informative visit times in marginal regression analyses of irregular longitudinal data. \tb{In particular,  we proposed a calibration procedure to anchor the range of the sensitivity parameter to the amount of variation in the visit process that could be additionally explained by the concurrent outcome given the observed history variables and time, which would help practitioners gauge the model specification in their sensitivity analyses to the observed information and facilitate the interpretation of the sensitivity parameter.  } Simulations showed that our balancing weights estimators of regression coefficients had improved robustness and efficiency than the IIWEs with the MLE weights, especially when the visit process was highly dependent on the observed time-varying covariates \tb{but not or weakly dependent} on the concurrent longitudinal outcome.  Our proposed sensitivity analysis approach is applicable to the existing IIWEs and can accommodate various types of outcomes within the IIW  frameworks. To facilitate implementation in practice,  we provided an \texttt{R Markdown} tutorial of the proposed methods.


The proposed sensitivity analysis approach can be extended to augmented weighted estimators  \citep{Pullenayegum2013, Smith2022}, where the standard IIWEs with the MLE weights or the balancing weights estimators can be combined with an outcome imputation model. Apart from the selection function included in the model for the visit process, the outcome imputation model for the augmented weighted estimators needs to be fitted by weighting each outcome observation with $Q(t ;\bm{\phi})$. This is similar to the outcome imputation included in the augmented inverse probability weighted estimators for non-ignorable non-monotone missing data \citep{Vansteelandt2007,Wen2018}. 

In addition, there are several future research directions for the proposed methods.  First, \tb{following \cite{Buzkova2008} and \cite{Pullenayegum2013}, we made the non-informative censoring assumption for the at-risk process $\xi(t)$. It is of interest to extend the proposed methods to accommodate an informative at-risk process that depends on the unobserved outcomes conditional on observed information because such a process would affect the parameter estimation in both the visit process model and the marginal regression model of the outcome.  }
\tb{Second, following \cite{Zhu2017}, we focused on assessing the effect of baseline biologics use and did not adjust for treatment switching that occurred post-baseline in the PsA clinic data analysis. For estimating causal treatment effects while correcting for informative monitoring/visit times and treatment switching, \cite{Coulombe2021} recently proposed two weighted estimators of the causal effect of a binary time-varying treatment on a longitudinal outcome. However, \cite{Coulombe2021} also made the visiting at random assumption when developing their estimators.  It would be of interest to extend our methods to the causal effect estimation settings where both treatment switching and informative visit times are present. }
Third, our methods might be computationally infeasible for large data sets because it requires a Cox model to be fitted. Moreover, it is time-consuming to use resampling methods such as bootstrap and jackknife for inference and repeat analyses at different values of the sensitivity parameter.  Therefore, it is important to improve the computational efficiency of the proposed methods in future work.  A recent proposal for fitting Cox models using optimal sub-sampling probabilities to improve computational efficiency \citep{Keret2023} may shed some light on this issue.
Fourth, 
aiming to reduce bias,  the balancing weights estimators could become inefficient when a large number of variables are balanced, some of which explain a small amount of variation of the outcome process. Therefore it would be worth considering a more parsimonious set of variables to balance, so as to trade off a small increase in bias for a large increase in efficiency. 
This is sensible in our sensitivity analysis setting because the balancing weights are for addressing the selection bias from the \emph{observed variables} only and cannot handle the selection bias from the possibly unobserved concurrent outcome. 
We could consider an approach similar to the recently proposed approximate balancing weights methods  \citep{Chattopadhyay2020} or we could balance a summary score of the variables that are predictive of the outcome, e.g., the predicted outcome from an outcome imputation model; see \cite{Han2016} for a similar approach for longitudinal data with non-informative dropout.





\section*{Acknowledgements}
The authors would like to thank Dr Dafna Gladman for providing the PsA clinic data,  Dr Brian Tom for helpful discussions, 
 Dr Yayuan Zhu for providing the \texttt{R} code to create the sub-cohort of the PsA clinic data,  three referees and the associate editor for constructive comments and suggestions. This research was funded by the Medical Research Council  [Unit programme numbers: MC\_UU\_00002/8,  MC\_UU\_00002/10, MC\_UU\_00002/15].  For the purpose of open access, the author has applied a Creative Commons Attribution (CC BY) licence to any Author Accepted Manuscript version arising from this submission.

\bibliographystyle{rss}
\bibliography{iiwref}

\end{document}


\title{Supplementary Material for ``Accommodating  informative visit times for analysing irregular longitudinal data: a sensitivity analysis approach with balancing weights estimators''}
\author{Sean Yiu~  and Li Su\thanks{E-mail address: \texttt{li.su@mrc-bsu.cam.ac.uk}; corresponding author}}
\affil{MRC Biostatistics Unit, School of Clinical Medicine, University of Cambridge,  Cambridge, CB2 0SR, UK}
\date{}
\maketitle

\baselineskip=24pt
\setlength{\parindent}{.25in}

\section{Performance of the MLE weights in large samples}

\tb{In this section, we take an asymptotic viewpoint to demonstrate that using the MLE weights for IIW can still be problematic in large samples when there is model misspecification. For simplicity, we focus on the scenario where the visiting at random assumption is satisfied,  since, with informative visit times,  the selection function specifications are the same when estimating the MLE weights and the balancing weights.}

\tb{When the visiting at random assumption is satisfied such that  ${\lambda}\{t,O(t-), Y(t)\}= {\lambda}\{t, O(t-)\}$, the identity in (4) of the main text becomes 
\begin{equation}\label{IIW_id}
\begin{split}
&\Ex\left[\int_0^{\tau}U\{Y(t),\bm{X}(t);\bm{\beta}_0\}\frac{s(t)}{\lambda\{t, O(t-)\}}dN^*(t)\right]\\
=&\Ex\left(\int_0^{\tau} \Ex \left[U\{Y(t),\bm{X}(t);\bm{\beta}_0\}\frac{s(t)}{\lambda\{t, O(t-)\}}dN^*(t) \,\middle\vert\, O(t-)\right]\right)\\
=&\Ex\left[\int_0^{\tau}U\{Y(t),\bm{X}(t);\bm{\beta}_0\}\frac{s(t)}{\lambda\{t, O(t-)\}}\,\Ex\left\{dN^*(t) \mid O(t-)\right\}\right]\\
=&\Ex\left[\int_0^{\tau}U\{Y(t),\bm{X}(t);\bm{\beta}_0\}s(t) \xi(t)\,dt\right]=0,
\end{split} 
\end{equation} which is equivalent to
\begin{eqnarray}\label{identity}
 & \Ex\left\{\int_{0}^{\tau}U\{Y(t),\bm{X}(t);\bm{\beta}_0\}s(t)\left[\frac{\Ex\{dN^*(t)\mid O(t-)\}}{\lambda\{t,O(t-)\}}-\xi(t)dt\right]\right\}=0.
\end{eqnarray}}

\tb{Suppose that the true visit intensity function is ${\lambda}\{t,O(t-)\}={\lambda}_0(t)\exp\{{\bm{\gamma}}^{\text T}_0 \widetilde{\bm{Z}}_{0}(t)\}$, where $\widetilde{\bm{Z}}_{0}(t)$ are functions of $O(t-)$ and ${\bm{\gamma}}_0$ is the true regression coefficient vector. 
If $\lambda\{t,O(t-)\}$ in~\eqref{identity} is replaced by  ${\lambda}_0^p(t)\exp\{\bm{\bm{\gamma}}_p^{\text T} \widetilde{\bm{Z}}(t)\}$, where $\bm{\bm{\gamma}}_p$ and ${\lambda}_0^p(t)$ are the probability limits of $\hat{\bm{\bm{\gamma}}}$ from maximum partial likelihood estimation and $\hat{\lambda}_0(t)$ from the Breslow estimator (by solving (5) and (6)  in the main text), respectively, and 
$\Ex\{dN^*(t)\mid O(t-)\}$ is replaced by $\xi(t){\lambda}_0(t)\exp\{{\bm{\gamma}}^{\text T}_0 \widetilde{\bm{Z}}_{0}(t)\} dt$, 
the left hand side of \eqref{identity} becomes
\begin{equation}\label{IIW_id_est}
\Ex\left\{\int_{0}^{\tau}U\{Y(t),\bm{X}(t);\bm{\beta}_0\}s(t)\left[\frac{{\lambda}_0(t)\exp\{{\bm{\gamma}}^{\text T}_0 \widetilde{\bm{Z}}_{0}(t)\}}{{\lambda}_0^p(t)\exp\{\bm{\bm{\gamma}}_p^{\text T} \widetilde{\bm{Z}}(t)\}}-1\right]\xi(t)dt\right\}.
\end{equation}
It is therefore clear that the degree to which the identity in~\eqref{IIW_id}  is unsatisfied, i.e., by how much \eqref{IIW_id_est} departs from zero, depends crucially on the relative error of the visit intensity estimation  $$\left|\frac{{\lambda}_0(t)\exp\{\bm{\bm{\gamma}}^{\text T}_0 \widetilde{\bm{Z}}_{0}(t)\}}{{\lambda}_0^p(t)\exp\{\bm{\bm{\gamma}}_p^{\text T} \widetilde{\bm{Z}}(t)\}}-1\right|.$$ In particular, as the relative error decreases so does the upper bound of the absolute value of \eqref{IIW_id_est}.}


\tb{When the Cox model for visit intensity is correctly specified, i.e., we have $\widetilde{\bm{Z}}_{0}(t)=\widetilde{\bm{Z}}(t)$,  the following statements all hold asymptotically: (i) $\bm{\bm{\gamma}}_p=\bm{\bm{\gamma}}_0$, ${\lambda}_0^p(t)={\lambda}_0(t)$; (ii) the relative error disappears; (iii) the identity in~\eqref{IIW_id} is satisfied with the MLE weights in place of true inverse intensity weights; (iv) the IIWE in (3) of the main text is consistent. Thus with large samples, the finite-sample bias problem of the MLE weights described in Section 2.4.1 of the main text is less of a concern. However, when the visit intensity model is {\it{misspecified}}, then the identity in~\eqref{IIW_id} with the MLE weights in place of the  true weights can be far from satisfied, even with large samples. This is because, by minimising the Kullback-Leibler divergence, the maximum partial likelihood approach to weight estimation can ensure that the absolute errors of the conditional probabilities of visiting are small, i.e., $\left|\exp\{\bm{\bm{\gamma}}^{\text T}_0 \widetilde{\bm{Z}}_{0i}(t)\}\right./[\sum_{l=1}^n\xi_l(t)\exp\{\bm{\bm{\gamma}}^{\text T}_0 \widetilde{\bm{Z}}_{0l}(t)\}]-\left.\exp\{\bm{\bm{\gamma}}_p^{\text T} \widetilde{\bm{Z}}_i(t)\}/[\sum_{l=1}^n\xi_l(t)\exp\{\bm{\bm{\gamma}}_p^{\text T} \widetilde{\bm{Z}}_l(t)\}\right|\approx0$, so the Cox model for visit intensity could be mildly misspecified in the usual way of model assessment \cite[]{Kang2007}. But
this does not guarantee that the relative errors $\left|{\lambda}_0(t)\exp\{\bm{\bm{\gamma}}^{\text T}_0 \widetilde{\bm{Z}}_{0i}(t)\}/\left[{\lambda}_0^p(t)\exp\{\bm{\bm{\gamma}}_p^{\text T} \widetilde{\bm{Z}}_i(t)\}\right]-1\right|$ are small. For example, when the true visit intensity is small and underestimated, but both true and estimated intensities are small relative to the sums of the true and estimated intensities for the risk set, respectively, the relative error, but not the absolute error, can be large. Thus, even with large samples and mild misspecification of the Cox model for visit intensity, the IIWE with the MLE weights can perform poorly.  This phenomenon is in line with the theoretical explanation in \cite{Tan2017} for the poor performance of inverse probability weighted estimators for propensity score weighting when weights are estimated by maximum likelihood in logistic regression.}

\section{Proof that the  true inverse intensity weights satisfy the balancing weights conditions in (18) of the main text}

\tb{We assume that the true visit intensity follows the model in (7) of the main text, 
\begin{equation} \label{sfmodel}
{\lambda}\{t, O(t-), Y(t);\bm{\gamma}_{s0}, \bm{\phi}\}={\lambda}_0(t)\exp[\bm{\bm{\gamma}}_{s0}^{\text T} \widetilde{\bm{Z}}(t)+q\{O(t-), Y(t);\bm{\phi}\} ],
\end{equation} where $q\{O(t-), Y(t);\bm{\phi}\}$ is a \textit{known} selection function with a \textit{known} sensitivity parameter vector $\bm{\phi}$. Let $Q(t ;\bm{\phi})=\exp[-q\{O(t-), Y(t);\bm{\phi}\}]$. We have
\[
\Ex\left\{dN^*(t) \mid O(t-), Y(t) \right\}= \xi(t) {\lambda}_0(t)\exp[\bm{\bm{\gamma}}_{s0}^{\text T} \widetilde{\bm{Z}}(t)\} \{Q(t;\bm{\phi})\}^{-1} dt.
\]}

\tb{By setting $s(t)={\lambda}_0(t)$ in (4) of the main text,   the true inverse intensity weight is
\[
\frac{{\lambda}_0(t)}{{\lambda}_0(t)\exp[\bm{\bm{\gamma}}_{s0}^{\text T} \widetilde{\bm{Z}}(t)+q\{O(t-), Y(t);\bm{\phi}\} ],}= \exp\{-{\bm{\gamma}}_{s0}^{\text T} \widetilde{\bm{Z}}(t)\}Q(t ;\bm{\phi}).
\]}

\tb{We replace the balancing weights $W_i(t;\bm{\gamma}_\text{b}, \bm{\phi})$ by the true inverse intensity weights \\$\exp[-{\bm{\gamma}}_{s0}^{\text T} \widetilde{\bm{Z}}_i(t)]Q_i(t;\bm{\phi})$ and  $\hat{\bm{\gamma}}_s$ by the true regression parameter vector $\bm{\gamma}_{s0}$, respectively, in the left-hand side of the balancing weights conditions:
\begin{eqnarray}\label{bal_eq_m}
&&\sum_{i=1}^n\int_{0}^{\tau}h\{t, O_i(t-)\}\left\{W_i(t;\bm{\gamma}_\text{b}, \bm{\phi})dN^*_i(t) 
- \frac{\xi_i(t)\sum_{l=1}^n Q_l(t ;\bm{\phi})dN^*_l(t)}{\sum_{l=1}^n\xi_l(t)\exp\{\hat{\bm{\gamma}}_s^{\text T} \widetilde{\bm{Z}}_l(t)\}}\right\}\\
&=& \sum_{i=1}^n\int_{0}^{\tau}h\{t, O_i(t-)\}\left\{\exp[-{\bm{\gamma}}_{s0}^{\text T} \widetilde{\bm{Z}}_i(t)]Q_i(t;\bm{\phi})dN^*_i(t) 
- \frac{\xi_i(t)\sum_{l=1}^n Q_l(t ;\bm{\phi})dN^*_l(t)}{\sum_{l=1}^n\xi_l(t)\exp\{\bm{\gamma}_{s0}^{\text T} \widetilde{\bm{Z}}_l(t)\}}\right\} \nonumber
\end{eqnarray}}

\tb{Taking expectations of~\eqref{bal_eq_m} given $ O(t-)$ and $Y(t)$ and noting that $\Ex\left\{dN^*(t) \mid O(t-), Y(t) \right\}= \xi(t) {\lambda}_0(t)\exp[\bm{\bm{\gamma}}_{s0}^{\text T} \widetilde{\bm{Z}}(t)\} \{Q(t;\bm{\phi})\}^{-1} dt$, we have
\begin{eqnarray}
&& \sum_{i=1}^n\int_{0}^{\tau}h\{t, O_i(t-)\}\left\{\exp[-{\bm{\gamma}}_{s0}^{\text T} \widetilde{\bm{Z}}_i(t)]Q_i(t;\bm{\phi})\Ex\left\{dN_i^*(t) \mid O_i(t-), Y_i(t) \right\}\right. \nonumber \\
&&- \left.\frac{\xi_i(t)\sum_{l=1}^n Q_l(t ;\bm{\phi})\Ex\left\{dN_l^*(t) \mid O_l(t-), Y_l(t) \right\}}{\sum_{l=1}^n\xi_l(t)\exp\{\bm{\gamma}_{s0}^{\text T} \widetilde{\bm{Z}}_l(t)\}}\right\} \nonumber\\
&=& \sum_{i=1}^n\int_{0}^{\tau}h\{t, O_i(t-)\}\left\{\exp[-{\bm{\gamma}}_{s0}^{\text T} \widetilde{\bm{Z}}_i(t)]Q_i(t;\bm{\phi}) \xi_i(t) {\lambda}_0(t)\exp[\bm{\bm{\gamma}}_{s0}^{\text T} \widetilde{\bm{Z}}_i(t)\} \{Q_i(t;\bm{\phi})\}^{-1} dt  \right. \nonumber \\
&&- \left.\frac{\xi_i(t)\sum_{l=1}^n Q_l(t ;\bm{\phi})\xi_l(t) {\lambda}_0(t)\exp[\bm{\bm{\gamma}}_{s0}^{\text T} \widetilde{\bm{Z}}_l(t)\} \{Q_l(t;\bm{\phi})\}^{-1}dt}{\sum_{l=1}^n\xi_l(t)\exp\{\bm{\gamma}_{s0}^{\text T} \widetilde{\bm{Z}}_l(t)\}}\right\} \nonumber\\
&=& \sum_{i=1}^n\int_{0}^{\tau}h\{t, O_i(t-)\}\left\{ \xi_i(t) {\lambda}_0(t)-  \xi_i(t) {\lambda}_0(t)\right\}dt \nonumber\\
&=&0 \nonumber
\end{eqnarray}
Thus the estimator of the balancing weights in (18) of the main text is consistent for the true inverse intensity weights if the weight model in~\eqref{sfmodel}  includes the correct observed history variables $\widetilde{\bm{Z}}_i(t)$,  the selection function is correctly specified and the Cox model with estimates  used in the modified Breslow estimator is correctly specified.  As a result,  the estimators in (3) of the main text using the balancing weights are consistent for $\bm{\beta}_0$. }

\section{Simulation}

We conduct a simulation study to assess the performance of the proposed methods in finite samples. Specifically, we compare the IIWEs with the MLE weights and the balancing weights estimators under the combinations of the following scenarios: (1) The true visit process depends on baseline and time-varying covariates  \tb{only or additionally on} the current outcome, but the selection function is omitted (i.e., assuming visiting at random) or included (i.e., accommodating informative visit times) when estimating the weights. (2) The visit process is highly or moderately dependent on time-varying covariates.  (3) Correct or incorrect (transformed) covariates are included in the visit process models for estimating the weights.



\subsection{Data generating mechanism}\label{DGM}
The data generating mechanism for one patient is summarised in Table~\ref{simset}. We omit the subscript $i$ for clearer presentation. 
Both longitudinal count and continuous outcomes are considered as the analysis of the  PsA clinic cohort data in Section~4 of the main text focuses on a longitudinal count outcome.

\begin{table}[htp]
\caption{Data generating mechanism for the simulations\label{simset}} 
\centering
\footnotesize{
\begin{tabular}{llll}
\hline
\hline
\multicolumn{1}{l}{\footnotesize\textit{Possible visit times}:} & \multicolumn{3}{l}{ \footnotesize $t=0.01,0.02,\ldots,5$} \\ 
\multicolumn{1}{l}{\footnotesize \textit{Baseline covariates}:} & \multicolumn{3}{l}{\footnotesize $X \stackrel{i.i.d}{\sim} Bernoulli(0.5)$}\\ 		
\multicolumn{1}{l}{\footnotesize \textit{Time varying covariates}:} & \multicolumn{3}{l}{ \footnotesize $Z_1(t), Z_2(t) \stackrel{i.i.d}{\sim} N(-X,1)$} \\
&&&\\
\multicolumn{1}{l}{\footnotesize \textit{Continuous outcome}:} & \multicolumn{3}{l}{ \footnotesize $Y(t)= 5+Z_1(t)+Z_2(t)-0.5Z_1(t)Z_2(t)-2X-0.5t+\epsilon(t)$, $\epsilon(t) \stackrel{i.i.d}{\sim} N(0,0.25)$ } \\
&&&\\
\multicolumn{1}{l}{\footnotesize \textit{Count outcome}:~~~ }& \multicolumn{3}{l}{\footnotesize $Y(t)\stackrel{i.i.d}{\sim} \mbox{NegBin}\{\mu(t),\theta\}$ with p.m.f. 
\(\displaystyle \frac{\bm{\Gamma}(y+1/\theta)}{\bm{\Gamma}(1/\theta)y!}\left\{\frac{\theta\mu(t)}{1+\theta\mu(t)}\right\}\left\{\frac{1}{1+\theta\mu(t)}\right\}^{1/\theta}, \) }
 \\
\multicolumn{1}{l}{\footnotesize }& \multicolumn{3}{l}{\footnotesize mean $\mu(t)=\exp\{1.69+Z_1(t)+Z_2(t)-0.5Z_1(t)Z_2(t)+0.67X-0.5t\}$,}\\
\multicolumn{1}{l}{\footnotesize }& \multicolumn{3}{l}{\footnotesize over-dispersion parameter $\theta=0.5$}\\
					&&&\\
\multicolumn{1}{l}{\footnotesize \textit{Visiting process}:~} &  \multicolumn{3}{l}{{\footnotesize $dN(t)  \sim \text{Bernoulli}\{\pi (t)\}$,}}   \\			
			& \multicolumn{3}{l}{{\footnotesize $\pi (t)= \min\left(1,\exp\left[-3.05-2t+\gamma_z Z_1(t)+ \gamma_z Z_2(t) + 0.5 Z_1(t)Z_2(t) + X+ \phi S\{Y(t)\}\right]\right)$},} \\
& \multicolumn{3}{l}{\footnotesize $S\{Y(t)\}= Y(t)$ if $Y(t)$ is continuous,} \\
& \multicolumn{3}{l}{\footnotesize $S\{Y(t)\}= \log\{ Y(t) +1\} $ if $Y(t)$ is the count outcome.} \\
   	&&&\\
&\multicolumn{3}{l}{\footnotesize \textit{{Sensitivity parameter}:} \tb{$\phi=0$ (visiting at random assumption is satisfied)}} \\
&\multicolumn{3}{l}{~~~~~~~~~~~~~~~~~~~~~~~~~~~~~~$\phi=0.15, 0.3, 0.45, 0.6$} \\

&\multicolumn{3}{l}{\footnotesize \textit{{Dependence on covariates}:} $\gamma_z=0.5, 1.25$} \\
&&&\\
\multicolumn{1}{l}{\footnotesize \tb{\textit{Observed outcomes}:~}} & \multicolumn{3}{l}{{\footnotesize \tb{$Y(t) \mid d N(t)=1$; $Y(t)$ is only observable when $d N(t)=1$.}}}   \\
			&&&\\
\multicolumn{1}{l}{\footnotesize \textit{Transformed covariates}:} &    \multicolumn{3}{l}{\footnotesize $Z^*_1(t)=Z_1(t)-Z_2(t)$} \\
&    \multicolumn{3}{l}{\footnotesize $Z^*_2(t)=Z_2(t)+e(t)$, $e(t)$ $\stackrel{i.i.d}{\sim}$ $N(0, 0.01)$} \\
\hline
\end{tabular}}
\end{table}

In this simulation setup, we have one baseline covariate $X$,  two time-varying covariates $Z_1(t)$ and $Z_2(t)$ as well as the time variable $t$, which affect visiting at $t$ and also the outcome  $Y(t)$. The dependence of the visit process on the time-varying covariates is characterised by the parameter $\gamma_z$ in the visit process model, where we set $\gamma_z=0.5, 1.25$ to allow moderate and high levels of such dependence. We use the Bernoulli distribution to approximate simulated data from a Cox model for the visit process since the visit/event intensity is low. 

In addition, the concurrent outcome \tb{is allowed to affect} the visit process. We let $S\{Y(t)\}=Y(t)$ if the outcome is continuous and  $S\{Y(t)\}=\log\{Y(t)+1\}$  for the case of the longitudinal count outcome. The dependence on $S\{Y(t)\}$    is characterised by the sensitivity parameter $\phi$. 
\tb{When we set $\phi=0$, the visiting at random assumption is satisfied. } We also set $\phi =0.15, 0.3, 0.45, 0.6$ such that those patients with larger outcome values are more likely to make a visit regardless of their observed histories.  

For the longitudinal continuous outcome, our aim is to estimate the regression parameters $\beta_1$ and $\beta_2$ in the marginal regression model  $\Ex\{Y(t) \mid X\}=\beta_0+\beta_1 X+ \beta_2 t$. The true values of $\beta_1$ and $\beta_2$ are $-4.5$ and $-0.5$, respectively. 
For the longitudinal count outcome, we are interested in the parameters $\Tilde{\beta}_1$ and $\Tilde{\beta}_2$ in the log-linear model for the outcome mean $\log [\Ex\{Y(t) \mid X\}] =\Tilde{\beta}_0+\Tilde{\beta}_1 X+ \Tilde{\beta}_2 t$. 
The true values of $\Tilde{\beta}_1$ and $\Tilde{\beta}_2$ are $-1$ and $-0.5$, respectively. \tb{Note that the $Y(t)$ is only observable when $d N(t) =1 $, while here we are trying to estimate the parameters for the marginal regression model for $Y(t)$ without conditioning on $d N(t) =1$.}

Data from each patient are generated
independently. We simulate 1000 data sets with different sample sizes ($n = 200, 500, 1000$) for each scenario with different values of $\gamma_z$ and $\phi$ for both types of outcomes.

\subsection{Model specification and estimators}
We consider the performance of the IIWEs with the MLE weights and the balancing weights estimators under both correct and incorrect model specifications for the visit process. There are two types of model misspecification. The first is to assume visiting at random but in fact, the visit times can be informative with $\phi > 0$. The second is functional form misspecification by including transformed time-varying covariates $Z^*_1(t)$,  $Z^*_2(t)$ and their interaction (see Table~\ref{simset}), instead of $Z_1(t)$, $Z_2(t)$ and $Z_1(t)Z_2(t)$,  in the models for estimating the weights.

Specifically, we will evaluate the performance of the IIWEs using both the MLE weights and the balancing weights estimators, under the following scenarios:
\begin{enumerate}
    \item[1)] The selection function $\phi \cdot S\{Y(t)\}$ is omitted,  correct covariates $Z_1(t)$, $Z_2(t)$, $Z_1(t)Z_2(t)$, and  $X$ are included.
    \item[2)] The selection function $\phi \cdot S\{Y(t)\}$ is omitted,  transformed covariates $Z^*_1(t)$, $Z^*_2(t)$,  $Z^*_1(t)Z_2^*(t)$, and  $X$ are included.
      \item[3)] The selection function $\phi \cdot S\{Y(t)\}$,  correct covariates $Z_1(t)$, $Z_2(t)$, $Z_1(t)Z_2(t)$,  and  $X$ are included.
    \item[4)] The selection function $\phi \cdot S\{Y(t)\}$, transformed covariates $Z^*_1(t)$, $Z^*_2(t)$,  $Z^*_1(t)Z_2^*(t)$, and  $X$ are included.
\end{enumerate}

The MLE weights and the balancing weights are estimated using the methods described in Section~2 of the main text. When the concurrent outcome $S\{Y(t)\}$ is accommodated when estimating the weights, we use the true value of the sensitivity parameter $\phi$ if the correct covariates are included. \tb{Note that when $\phi=0$, scenarios 1) and 3) are equivalent.   } When the transformed covariates are instead used, we fix $\phi$ at the limiting value calculated by fitting a Cox model for the visit process with transformed covariates and $S\{Y(t)\}$ ($\forall t$) to a large data set with $n=1 \times 10^5$.
For estimating the balancing weights, we include $1$,  the visit times $t$ (treated as a continuous variable) and their interactions with the baseline and time-varying covariates in the set of variables for balancing.  

Finally, the models for the marginal mean of $Y(t)$ are correctly specified, where the weighted GEEs in equation~(3) of the main text with different sets of weights are used to estimate the parameters $\beta_0$, $\beta_1$,  $\beta_2$,  $\Tilde{\beta}_0$, $\Tilde{\beta}_1$ and $\Tilde{\beta}_2$. For comparison, we also perform the analysis that uses the complete outcome data at all possible visit times and the \tb{naive} analysis that uses the observed data without weighting.

\subsection{Results}

\begin{sidewaystable}
\caption{\label{Tab2}\footnotesize{Empirical bias, empirical standard deviation (SD), and root mean squared error (RMSE) of the IIWEs with the MLE weights and the balancing weights estimators for $\beta_1$ (group effect) and $\beta_2$ (time effect)  in the marginal regression model of a longitudinal \textbf{continuous} outcome, when the sample size $\bm{n=200}$, the sensitivity parameter $\phi$ is set at $0, 0.15, 0.30, 0.45, 0.60$ and the visit process is \textbf{\emph{highly}} dependant on time-varying covariates with $\gamma_z=1.25$. The naive analysis without weighting and the analysis based on complete data are also presented. }}
\centering
\scriptsize{
\begin{tabular}{llrrrrrrrrrrrrrrr}
  \hline
                                                &&\multicolumn{3}{c}{$\phi=0$}&\multicolumn{3}{c}{$\phi=0.15$}&\multicolumn{3}{c}{$\phi=0.30$}&\multicolumn{3}{c}{$\phi=0.45$}&\multicolumn{3}{c}{$\phi=0.60$}\\
                                                  && Bias & SD & RMSE & Bias & SD & RMSE & Bias & SD & RMSE & Bias & SD & RMSE& Bias & SD & RMSE \\ 
  \hline
  Complete data & $\beta_1$                      & -0.00 & 0.01 & 0.01 & -0.00 & 0.01 & 0.01 & 0.00 & 0.01 & 0.01 & 0.00 & 0.01 & 0.01 & -0.00 & 0.01 & 0.01 \\ 
               & $\beta_2$                       & 0.00 & 0.00 & 0.00 & -0.00 & 0.00 & 0.00 & -0.00 & 0.00 & 0.00 & 0.00 & 0.00 & 0.00 & 0.00 & 0.00 & 0.00 \\ 
 Naive analysis & $\beta_1$                      & 0.99 & 0.12 & 1.00 & 1.32 & 0.08 & 1.32 & 1.57 & 0.06 & 1.57 & 1.75 & 0.05 & 1.75 & 1.90 & 0.04 & 1.90 \\ 
               & $\beta_2$                       & 0.20 & 0.04 & 0.20 & 0.26 & 0.03 & 0.26 & 0.32 & 0.02 & 0.32 & 0.38 & 0.02 & 0.38 & 0.43 & 0.01 & 0.43 \\ 
  $S\{Y(t)\}$ not included    &                       &  &  &  &  &  &  &  &  &  &  &  &  &  &  &  \\ 
  ~~~~~~~~correct $Z(t)$  &                      &  &  &  &  &  &  &  &  &  &  &  &  &  &  &  \\ 
 ~~~~~~~~~~~~~~~~MLE weights& $\beta_1$           & -1.49 & 1.39 & 2.04 & -1.08 & 1.17 & 1.59 & -0.39 & 1.02 & 1.10 & 0.37 & 0.68 & 0.78 & 1.05 & 0.46 & 1.14 \\ 
 ~~~~~~~~~~~~~~~~           & $\beta_2$           & 0.23 & 1.29 & 1.31 & 0.31 & 1.04 & 1.08 & 0.36 & 0.83 & 0.90 & 0.44 & 0.41 & 0.60 & 0.56 & 0.19 & 0.59 \\ 
 ~~~~~~~~~~~~~~~~balancing weights& $\beta_1$    & 0.02 & 0.41 & 0.41 & 0.03 & 0.40 & 0.40 & 0.16 & 0.45 & 0.48 & 0.31 & 0.51 & 0.60 & 0.55 & 0.55 & 0.78 \\ 
                               & $\beta_2$       & -0.00 & 0.09 & 0.09 & 0.00 & 0.09 & 0.09 & 0.00 & 0.09 & 0.09 & 0.01 & 0.09 & 0.09 & 0.01 & 0.10 & 0.10 \\ 
  ~~~~~~~~incorrect $Z(t)$  &                     &  &  &  &  &  &  &  &  &  &  &  &  &  &  &  \\ 
 ~~~~~~~~~~~~~~~~MLE weights& $\beta_1$          & -1.81 & 1.68 & 2.47 & -0.75 & 0.95 & 1.21 & 0.19 & 0.62 & 0.65 & 0.94 & 0.28 & 0.98 & 1.45 & 0.15 & 1.46 \\ 
                             & $\beta_2$          & 0.30 & 1.62 & 1.65 & 0.26 & 0.80 & 0.84 & 0.30 & 0.47 & 0.56 & 0.41 & 0.17 & 0.45 & 0.51 & 0.07 & 0.52 \\ 
  ~~~~~~~~~~~~~~~~balancing weights& $\beta_1$     & -1.20 & 0.84 & 1.46 & -0.55 & 0.72 & 0.91 & 0.01 & 0.66 & 0.66 & 0.44 & 0.60 & 0.74 & 0.84 & 0.54 & 1.00 \\ 
                               & $\beta_2$        & 0.03 & 0.26 & 0.26 & 0.03 & 0.21 & 0.21 & 0.03 & 0.18 & 0.18 & 0.02 & 0.15 & 0.15 & 0.03 & 0.13 & 0.14 \\ 
  $S\{Y(t)\}$  included      &                          &  &  &  &  &  &  &  &  &  &  &  &  &  &  &  \\ 
   ~~~~~~~~correct $Z(t)$  &                      &  &  &  &  &  &  &  &  &  &  &  &  &  &  &  \\ 
  ~~~~~~~~~~~~~~~~MLE weights& $\beta_1$          & -1.49 & 1.39 & 2.04 & -1.03 & 1.10 & 1.51 & -0.50 & 0.94 & 1.07 & -0.03 & 0.78 & 0.78 & 0.44 & 0.65 & 0.79 \\ 
  ~~~~~~~~~~~~~~~~           & $\beta_2$          & 0.23 & 1.29 & 1.31 & 0.31 & 0.97 & 1.02 & 0.34 & 0.77 & 0.85 & 0.42 & 0.55 & 0.69 & 0.54 & 0.43 & 0.69 \\ 
  ~~~~~~~~~~~~~~~~balancing weights& $\beta_1$    & 0.02 & 0.41 & 0.41 & 0.03 & 0.40 & 0.40 & 0.17 & 0.45 & 0.48 & 0.31 & 0.50 & 0.59 & 0.54 & 0.53 & 0.76 \\ 
                                & $\beta_2$        & -0.00 & 0.09 & 0.09 & 0.01 & 0.09 & 0.09 & 0.01 & 0.09 & 0.09 & 0.02 & 0.09 & 0.10 & 0.03 & 0.10 & 0.11 \\ 
 ~~~~~~~~incorrect $Z(t)$  &                      &  &  &  &  &  &  &  &  &  &  &  &  &  &  &  \\ 
  ~~~~~~~~~~~~~~~~MLE weights& $\beta_1$           & -2.10 & 1.86 & 2.81 & -1.40 & 1.36 & 1.95 & -0.71 & 1.13 & 1.33 & -0.14 & 0.95 & 0.96 & 0.45 & 0.80 & 0.92 \\ 
                              & $\beta_2$         & 0.39 & 1.80 & 1.84 & 0.40 & 1.11 & 1.18 & 0.38 & 0.85 & 0.94 & 0.44 & 0.61 & 0.75 & 0.54 & 0.44 & 0.69 \\ 
   ~~~~~~~~~~~~~~~~balancing weights& $\beta_1$   & -1.44 & 0.93 & 1.71 & -0.94 & 0.88 & 1.29 & -0.36 & 0.83 & 0.90 & 0.12 & 0.76 & 0.77 & 0.58 & 0.67 & 0.88 \\ 
                                & $\beta_2$       & 0.04 & 0.29 & 0.29 & 0.06 & 0.25 & 0.26 & 0.06 & 0.23 & 0.23 & 0.06 & 0.19 & 0.20 & 0.06 & 0.16 & 0.17 \\ 
 \hline
\end{tabular}}
\end{sidewaystable}

\begin{figure}[htp]
\centering\includegraphics[scale=0.5]{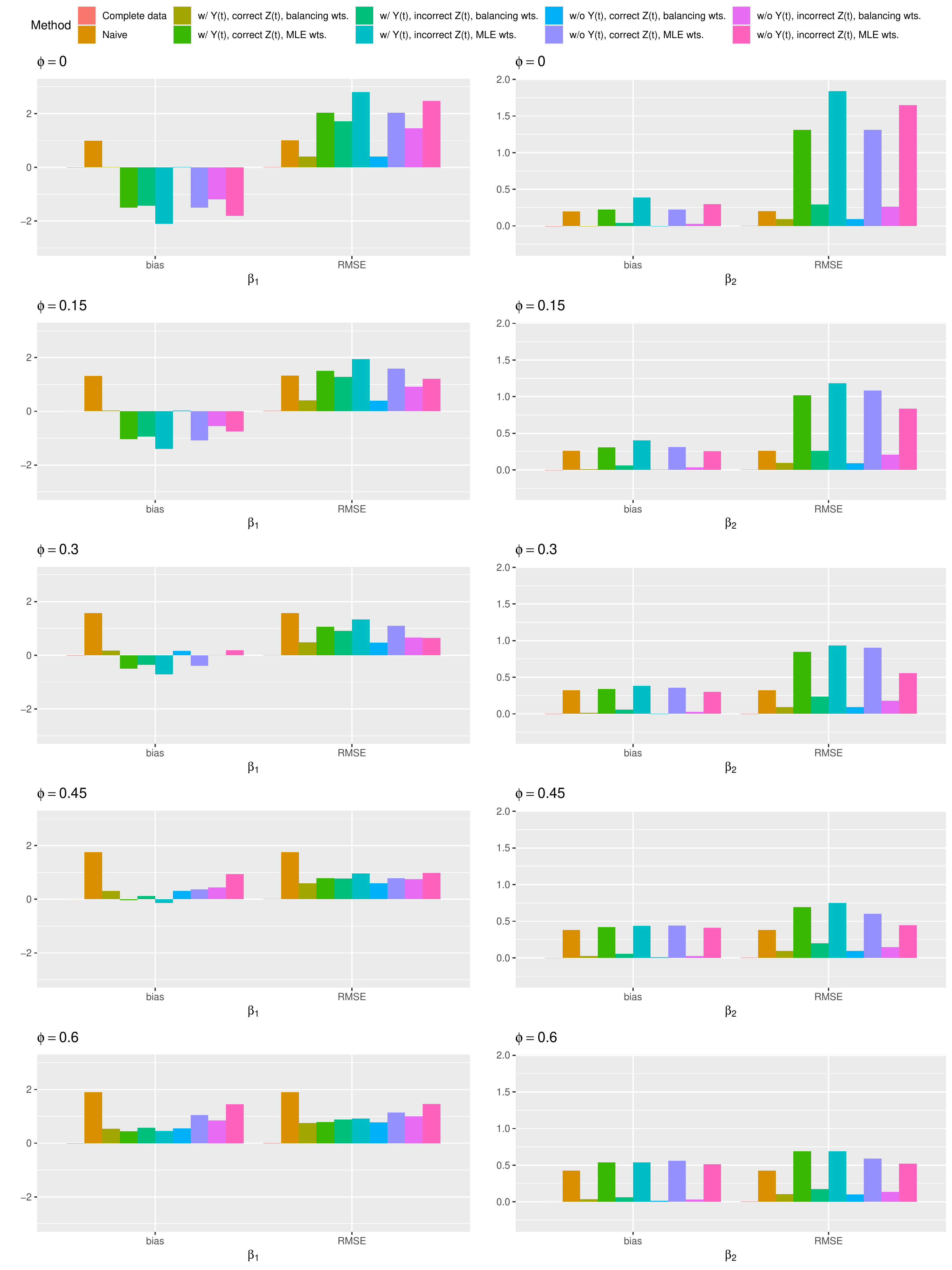}
\caption{Empirical bias and root mean squared error (RMSE) of the IIWEs with the MLE weights and the balancing weights estimators for $\beta_1$ (group effect) and $\beta_2$ (time effect)  in the marginal regression model of a longitudinal \textbf{continuous} outcome, when the sample size $\bm{n=200}$, the sensitivity parameter $\phi$ is set at $0, 0.15, 0.30, 0.45, 0.60$ and the visit process is\textbf{ \emph{highly}} dependant on time-varying covariates with $\gamma_z=1.25$. The naive analysis without weighting and the analysis based on complete data are also presented (note that the results from the complete data analyses are not visible due to the small values close to zeros). }
 \label{Fig1}
\end{figure}

Table~\ref{Tab2} and Figure~\ref{Fig1} present the empirical bias, empirical standard deviation and root mean squared error (RMSE) of the IIWEs with the MLE weights and the balancing weights  estimators for $\beta_1$ and $\beta_2$ in the marginal regression model of a longitudinal continuous outcome when the sample size $n=200$, the sensitivity parameter $\phi$ is set at various values and the visit process is \emph{highly} dependant on time-varying covariates.  \tb{When $\phi=0$ and correct covariates are included (scenarios 1) and 3)), the balancing weights estimators of $\beta_1$ and $\beta_2$ have negligible biases and much smaller RMSEs than their counterparts with the MLE weights. Notably, there are large finite-sample biases (relative to the empirical standard deviation) for the estimators with the MLE weights. When incorrect covariates are included,  the balancing weights estimators still have smaller biases and RMSEs than the estimators with the MLE weights. For both estimators, incorporating an incorrectly specified selection function in scenario 4) seems to have increased the biases and RMSEs observed in scenario 2) induced by the incorrect covariates. }

\tb{For different positive values of $\phi$,} it is also clear that the balancing weights estimators for $\beta_1$ have smaller  RMSEs than their counterparts with the MLE weights in all scenarios compared, while they also have smaller biases (in most of the scenarios) or similar biases. 
The contrast of the performance of the IIWEs with the MLE weights and the balancing weights estimators is more striking for $\beta_2$, where the balancing weights estimators perform consistently better in terms of bias and RMSE than their counterparts with the MLE weights.

We also note that the estimators with the MLE weights have large biases and RMSEs even if the visit model is \emph{correctly} specified with correct covariates and true selection function.  This is not surprising because the visit process is highly dependent on time-varying covariates, thus extreme MLE weights are more likely to occur (especially under model misspecification), which exacerbates the performance of the corresponding IIWEs. Balancing weights have been shown that they can bound the RMSEs of inverse probability weighted estimators under model misspecification in the settings with point treatments or missing data in cross-sectional studies \citep{Tan2017}. Our simulation results confirm that the proposed balancing weights also reduce the RMSEs of the weighted estimators when there is model misspecification and the visit process is highly dependent on the time-varying covariates. 

Interestingly, for both $\beta_1$ and $\beta_2$, the large RMSEs of the estimators with the MLE weights are increasing as the value of the sensitivity parameter $\phi$ decreased. This could be due to that the variation in the visit process is dominated by the time-varying covariates  \tb{when  $\phi$ is small or completely determined by the time-varying covariates when $\phi=0$}, thus leading to more extreme MLE weights. This finding also suggests that the balancing weights estimators have the most potential for improved performance when the visit process is highly dependent on time-varying covariates \tb{but not dependent or weakly dependent on the concurrent outcome (i.e.,  when the visiting at random is  satisfied or nearly satisfied).}

\begin{figure}[htp]
\centering\includegraphics[scale=0.5]{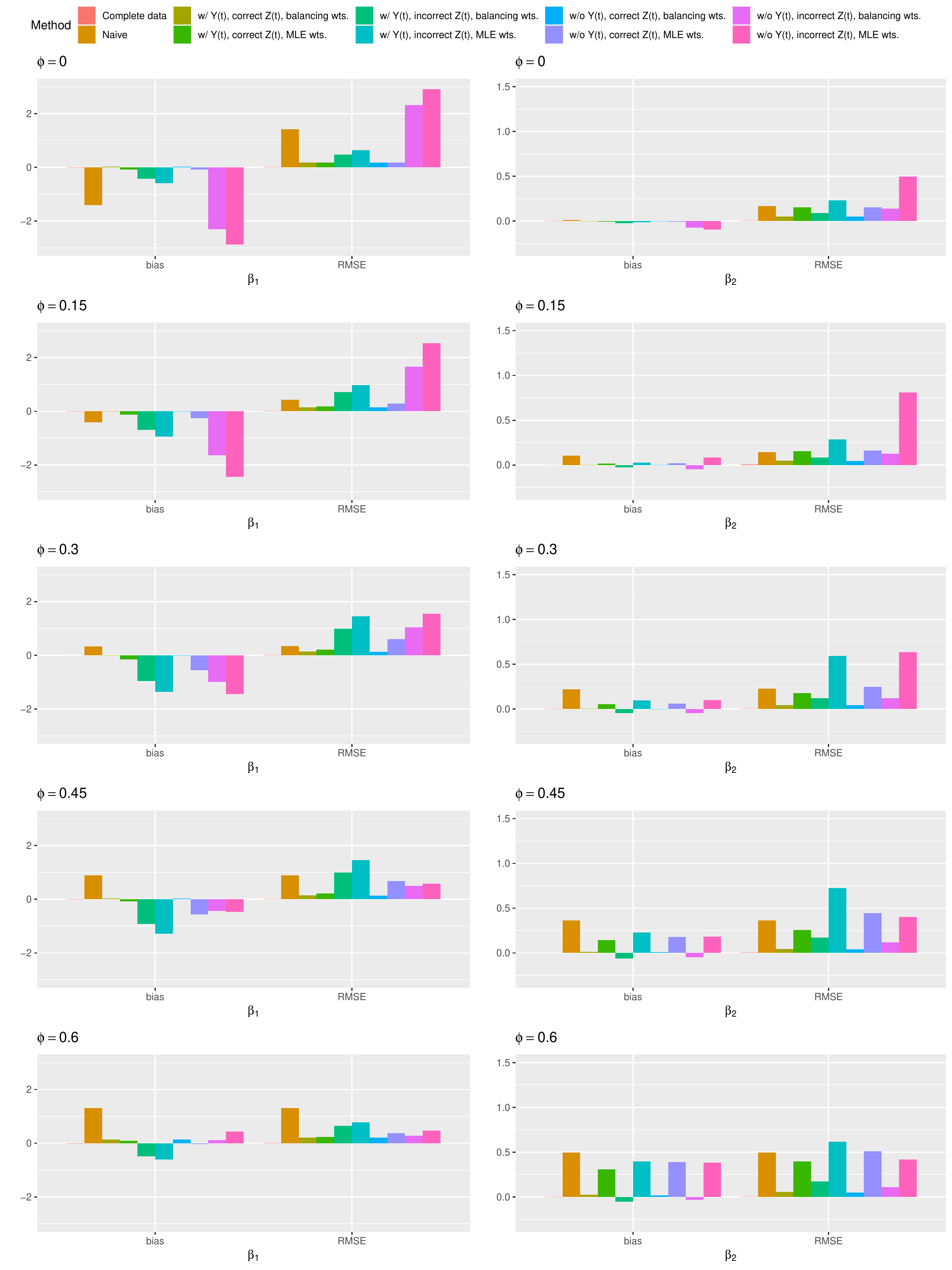}
\caption{Empirical bias and root mean squared error (RMSE) of the IIWEs with the MLE weights and the balancing weights estimators for $\beta_1$ (group effect)  and $\beta_2$ (time effect) in the marginal regression model of a longitudinal \textbf{continuous} outcome,  when the sample size $\bm{n=200}$,  the sensitivity parameter $\phi$ is set at $0.15, 0.30, 0.45, 0.60$ and the visit process is \textbf{\emph{moderately}} dependant on time-varying covariates with $\gamma_z=0.5$. The naive analysis without weighting and the analysis based on complete data are also presented. }
 \label{Fig2}
\end{figure}

Table~\ref{Tab3} and Figure~\ref{Fig2} present the results when the visit process is \emph{moderately} dependent on the time-varying covariates with the sample size $n=200$.  We also note that when correct covariates are included,  the biases and RMSEs of the estimators with the MLE weights are much reduced compared to the setting with the visit process highly dependent on time-varying covariates. But when there is model misspecification (omission of the selection function and inclusion of incorrect covariates), the reduction of the biases and RMSEs of the estimators with the MLE weights is not obvious. On the other hand,  the balancing weights estimators still perform consistently better regardless of the associations of the visit process with the time-varying covariates. 
This suggests that the performance of the balancing weights estimators is less sensitive to the true setup of the visit process.

\begin{figure}[!p]
\centering\includegraphics[scale=0.5]{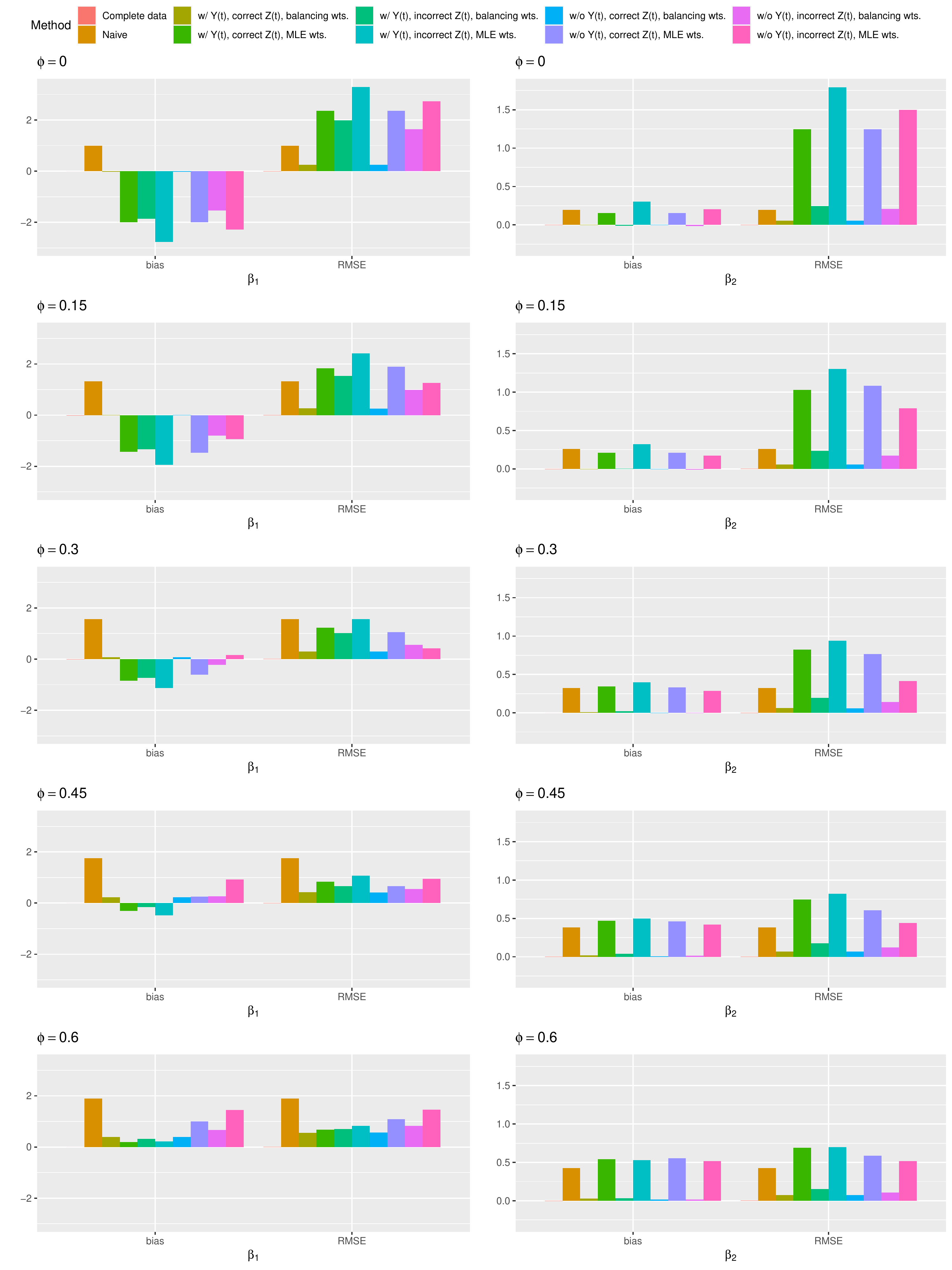}
\caption{Empirical bias and root mean squared error (RMSE) of the IIWEs with the MLE weights and the stable weights for $\beta_1$ and $\beta_2$ in the marginal regression model of a longitudinal \textbf{continuous} outcome, when the sample size $\bm{n=500}$, the sensitivity parameter $\gamma_y$ is set at $0, 0.15, 0.30, 0.45, 0.60$ and the visit process is \textbf{\emph{highly}} dependant on time-varying covariates with $\gamma_z=1.25$. The naive analysis without weighting and the analysis based on complete data are also presented. }
 \label{Fig3}
\end{figure}

\begin{figure}[!p]
\centering\includegraphics[scale=0.5]{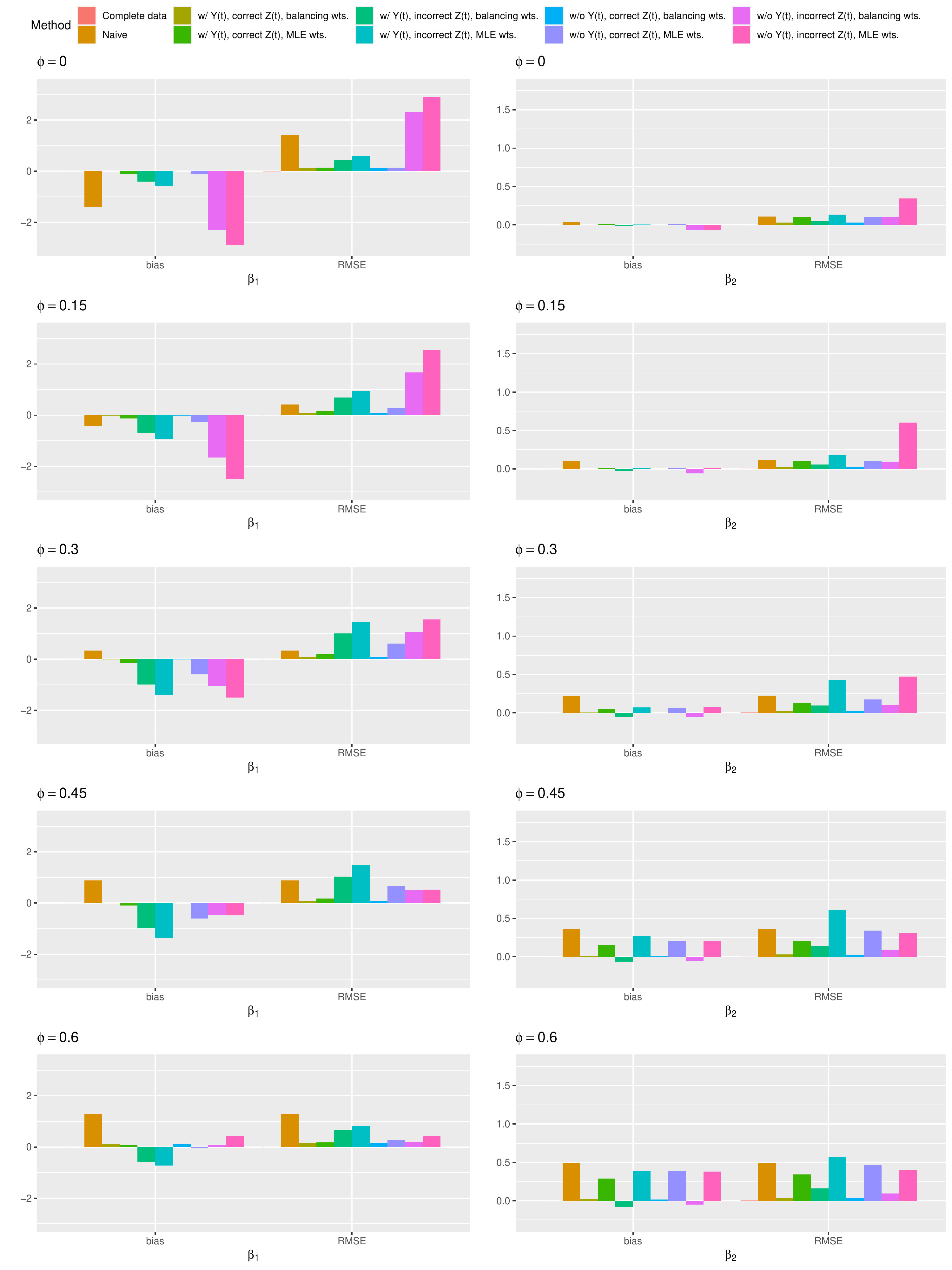}
\caption{Empirical bias and root mean squared error (RMSE) of the IIWEs with the MLE weights and the stable weights for $\beta_1$ and $\beta_2$ in the marginal regression model of a longitudinal \textbf{continuous} outcome,  when the sample size $\bm{n=500}$, when the sensitivity parameter $\gamma_y$ is set at $0, 0.15, 0.30, 0.45, 0.60$ and the visit process is \textbf{\emph{moderately}} dependant on time-varying covariates with $\gamma_z=0.5$. The naive analysis without weighting and the analysis based on complete data are also presented. }
 \label{Fig4}
\end{figure}

\begin{figure}[!p]
\centering\includegraphics[scale=0.5]{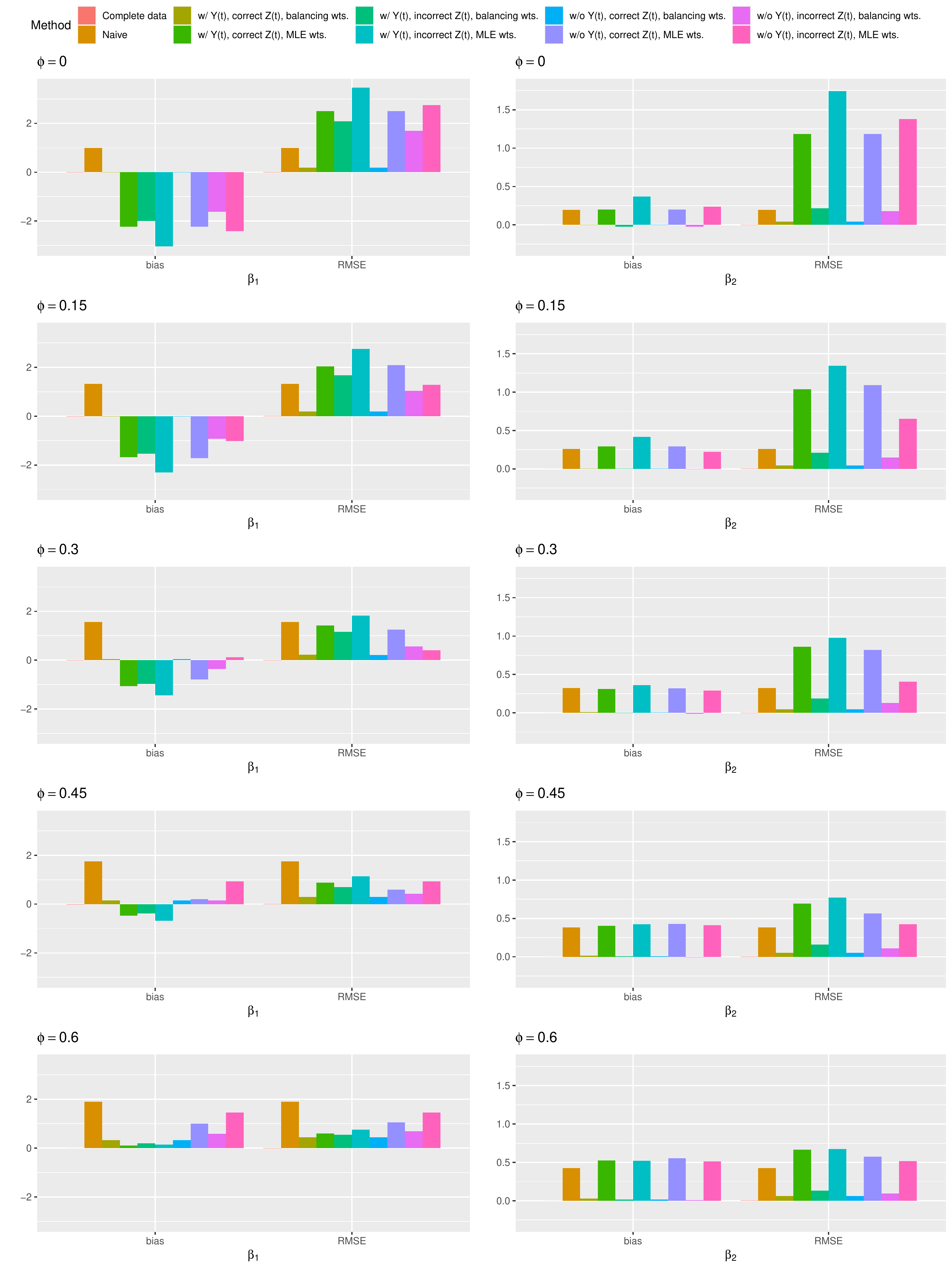}
\caption{Empirical bias and root mean squared error (RMSE) of the IIWEs with the MLE weights and the stable weights for $\beta_1$ and $\beta_2$ in the marginal regression model of a longitudinal \textbf{continuous} outcome, when the sample size $\bm{n=1000}$, the sensitivity parameter $\gamma_y$ is set at $0, 0.15, 0.30, 0.45, 0.60$ and the visit process is \textbf{\emph{highly}} dependant on time-varying covariates with $\gamma_z=1.25$. The naive analysis without weighting and the analysis based on complete data are also presented. }
 \label{Fig5}
\end{figure}

\begin{figure}[!p]
\centering\includegraphics[scale=0.5]{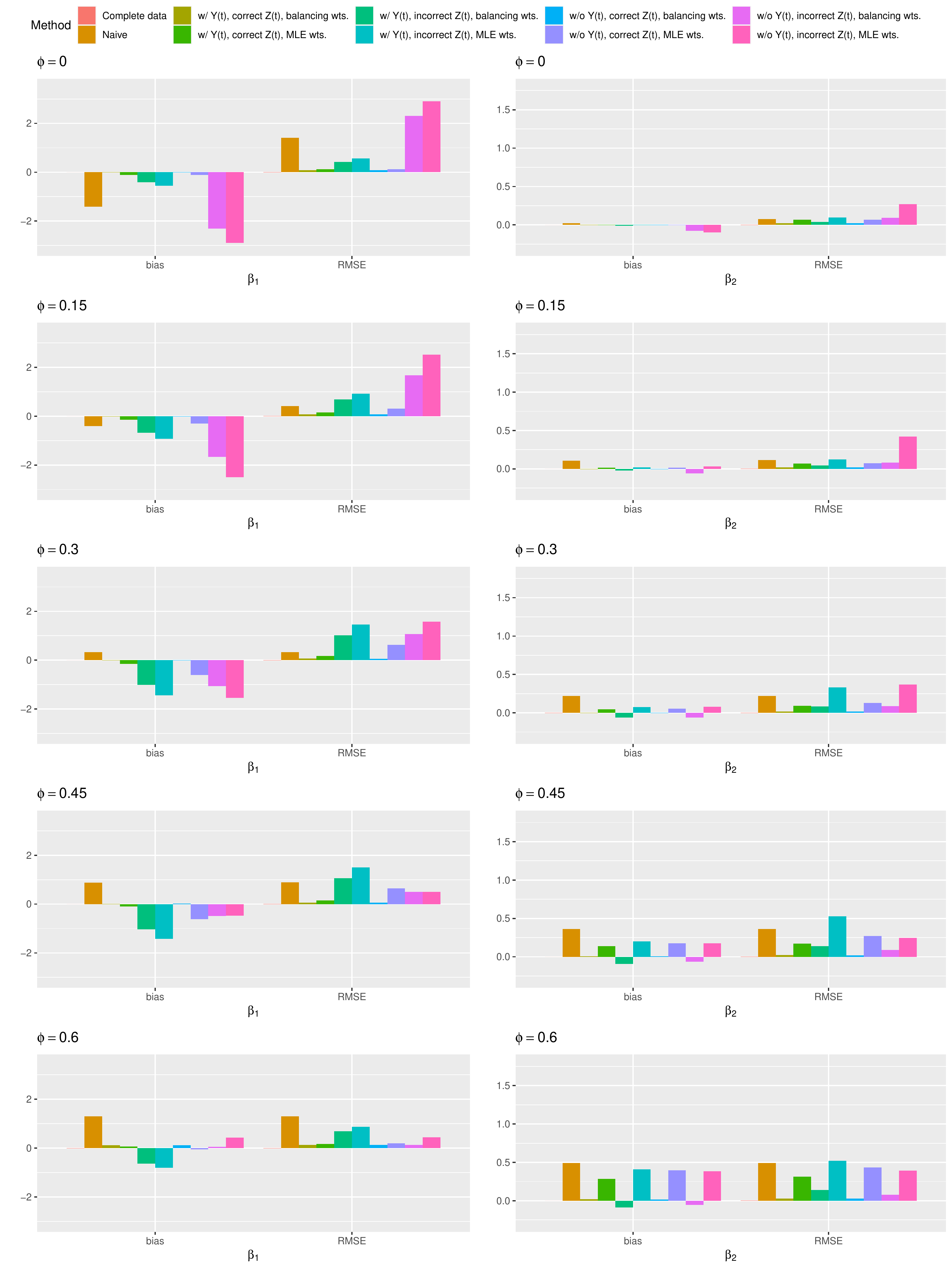}
\caption{Empirical bias and root mean squared error (RMSE) of the IIWEs with the MLE weights and the stable weights for $\beta_1$ and $\beta_2$ in the marginal regression model of a longitudinal \textbf{continuous} outcome, when the sample size $\bm{n=1000}$, the sensitivity parameter $\gamma_y$ is set at $0, 0.15, 0.30, 0.45, 0.60$ and the visit process is \textbf{\emph{moderately}} dependant on time-varying covariates with $\gamma_z=0.5$. The naive analysis without weighting and the analysis based on complete data are also presented. }
 \label{Fig6}
\end{figure}

Similar results for the longitudinal continuous outcome with the sample size $n=500$ and $n=1000$ can be found in  Tables~\ref{econt500}-\ref{ncont1000} and  
Figures~\ref{Fig3}-\ref{Fig6}.  \tb{Remarkably, the large  RMSEs of the IIWEs with the MLE weights under the correct model specification does not improve by the larger sample sizes, while the balancing weights estimators have reduced biases and RMSEs as sample sizes increase.}

\begin{figure}[!p]
\centering\includegraphics[scale=0.5]{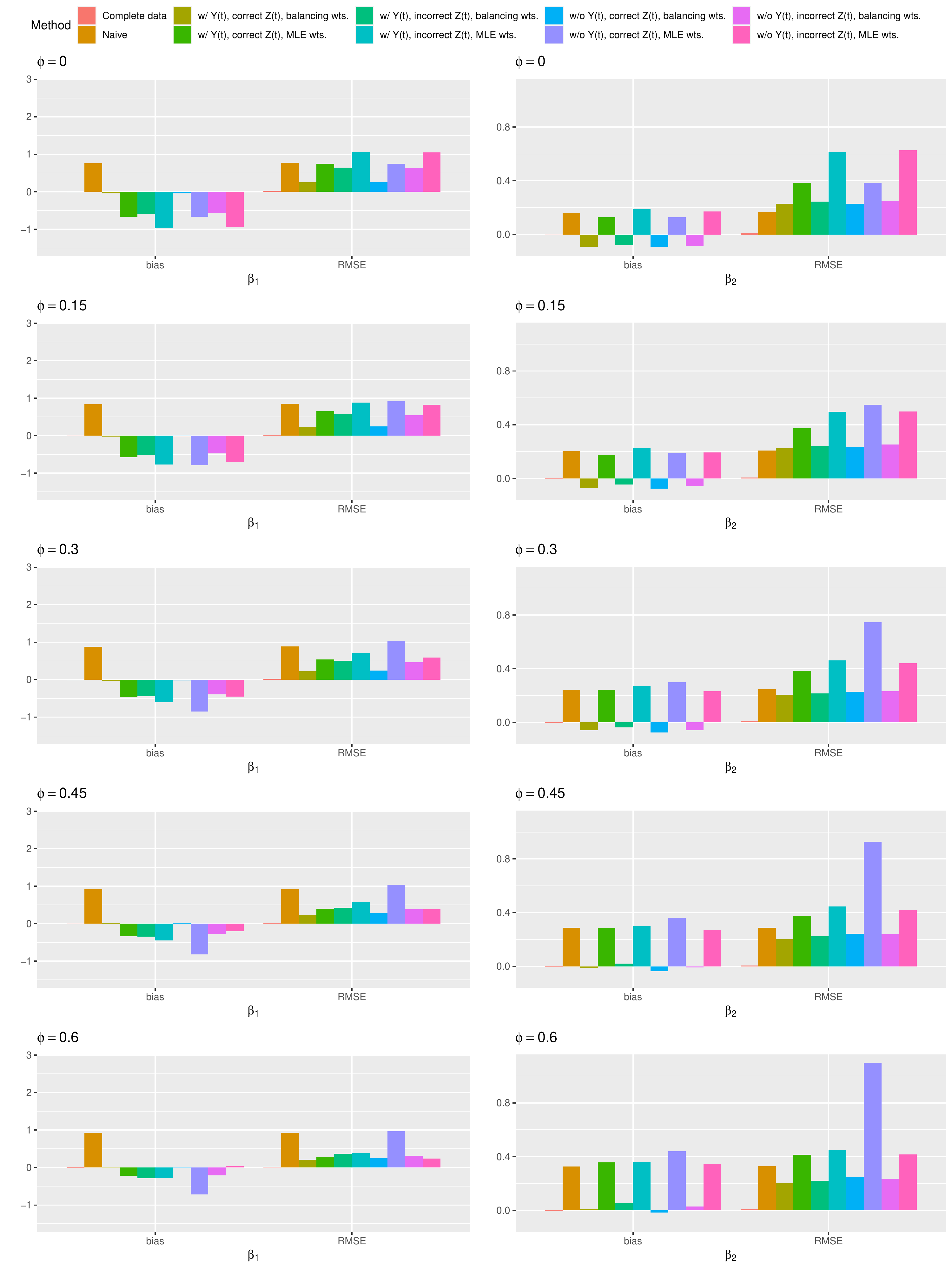}
\caption{Empirical bias and root mean squared error (RMSE) of the IIWEs with the MLE weights and the stable weights for $\tilde{\beta}_1$ and $\tilde{\beta}_2$ in the marginal regression model of a longitudinal \textbf{ count }outcome, when the sample size $\bm{n=200}$, the sensitivity parameter $\gamma_y$ is set at $0, 0.15, 0.30, 0.45, 0.60$ and the visit process is \textbf{\emph{highly}} dependant on time-varying covariates with $\gamma_z=1.25$. The naive analysis without weighting and the analysis based on complete data are also presented. }
 \label{Fig7}
\end{figure}

\begin{figure}[!p]
\centering\includegraphics[scale=0.5]{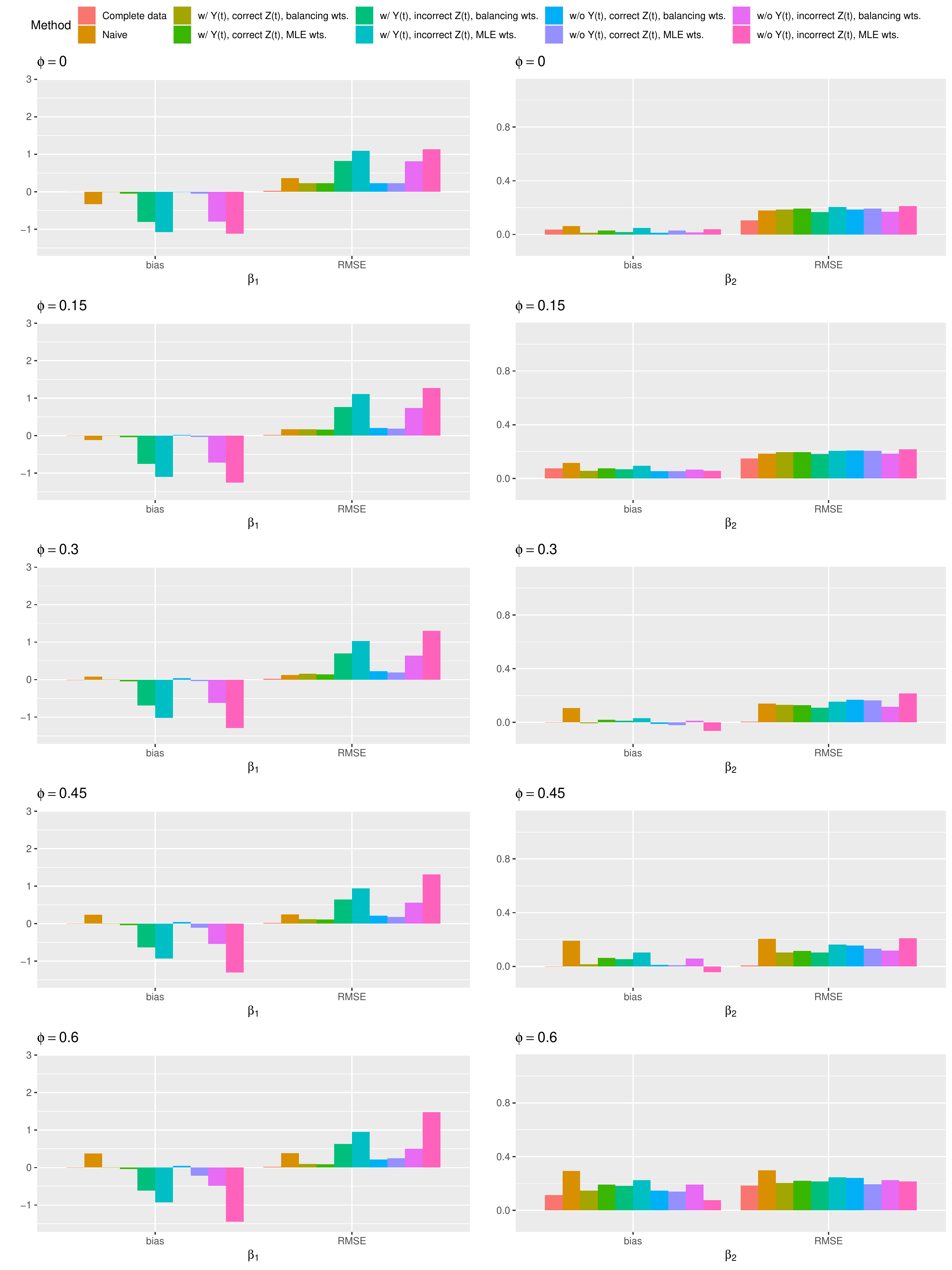}
\caption{Empirical bias and root mean squared error (RMSE) of the IIWEs with the MLE weights and the stable weights for $\tilde{\beta}_1$ and $\tilde{\beta}_2$ in the marginal regression model of a longitudinal\textbf{ count} outcome,  when the sample size $\bm{n=200}$, when the sensitivity parameter $\gamma_y$ is set at $0, 0.15, 0.30, 0.45, 0.60$ and the visit process is \textbf{\emph{moderately}} dependant on time-varying covariates with $\gamma_z=0.5$. The naive analysis without weighting and the analysis based on complete data are also presented. }
 \label{Fig8}
\end{figure}

\begin{figure}[!p]
\centering\includegraphics[scale=0.5]{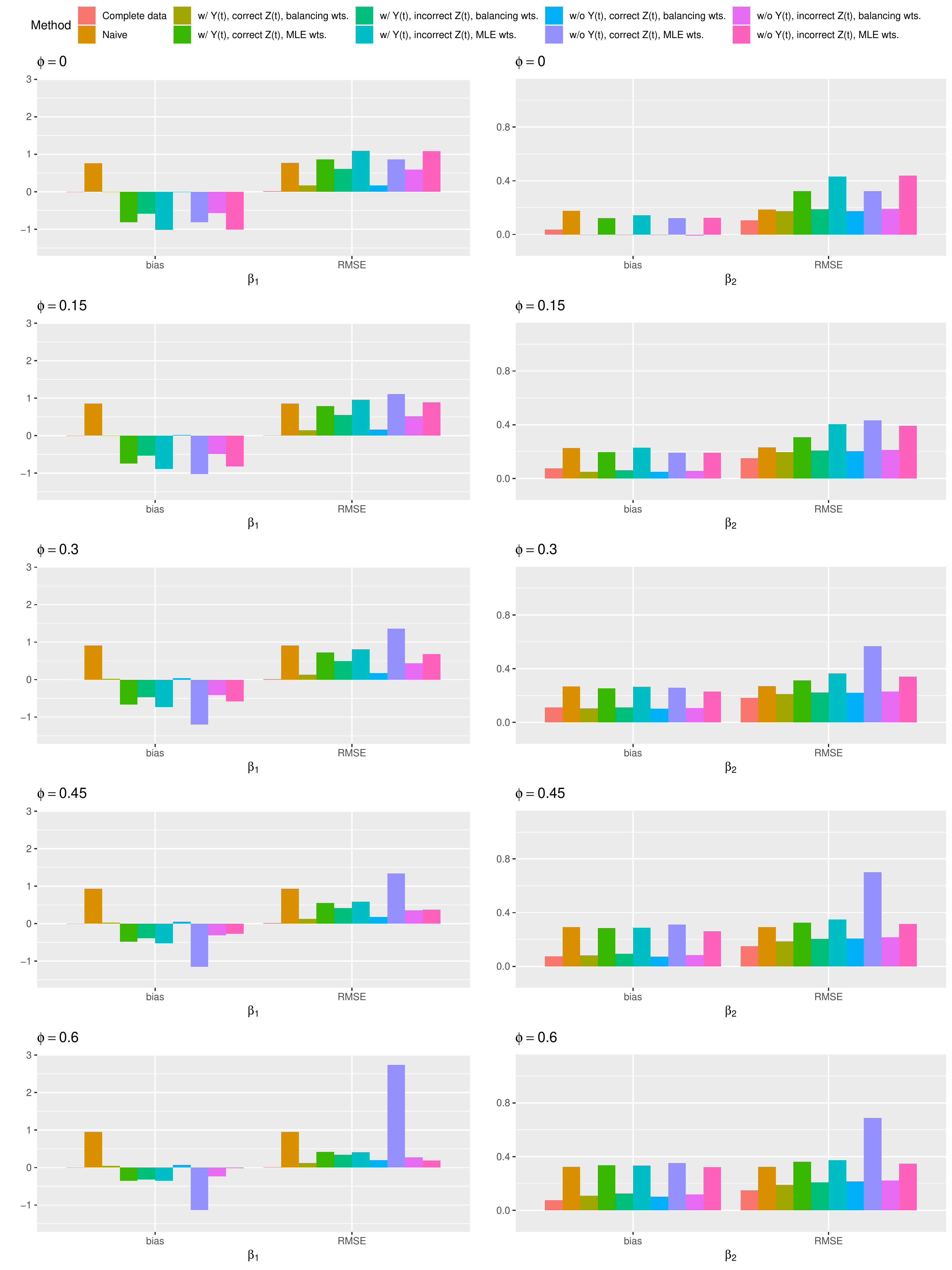}
\caption{Empirical bias and root mean squared error (RMSE) of the IIWEs with the MLE weights and the stable weights for $\tilde{\beta}_1$ and $\tilde{\beta}_2$ in the marginal regression model of a longitudinal \textbf{count} outcome, when the sample size $\bm{n=500}$, the sensitivity parameter $\gamma_y$ is set at $0, 0.15, 0.30, 0.45, 0.60$ and the visit process is \textbf{\emph{highly}} dependant on time-varying covariates with $\gamma_z=1.25$. The naive analysis without weighting and the analysis based on complete data are also presented. }
 \label{Fig9}
\end{figure}

\begin{figure}[!p]
\centering\includegraphics[scale=0.5]{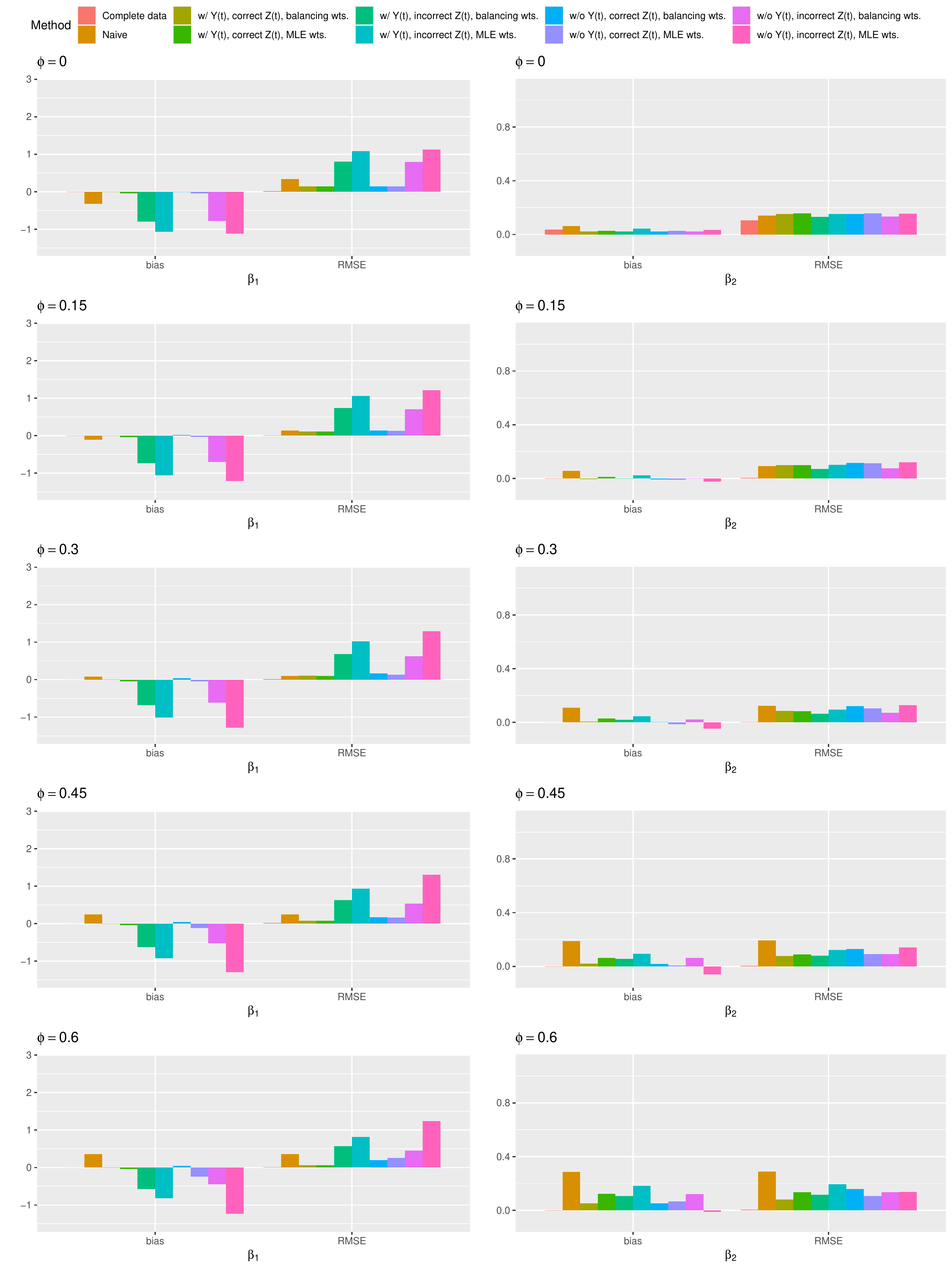}
\caption{Empirical bias and root mean squared error (RMSE) of the IIWEs with the MLE weights and the stable weights for $\tilde{\beta}_1$ and $\tilde{\beta}_2$ in the marginal regression model of a longitudinal\textbf{ count} outcome,  when the sample size $\bm{n=500}$, when the sensitivity parameter $\gamma_y$ is set at $0, 0.15, 0.30, 0.45, 0.60$ and the visit process is \textbf{\emph{moderately}} dependant on time-varying covariates with $\gamma_z=0.5$. The naive analysis without weighting and the analysis based on complete data are also presented. }
 \label{Fig10}
\end{figure}

\begin{figure}[!p]
\centering\includegraphics[scale=0.5]{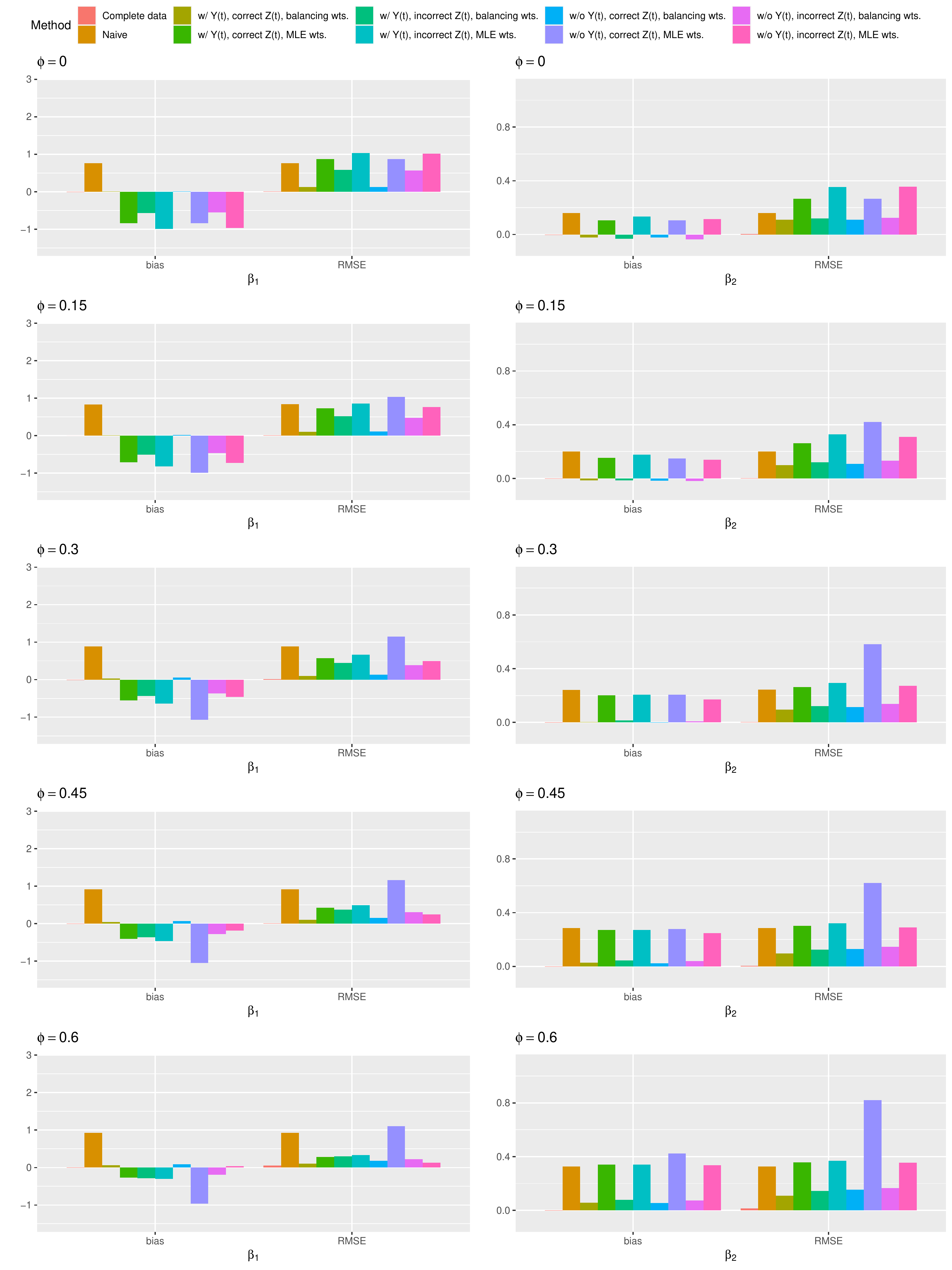}
\caption{Empirical bias and root mean squared error (RMSE) of the IIWEs with the MLE weights and the stable weights for $\tilde{\beta}_1$ and $\tilde{\beta}_2$ in the marginal regression model of a longitudinal \textbf{count} outcome, when the sample size $\bm{n=1000}$, the sensitivity parameter $\gamma_y$ is set at $0, 0.15, 0.30, 0.45, 0.60$ and the visit process is \textbf{\emph{highly}} dependant on time-varying covariates with $\gamma_z=1.25$. The naive analysis without weighting and the analysis based on complete data are also presented. }
 \label{Fig11}
\end{figure}

\begin{figure}[!p]
\centering\includegraphics[scale=0.5]{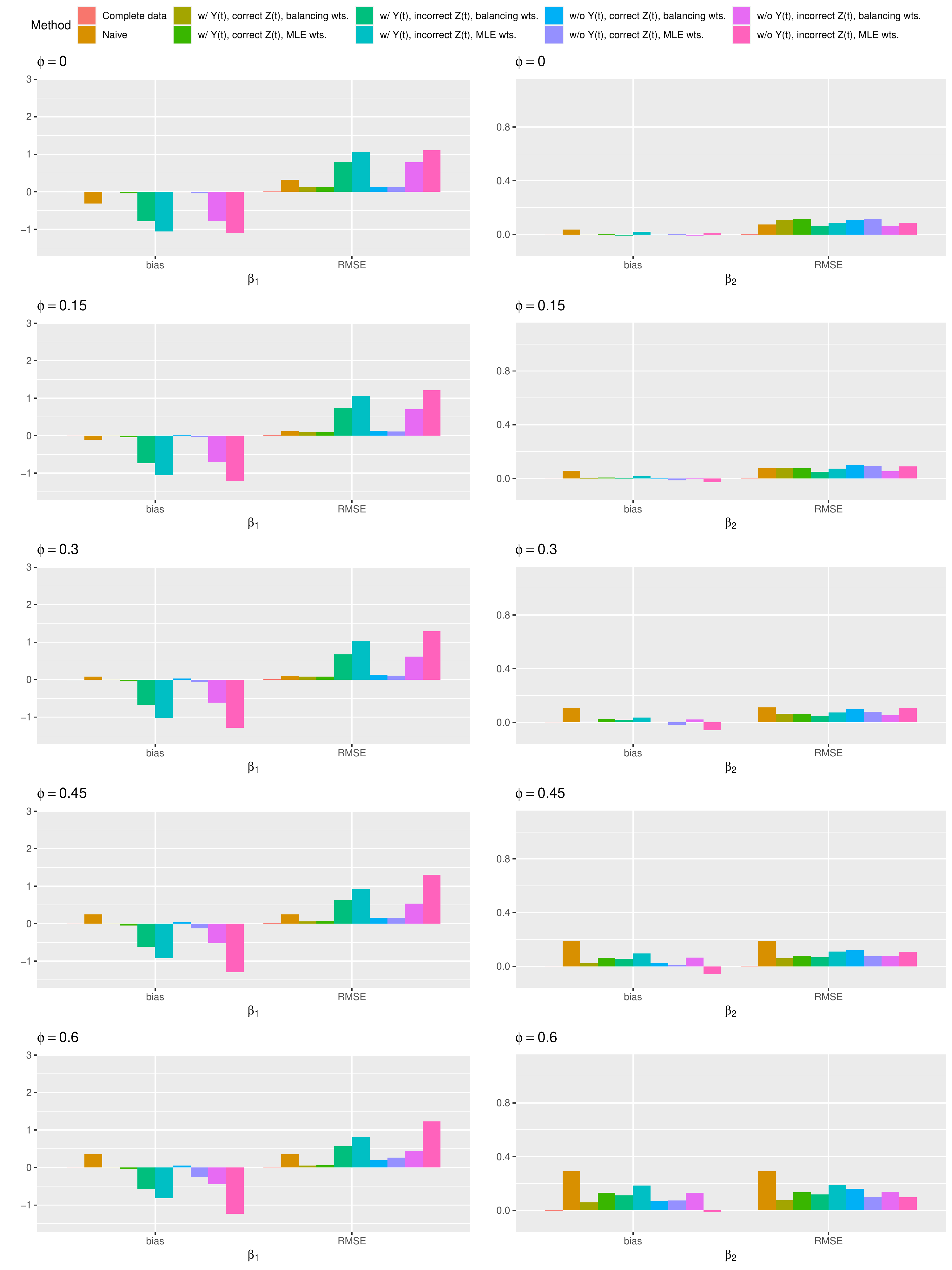}
\caption{Empirical bias and root mean squared error (RMSE) of the IIWEs with the MLE weights and the stable weights for $\tilde{\beta}_1$ and $\tilde{\beta}_2$ in the marginal regression model of a longitudinal \textbf{count} outcome,  when the sample size $\bm{n=1000}$, when the sensitivity parameter $\gamma_y$ is set at $0, 0.15, 0.30, 0.45, 0.60$ and the visit process is \textbf{\emph{moderately} }dependant on time-varying covariates with $\gamma_z=0.5$. The naive analysis without weighting and the analysis based on complete data are also presented. }
 \label{Fig12}
\end{figure}

Results for the longitudinal count outcome can be found in  Tables~\ref{ecount200}-\ref{ncount1000} and  Figures~\ref{Fig7}-\ref{Fig12}. When the visit process is highly dependent on time-varying covariates, the balancing weights estimators still have much better performance than their counterparts with the MLE weights. When the visit process is moderately dependent on time-varying covariates, the improvements brought by the balancing weights are less obvious.

\subsection{Summary}
  
 Overall, our simulations suggest that the balancing weights estimators have better performance than the IIWEs with the MLE weights under both correct and incorrect model specifications, demonstrating their improved robustness and efficiency. The balancing weights have the most potential for improving IIWEs when the true visit process is highly dependent on time-varying covariates but \tb{not dependent or weakly dependent}on the concurrent outcome.


\begin{sidewaystable}
\caption{\label{Tab3}\footnotesize{Empirical bias, empirical standard deviation (SD), and root mean squared error (RMSE) of the IIWEs with the MLE weights and the balancing weights estimators for $\beta_1$ (group effect) and $\beta_2$ (time effect)  in the marginal regression model of a longitudinal \textbf{continuous }outcome, when the sample size $\bm{n=200}$, the sensitivity parameter $\phi$ is set at $0, 0.15, 0.30, 0.45, 0.60$ and the visit process is \textbf{\emph{moderately} }dependant on time-varying covariates with $\gamma_z=0.5$. The naive analysis without weighting and the analysis based on complete data are also presented. }}
\centering
\scriptsize{
\begin{tabular}{llrrrrrrrrrrrrrrr}
  \hline
                                                &&\multicolumn{3}{c}{$\phi=0$}&\multicolumn{3}{c}{$\phi=0.15$}&\multicolumn{3}{c}{$\phi=0.30$}&\multicolumn{3}{c}{$\phi=0.45$}&\multicolumn{3}{c}{$\phi=0.60$}\\
                                                  && Bias & SD & RMSE & Bias & SD & RMSE & Bias & SD & RMSE & Bias & SD & RMSE& Bias & SD & RMSE \\ 
  \hline
  Complete data & $\beta_1$                      & -0.00 & 0.01 & 0.01 & 0.00 & 0.01 & 0.01 & -0.00 & 0.01 & 0.01 & -0.00 & 0.01 & 0.01 & -0.00 & 0.01 & 0.01 \\ 
               & $\beta_2$                       & 0.00 & 0.00 & 0.00 & 0.00 & 0.00 & 0.00 & -0.00 & 0.00 & 0.00 & 0.00 & 0.00 & 0.00 & -0.00 & 0.00 & 0.00 \\ 
 Naive analysis & $\beta_1$                      & -1.40 & 0.16 & 1.41 & -0.41 & 0.13 & 0.43 & 0.33 & 0.09 & 0.34 & 0.89 & 0.07 & 0.89 & 1.31 & 0.05 & 1.31 \\ 
               & $\beta_2$                       & 0.01 & 0.17 & 0.17 & 0.11 & 0.10 & 0.14 & 0.22 & 0.06 & 0.23 & 0.36 & 0.03 & 0.36 & 0.49 & 0.02 & 0.49 \\ 
  $S\{Y(t)\}$ not included    &                       &  &  &  &  &  &  &  &  &  &  &  &  &  &  &  \\ 
  ~~~~~~~~correct $Z(t)$  &                      &  &  &  &  &  &  &  &  &  &  &  &  &  &  &  \\ 
 ~~~~~~~~~~~~~~~~MLE weights& $\beta_1$           & -0.08 & 0.16 & 0.18 & -0.25 & 0.14 & 0.29 & -0.56 & 0.22 & 0.60 & -0.57 & 0.36 & 0.68 & -0.03 & 0.37 & 0.37 \\ 
 ~~~~~~~~~~~~~~~~           & $\beta_2$           & -0.01 & 0.15 & 0.15 & 0.02 & 0.16 & 0.16 & 0.06 & 0.24 & 0.25 & 0.18 & 0.41 & 0.45 & 0.39 & 0.33 & 0.51 \\ 
 ~~~~~~~~~~~~~~~~balancing weights& $\beta_1$    & 0.03 & 0.18 & 0.18 & 0.00 & 0.15 & 0.15 & -0.01 & 0.13 & 0.13 & 0.03 & 0.14 & 0.14 & 0.14 & 0.16 & 0.21 \\ 
                               & $\beta_2$       & -0.00 & 0.05 & 0.05 & 0.00 & 0.05 & 0.05 & 0.00 & 0.04 & 0.04 & 0.01 & 0.04 & 0.04 & 0.02 & 0.05 & 0.05 \\ 
  ~~~~~~~~incorrect $Z(t)$  &                     &  &  &  &  &  &  &  &  &  &  &  &  &  &  &  \\ 
 ~~~~~~~~~~~~~~~~MLE weights& $\beta_1$          & -2.88 & 0.44 & 2.91 & -2.44 & 0.67 & 2.53 & -1.45 & 0.55 & 1.55 & -0.47 & 0.34 & 0.58 & 0.43 & 0.19 & 0.47 \\ 
                             & $\beta_2$          & -0.10 & 0.49 & 0.50 & 0.08 & 0.81 & 0.81 & 0.10 & 0.63 & 0.64 & 0.18 & 0.36 & 0.40 & 0.38 & 0.17 & 0.42 \\ 
  ~~~~~~~~~~~~~~~~balancing weights& $\beta_1$     & -2.30 & 0.26 & 2.32 & -1.64 & 0.25 & 1.65 & -1.00 & 0.26 & 1.03 & -0.43 & 0.25 & 0.50 & 0.12 & 0.25 & 0.28 \\ 
                               & $\beta_2$        & -0.07 & 0.12 & 0.14 & -0.05 & 0.12 & 0.13 & -0.05 & 0.11 & 0.12 & -0.05 & 0.11 & 0.12 & -0.03 & 0.10 & 0.11 \\ 
  $S\{Y(t)\}$  included &                               &  &  &  &  &  &  &  &  &  &  &  &  &  &  &  \\ 
   ~~~~~~~~correct $Z(t)$  &                      &  &  &  &  &  &  &  &  &  &  &  &  &  &  &  \\ 
  ~~~~~~~~~~~~~~~~MLE weights& $\beta_1$          & -0.08 & 0.16 & 0.18 & -0.13 & 0.13 & 0.19 & -0.16 & 0.15 & 0.22 & -0.09 & 0.20 & 0.22 & 0.08 & 0.21 & 0.23 \\ 
  ~~~~~~~~~~~~~~~~           & $\beta_2$          & -0.01 & 0.15 & 0.15 & 0.02 & 0.15 & 0.16 & 0.05 & 0.17 & 0.18 & 0.14 & 0.21 & 0.26 & 0.31 & 0.25 & 0.40 \\ 
  ~~~~~~~~~~~~~~~~balancing weights& $\beta_1$    & 0.03 & 0.18 & 0.18 & 0.00 & 0.15 & 0.15 & -0.01 & 0.14 & 0.14 & 0.02 & 0.14 & 0.14 & 0.14 & 0.16 & 0.21 \\ 
                                & $\beta_2$        & -0.00 & 0.05 & 0.05 & 0.00 & 0.05 & 0.05 & 0.00 & 0.04 & 0.04 & 0.01 & 0.04 & 0.04 & 0.03 & 0.05 & 0.06 \\ 
 ~~~~~~~~incorrect $Z(t)$  &                      &  &  &  &  &  &  &  &  &  &  &  &  &  &  &  \\ 
  ~~~~~~~~~~~~~~~~MLE weights& $\beta_1$           & -0.59 & 0.24 & 0.64 & -0.95 & 0.25 & 0.98 & -1.36 & 0.51 & 1.45 & -1.28 & 0.69 & 1.45 & -0.61 & 0.48 & 0.78 \\ 
                              & $\beta_2$         & -0.01 & 0.23 & 0.23 & 0.03 & 0.28 & 0.28 & 0.09 & 0.58 & 0.59 & 0.23 & 0.69 & 0.73 & 0.40 & 0.47 & 0.62 \\ 
   ~~~~~~~~~~~~~~~~balancing weights& $\beta_1$   & -0.43 & 0.20 & 0.47 & -0.69 & 0.18 & 0.71 & -0.96 & 0.26 & 0.99 & -0.93 & 0.38 & 1.00 & -0.48 & 0.43 & 0.64 \\ 
                                & $\beta_2$       & -0.02 & 0.09 & 0.09 & -0.03 & 0.08 & 0.09 & -0.05 & 0.11 & 0.12 & -0.06 & 0.16 & 0.17 & -0.05 & 0.17 & 0.17 \\ 
 \hline
\end{tabular}}
\end{sidewaystable}

\begin{sidewaystable}
\caption{\label{econt500}\footnotesize{Empirical bias, empirical standard deviation (SD), and root mean squared error (RMSE) of the IIWEs with the MLE weights and the balancing weights estimators for $\beta_1$ (group effect) and $\beta_2$ (time effect)  in the marginal regression model of a longitudinal \textbf{continuous} outcome, when the sample size $\bm{n=500}$, the sensitivity parameter $\phi$ is set at $0, 0.15, 0.30, 0.45, 0.60$ and the visit process is \textbf{\emph{highly}} dependant on time-varying covariates with $\gamma_z=1.25$. The naive analysis without weighting and the analysis based on complete data are also presented. }}
\centering
\scriptsize{
\begin{tabular}{llrrrrrrrrrrrrrrr}
  \hline
                                                &&\multicolumn{3}{c}{$\phi=0$}&\multicolumn{3}{c}{$\phi=0.15$}&\multicolumn{3}{c}{$\phi=0.30$}&\multicolumn{3}{c}{$\phi=0.45$}&\multicolumn{3}{c}{$\phi=0.60$}\\
                                                  && Bias & SD & RMSE & Bias & SD & RMSE & Bias & SD & RMSE & Bias & SD & RMSE& Bias & SD & RMSE \\ 
  \hline
  Complete data & $\beta_1$                     & 0.00 & 0.01 & 0.01 & 0.00 & 0.01 & 0.01 & 0.00 & 0.01 & 0.01 & 0.00 & 0.01 & 0.01 & -0.00 & 0.01 & 0.01 \\ 
               & $\beta_2$                      & -0.00 & 0.00 & 0.00 & -0.00 & 0.00 & 0.00 & -0.00 & 0.00 & 0.00 & 0.00 & 0.00 & 0.00 & 0.00 & 0.00 & 0.00 \\ 
 Naive analysis & $\beta_1$                     & 0.99 & 0.07 & 1.00 & 1.32 & 0.05 & 1.32 & 1.57 & 0.04 & 1.57 & 1.75 & 0.03 & 1.75 & 1.90 & 0.02 & 1.90 \\ 
               & $\beta_2$                      & 0.19 & 0.03 & 0.20 & 0.26 & 0.02 & 0.26 & 0.32 & 0.01 & 0.32 & 0.38 & 0.01 & 0.38 & 0.43 & 0.01 & 0.43 \\ 
  $S\{Y(t)\}$ not included    &                      &  &  &  &  &  &  &  &  &  &  &  &  &  &  &  \\ 
  ~~~~~~~~correct $Z(t)$  &                     &  &  &  &  &  &  &  &  &  &  &  &  &  &  &  \\ 
 ~~~~~~~~~~~~~~~~MLE weights& $\beta_1$          & -1.99 & 1.26 & 2.36 & -1.47 & 1.21 & 1.90 & -0.62 & 0.86 & 1.06 & 0.25 & 0.62 & 0.66 & 1.00 & 0.41 & 1.08 \\ 
 ~~~~~~~~~~~~~~~~           & $\beta_2$          & 0.16 & 1.24 & 1.25 & 0.21 & 1.06 & 1.08 & 0.33 & 0.69 & 0.76 & 0.46 & 0.39 & 0.61 & 0.55 & 0.19 & 0.59 \\ 
 ~~~~~~~~~~~~~~~~balancing weights& $\beta_1$   & -0.03 & 0.25 & 0.25 & 0.01 & 0.26 & 0.26 & 0.07 & 0.29 & 0.30 & 0.22 & 0.35 & 0.42 & 0.40 & 0.41 & 0.57 \\ 
                               & $\beta_2$      & 0.00 & 0.06 & 0.06 & -0.00 & 0.06 & 0.06 & 0.00 & 0.06 & 0.06 & 0.01 & 0.07 & 0.07 & 0.02 & 0.07 & 0.07 \\ 
  ~~~~~~~~incorrect $Z(t)$  &                    &  &  &  &  &  &  &  &  &  &  &  &  &  &  &  \\ 
 ~~~~~~~~~~~~~~~~MLE weights& $\beta_1$         & -2.28 & 1.51 & 2.73 & -0.93 & 0.86 & 1.27 & 0.17 & 0.39 & 0.42 & 0.92 & 0.19 & 0.94 & 1.45 & 0.10 & 1.46 \\ 
                             & $\beta_2$         & 0.20 & 1.48 & 1.50 & 0.17 & 0.77 & 0.79 & 0.29 & 0.30 & 0.42 & 0.42 & 0.14 & 0.44 & 0.52 & 0.05 & 0.52 \\ 
  ~~~~~~~~~~~~~~~~balancing weights& $\beta_1$    & -1.53 & 0.59 & 1.64 & -0.80 & 0.57 & 0.98 & -0.23 & 0.51 & 0.56 & 0.27 & 0.48 & 0.55 & 0.67 & 0.48 & 0.82 \\ 
                               & $\beta_2$       & -0.02 & 0.21 & 0.21 & -0.01 & 0.17 & 0.17 & -0.00 & 0.14 & 0.14 & 0.02 & 0.12 & 0.12 & 0.02 & 0.11 & 0.11 \\ 
  $S\{Y(t)\}$  included    &                          &  &  &  &  &  &  &  &  &  &  &  &  &  &  &  \\ 
   ~~~~~~~~correct $Z(t)$  &                     &  &  &  &  &  &  &  &  &  &  &  &  &  &  &  \\ 
  ~~~~~~~~~~~~~~~~MLE weights& $\beta_1$         & -1.99 & 1.26 & 2.36 & -1.43 & 1.14 & 1.83 & -0.84 & 0.90 & 1.23 & -0.31 & 0.78 & 0.84 & 0.20 & 0.65 & 0.68 \\ 
  ~~~~~~~~~~~~~~~~           & $\beta_2$         & 0.16 & 1.24 & 1.25 & 0.21 & 1.01 & 1.03 & 0.34 & 0.75 & 0.82 & 0.47 & 0.58 & 0.75 & 0.54 & 0.43 & 0.69 \\ 
  ~~~~~~~~~~~~~~~~balancing weights& $\beta_1$   & -0.03 & 0.25 & 0.25 & 0.01 & 0.26 & 0.26 & 0.07 & 0.29 & 0.30 & 0.23 & 0.35 & 0.42 & 0.39 & 0.39 & 0.55 \\ 
                                & $\beta_2$       & 0.00 & 0.06 & 0.06 & 0.00 & 0.06 & 0.06 & 0.01 & 0.06 & 0.06 & 0.02 & 0.07 & 0.07 & 0.03 & 0.07 & 0.08 \\ 
 ~~~~~~~~incorrect $Z(t)$  &                     &  &  &  &  &  &  &  &  &  &  &  &  &  &  &  \\ 
  ~~~~~~~~~~~~~~~~MLE weights& $\beta_1$          & -2.76 & 1.78 & 3.29 & -1.94 & 1.44 & 2.41 & -1.13 & 1.09 & 1.57 & -0.48 & 0.96 & 1.07 & 0.22 & 0.80 & 0.83 \\ 
                              & $\beta_2$        & 0.31 & 1.77 & 1.79 & 0.32 & 1.26 & 1.30 & 0.40 & 0.85 & 0.94 & 0.50 & 0.65 & 0.82 & 0.53 & 0.46 & 0.70 \\ 
   ~~~~~~~~~~~~~~~~balancing weights& $\beta_1$  & -1.86 & 0.69 & 1.99 & -1.33 & 0.76 & 1.53 & -0.74 & 0.70 & 1.02 & -0.16 & 0.65 & 0.67 & 0.32 & 0.62 & 0.70 \\ 
                                & $\beta_2$      & -0.01 & 0.24 & 0.24 & 0.00 & 0.23 & 0.23 & 0.02 & 0.19 & 0.19 & 0.04 & 0.17 & 0.17 & 0.03 & 0.15 & 0.15 \\ 
 \hline
\end{tabular}}
\end{sidewaystable}

\begin{sidewaystable}
\caption{\label{ncont500}\footnotesize{Empirical bias, empirical standard deviation (SD), and root mean squared error (RMSE) of the IIWEs with the MLE weights and the balancing weights estimators for $\beta_1$ (group effect) and $\beta_2$ (time effect)  in the marginal regression model of a longitudinal \textbf{continuous} outcome, when the sample size $\bm{n=500}$, the sensitivity parameter $\phi$ is set at $0, 0.15, 0.30, 0.45, 0.60$ and the visit process is \textbf{\emph{moderately}} dependant on time-varying covariates with $\gamma_z=0.5$. The naive analysis without weighting and the analysis based on complete data are also presented. }}
\centering
\scriptsize{
\begin{tabular}{llrrrrrrrrrrrrrrr}
  \hline
                                                &&\multicolumn{3}{c}{$\phi=0$}&\multicolumn{3}{c}{$\phi=0.15$}&\multicolumn{3}{c}{$\phi=0.30$}&\multicolumn{3}{c}{$\phi=0.45$}&\multicolumn{3}{c}{$\phi=0.60$}\\
                                                  && Bias & SD & RMSE & Bias & SD & RMSE & Bias & SD & RMSE & Bias & SD & RMSE& Bias & SD & RMSE \\ 
  \hline
  Complete data & $\beta_1$                     &  -0.00 & 0.01 & 0.01 & 0.00 & 0.01 & 0.01 & -0.00 & 0.01 & 0.01 & 0.00 & 0.01 & 0.01 & -0.00 & 0.01 & 0.01 \\ 
               & $\beta_2$                      & 0.00 & 0.00 & 0.00 & -0.00 & 0.00 & 0.00 & -0.00 & 0.00 & 0.00 & 0.00 & 0.00 & 0.00 & 0.00 & 0.00 & 0.00 \\ 
 Naive analysis & $\beta_1$                     & -1.40 & 0.10 & 1.41 & -0.41 & 0.08 & 0.42 & 0.33 & 0.06 & 0.33 & 0.88 & 0.04 & 0.88 & 1.30 & 0.03 & 1.30 \\ 
               & $\beta_2$                      & 0.03 & 0.10 & 0.11 & 0.10 & 0.06 & 0.12 & 0.22 & 0.04 & 0.22 & 0.36 & 0.02 & 0.37 & 0.49 & 0.01 & 0.49 \\ 
  $S\{Y(t)\}$ not included    &                      &  &  &  &  &  &  &  &  &  &  &  &  &  &  &  \\ 
  ~~~~~~~~correct $Z(t)$  &                     &  &  &  &  &  &  &  &  &  &  &  &  &  &  &  \\ 
 ~~~~~~~~~~~~~~~~MLE weights& $\beta_1$          & -0.10 & 0.10 & 0.14 & -0.28 & 0.10 & 0.30 & -0.60 & 0.14 & 0.61 & -0.61 & 0.25 & 0.66 & -0.04 & 0.27 & 0.27 \\ 
 ~~~~~~~~~~~~~~~~           & $\beta_2$          & 0.01 & 0.10 & 0.10 & 0.01 & 0.11 & 0.11 & 0.06 & 0.16 & 0.18 & 0.20 & 0.27 & 0.34 & 0.39 & 0.26 & 0.47 \\ 
 ~~~~~~~~~~~~~~~~balancing weights& $\beta_1$   & 0.01 & 0.12 & 0.12 & -0.00 & 0.10 & 0.10 & -0.01 & 0.08 & 0.08 & 0.01 & 0.08 & 0.08 & 0.12 & 0.09 & 0.15 \\ 
                               & $\beta_2$      & 0.00 & 0.03 & 0.03 & 0.00 & 0.03 & 0.03 & 0.00 & 0.02 & 0.02 & 0.01 & 0.03 & 0.03 & 0.02 & 0.03 & 0.04 \\ 
  ~~~~~~~~incorrect $Z(t)$  &                    &  &  &  &  &  &  &  &  &  &  &  &  &  &  &  \\ 
 ~~~~~~~~~~~~~~~~MLE weights& $\beta_1$         & -2.89 & 0.27 & 2.91 & -2.49 & 0.47 & 2.53 & -1.51 & 0.37 & 1.55 & -0.48 & 0.23 & 0.53 & 0.43 & 0.12 & 0.45 \\ 
                             & $\beta_2$         & -0.06 & 0.34 & 0.35 & 0.02 & 0.60 & 0.60 & 0.08 & 0.47 & 0.47 & 0.20 & 0.23 & 0.31 & 0.38 & 0.11 & 0.40 \\ 
  ~~~~~~~~~~~~~~~~balancing weights& $\beta_1$    & -2.31 & 0.16 & 2.31 & -1.66 & 0.17 & 1.67 & -1.04 & 0.16 & 1.06 & -0.47 & 0.18 & 0.50 & 0.07 & 0.17 & 0.19 \\ 
                               & $\beta_2$       & -0.07 & 0.07 & 0.10 & -0.06 & 0.08 & 0.10 & -0.06 & 0.08 & 0.10 & -0.05 & 0.08 & 0.09 & -0.05 & 0.08 & 0.09 \\ 
  $S\{Y(t)\}$  included    &                          &  &  &  &  &  &  &  &  &  &  &  &  &  &  &  \\ 
   ~~~~~~~~correct $Z(t)$  &                     &  &  &  &  &  &  &  &  &  &  &  &  &  &  &  \\ 
  ~~~~~~~~~~~~~~~~MLE weights& $\beta_1$         & -0.10 & 0.10 & 0.14 & -0.13 & 0.09 & 0.16 & -0.16 & 0.11 & 0.19 & -0.09 & 0.14 & 0.17 & 0.07 & 0.16 & 0.18 \\ 
  ~~~~~~~~~~~~~~~~           & $\beta_2$         & 0.01 & 0.10 & 0.10 & 0.01 & 0.10 & 0.10 & 0.05 & 0.11 & 0.12 & 0.15 & 0.14 & 0.21 & 0.29 & 0.18 & 0.34 \\ 
  ~~~~~~~~~~~~~~~~balancing weights& $\beta_1$   & 0.01 & 0.12 & 0.12 & -0.00 & 0.10 & 0.10 & -0.01 & 0.08 & 0.08 & 0.01 & 0.08 & 0.09 & 0.12 & 0.10 & 0.15 \\ 
                                & $\beta_2$       & 0.00 & 0.03 & 0.03 & 0.00 & 0.03 & 0.03 & 0.00 & 0.03 & 0.03 & 0.01 & 0.03 & 0.03 & 0.02 & 0.03 & 0.04 \\ 
 ~~~~~~~~incorrect $Z(t)$  &                     &  &  &  &  &  &  &  &  &  &  &  &  &  &  &  \\ 
  ~~~~~~~~~~~~~~~~MLE weights& $\beta_1$          & -0.56 & 0.14 & 0.58 & -0.93 & 0.16 & 0.94 & -1.41 & 0.33 & 1.45 & -1.37 & 0.56 & 1.49 & -0.73 & 0.37 & 0.82 \\ 
                              & $\beta_2$        & 0.01 & 0.13 & 0.13 & 0.01 & 0.18 & 0.18 & 0.07 & 0.42 & 0.43 & 0.27 & 0.54 & 0.61 & 0.39 & 0.42 & 0.57 \\ 
   ~~~~~~~~~~~~~~~~balancing weights& $\beta_1$  & -0.41 & 0.12 & 0.43 & -0.68 & 0.12 & 0.69 & -0.99 & 0.16 & 1.00 & -0.99 & 0.28 & 1.03 & -0.58 & 0.33 & 0.66 \\ 
                                & $\beta_2$      & -0.02 & 0.05 & 0.05 & -0.03 & 0.05 & 0.06 & -0.05 & 0.08 & 0.09 & -0.07 & 0.12 & 0.14 & -0.08 & 0.14 & 0.16 \\ 
 \hline
\end{tabular}}
\end{sidewaystable}

\begin{sidewaystable}
\caption{\label{econt1000}\footnotesize{Empirical bias, empirical standard deviation (SD), and root mean squared error (RMSE) of the IIWEs with the MLE weights and the balancing weights estimators for $\beta_1$ (group effect) and $\beta_2$ (time effect)  in the marginal regression model of a longitudinal \textbf{continuous} outcome, when the sample size $\bm{n=1000}$, the sensitivity parameter $\phi$ is set at $0, 0.15, 0.30, 0.45, 0.60$ and the visit process is \textbf{\emph{highly}} dependant on time-varying covariates with $\gamma_z=1.25$. The naive analysis without weighting and the analysis based on complete data are also presented. }}
\centering
\scriptsize{
\begin{tabular}{llrrrrrrrrrrrrrrr}
  \hline
                                                &&\multicolumn{3}{c}{$\phi=0$}&\multicolumn{3}{c}{$\phi=0.15$}&\multicolumn{3}{c}{$\phi=0.30$}&\multicolumn{3}{c}{$\phi=0.45$}&\multicolumn{3}{c}{$\phi=0.60$}\\
                                                  && Bias & SD & RMSE & Bias & SD & RMSE & Bias & SD & RMSE & Bias & SD & RMSE& Bias & SD & RMSE \\ 
  \hline
  Complete data & $\beta_1$                      & 0.00 & 0.01 & 0.01 & -0.00 & 0.01 & 0.01 & 0.00 & 0.01 & 0.01 & 0.00 & 0.01 & 0.01 & -0.00 & 0.01 & 0.01 \\ 
               & $\beta_2$                       & 0.00 & 0.00 & 0.00 & 0.00 & 0.00 & 0.00 & 0.00 & 0.00 & 0.00 & -0.00 & 0.00 & 0.00 & 0.00 & 0.00 & 0.00 \\ 
 Naive analysis & $\beta_1$                      & 1.00 & 0.05 & 1.00 & 1.32 & 0.04 & 1.33 & 1.57 & 0.03 & 1.57 & 1.75 & 0.02 & 1.75 & 1.90 & 0.02 & 1.90 \\ 
               & $\beta_2$                       & 0.20 & 0.02 & 0.20 & 0.26 & 0.01 & 0.26 & 0.32 & 0.01 & 0.32 & 0.38 & 0.01 & 0.38 & 0.43 & 0.01 & 0.43 \\ 
  $S\{Y(t)\}$ not included    &                       &  &  &  &  &  &  &  &  &  &  &  &  &  &  &  \\ 
  ~~~~~~~~correct $Z(t)$  &                      &  &  &  &  &  &  &  &  &  &  &  &  &  &  &  \\ 
 ~~~~~~~~~~~~~~~~MLE weights& $\beta_1$           & -2.23 & 1.15 & 2.51 & -1.71 & 1.21 & 2.10 & -0.79 & 0.98 & 1.26 & 0.20 & 0.56 & 0.60 & 1.00 & 0.33 & 1.06 \\ 
 ~~~~~~~~~~~~~~~~           & $\beta_2$           & 0.20 & 1.17 & 1.18 & 0.29 & 1.05 & 1.09 & 0.32 & 0.76 & 0.82 & 0.43 & 0.36 & 0.56 & 0.55 & 0.15 & 0.57 \\ 
 ~~~~~~~~~~~~~~~~balancing weights& $\beta_1$    & -0.01 & 0.19 & 0.19 & -0.00 & 0.19 & 0.19 & 0.04 & 0.21 & 0.21 & 0.15 & 0.25 & 0.29 & 0.33 & 0.30 & 0.45 \\ 
                               & $\beta_2$       & 0.00 & 0.04 & 0.04 & 0.00 & 0.04 & 0.04 & 0.00 & 0.05 & 0.05 & 0.01 & 0.05 & 0.05 & 0.02 & 0.06 & 0.06 \\ 
  ~~~~~~~~incorrect $Z(t)$  &                     &  &  &  &  &  &  &  &  &  &  &  &  &  &  &  \\ 
 ~~~~~~~~~~~~~~~~MLE weights& $\beta_1$          & -2.41 & 1.32 & 2.75 & -1.01 & 0.80 & 1.29 & 0.12 & 0.39 & 0.41 & 0.92 & 0.15 & 0.94 & 1.45 & 0.07 & 1.46 \\ 
                             & $\beta_2$          & 0.24 & 1.36 & 1.38 & 0.22 & 0.61 & 0.65 & 0.29 & 0.29 & 0.41 & 0.41 & 0.10 & 0.42 & 0.51 & 0.04 & 0.52 \\ 
  ~~~~~~~~~~~~~~~~balancing weights& $\beta_1$     & -1.62 & 0.49 & 1.70 & -0.92 & 0.47 & 1.03 & -0.36 & 0.43 & 0.56 & 0.14 & 0.40 & 0.43 & 0.59 & 0.36 & 0.69 \\ 
                               & $\beta_2$        & -0.02 & 0.18 & 0.18 & -0.00 & 0.15 & 0.15 & -0.01 & 0.13 & 0.13 & -0.00 & 0.11 & 0.11 & 0.01 & 0.09 & 0.09 \\ 
  $S\{Y(t)\}$  included    &                           &  &  &  &  &  &  &  &  &  &  &  &  &  &  &  \\ 
   ~~~~~~~~correct $Z(t)$  &                      &  &  &  &  &  &  &  &  &  &  &  &  &  &  &  \\ 
  ~~~~~~~~~~~~~~~~MLE weights& $\beta_1$          & -2.23 & 1.15 & 2.51 & -1.68 & 1.15 & 2.04 & -1.06 & 0.95 & 1.43 & -0.48 & 0.74 & 0.88 & 0.10 & 0.58 & 0.59 \\ 
  ~~~~~~~~~~~~~~~~           & $\beta_2$          & 0.20 & 1.17 & 1.18 & 0.29 & 0.99 & 1.04 & 0.31 & 0.80 & 0.86 & 0.40 & 0.56 & 0.69 & 0.53 & 0.41 & 0.67 \\ 
  ~~~~~~~~~~~~~~~~balancing weights& $\beta_1$    & -0.01 & 0.19 & 0.19 & 0.00 & 0.19 & 0.19 & 0.04 & 0.21 & 0.22 & 0.15 & 0.25 & 0.29 & 0.33 & 0.29 & 0.44 \\ 
                                & $\beta_2$        & 0.00 & 0.04 & 0.04 & 0.00 & 0.04 & 0.04 & 0.01 & 0.05 & 0.05 & 0.01 & 0.05 & 0.05 & 0.03 & 0.06 & 0.06 \\ 
 ~~~~~~~~incorrect $Z(t)$  &                      &  &  &  &  &  &  &  &  &  &  &  &  &  &  &  \\ 
  ~~~~~~~~~~~~~~~~MLE weights& $\beta_1$           & -3.03 & 1.68 & 3.47 & -2.30 & 1.52 & 2.76 & -1.43 & 1.12 & 1.82 & -0.67 & 0.91 & 1.14 & 0.14 & 0.74 & 0.75 \\ 
                              & $\beta_2$         & 0.37 & 1.71 & 1.74 & 0.42 & 1.28 & 1.34 & 0.36 & 0.91 & 0.98 & 0.42 & 0.65 & 0.77 & 0.52 & 0.43 & 0.67 \\ 
   ~~~~~~~~~~~~~~~~balancing weights& $\beta_1$   & -2.00 & 0.59 & 2.09 & -1.54 & 0.65 & 1.67 & -0.97 & 0.63 & 1.16 & -0.38 & 0.58 & 0.69 & 0.19 & 0.50 & 0.54 \\ 
                                & $\beta_2$       & -0.02 & 0.22 & 0.22 & 0.00 & 0.21 & 0.21 & -0.00 & 0.19 & 0.19 & 0.01 & 0.16 & 0.16 & 0.02 & 0.13 & 0.13 \\ 
 \hline
\end{tabular}}
\end{sidewaystable}

\begin{sidewaystable}
\caption{\label{ncont1000}\footnotesize{Empirical bias, empirical standard deviation (SD), and root mean squared error (RMSE) of the IIWEs with the MLE weights and the balancing weights estimators for $\beta_1$ (group effect) and $\beta_2$ (time effect)  in the marginal regression model of a longitudinal \textbf{continuous} outcome, when the sample size $\bm{n=1000}$, the sensitivity parameter $\phi$ is set at $0, 0.15, 0.30, 0.45, 0.60$ and the visit process is \textbf{\emph{moderately} }dependant on time-varying covariates with $\gamma_z=0.5$. The naive analysis without weighting and the analysis based on complete data are also presented. }}
\centering
\scriptsize{
\begin{tabular}{llrrrrrrrrrrrrrrr}
  \hline
                                                &&\multicolumn{3}{c}{$\phi=0$}&\multicolumn{3}{c}{$\phi=0.15$}&\multicolumn{3}{c}{$\phi=0.30$}&\multicolumn{3}{c}{$\phi=0.45$}&\multicolumn{3}{c}{$\phi=0.60$}\\
                                                  && Bias & SD & RMSE & Bias & SD & RMSE & Bias & SD & RMSE & Bias & SD & RMSE& Bias & SD & RMSE \\ 
  \hline
  Complete data & $\beta_1$                     & 0.00 & 0.01 & 0.01 & -0.00 & 0.01 & 0.01 & -0.00 & 0.01 & 0.01 & -0.00 & 0.01 & 0.01 & 0.00 & 0.01 & 0.01 \\ 
               & $\beta_2$                      & 0.00 & 0.00 & 0.00 & -0.00 & 0.00 & 0.00 & 0.00 & 0.00 & 0.00 & 0.00 & 0.00 & 0.00 & -0.00 & 0.00 & 0.00 \\ 
 Naive analysis & $\beta_1$                     & -1.41 & 0.07 & 1.41 & -0.41 & 0.06 & 0.41 & 0.33 & 0.04 & 0.33 & 0.89 & 0.03 & 0.89 & 1.30 & 0.02 & 1.30 \\ 
               & $\beta_2$                      & 0.02 & 0.07 & 0.08 & 0.11 & 0.04 & 0.12 & 0.22 & 0.02 & 0.22 & 0.36 & 0.02 & 0.36 & 0.49 & 0.01 & 0.49 \\ 
  $S\{Y(t)\}$ not included    &                      &  &  &  &  &  &  &  &  &  &  &  &  &  &  &  \\ 
  ~~~~~~~~correct $Z(t)$  &                     &  &  &  &  &  &  &  &  &  &  &  &  &  &  &  \\ 
 ~~~~~~~~~~~~~~~~MLE weights& $\beta_1$          & -0.11 & 0.07 & 0.13 & -0.30 & 0.08 & 0.31 & -0.61 & 0.10 & 0.62 & -0.62 & 0.18 & 0.65 & -0.05 & 0.19 & 0.20 \\ 
 ~~~~~~~~~~~~~~~~           & $\beta_2$          & -0.00 & 0.07 & 0.07 & 0.02 & 0.07 & 0.08 & 0.06 & 0.12 & 0.13 & 0.18 & 0.20 & 0.27 & 0.40 & 0.18 & 0.44 \\ 
 ~~~~~~~~~~~~~~~~balancing weights& $\beta_1$   & 0.01 & 0.08 & 0.08 & -0.00 & 0.07 & 0.07 & -0.01 & 0.06 & 0.06 & 0.02 & 0.06 & 0.06 & 0.12 & 0.07 & 0.13 \\ 
                               & $\beta_2$      & -0.00 & 0.02 & 0.02 & 0.00 & 0.02 & 0.02 & 0.00 & 0.02 & 0.02 & 0.01 & 0.02 & 0.02 & 0.02 & 0.02 & 0.03 \\ 
  ~~~~~~~~incorrect $Z(t)$  &                    &  &  &  &  &  &  &  &  &  &  &  &  &  &  &  \\ 
 ~~~~~~~~~~~~~~~~MLE weights& $\beta_1$         & -2.89 & 0.19 & 2.90 & -2.50 & 0.33 & 2.52 & -1.54 & 0.31 & 1.57 & -0.48 & 0.16 & 0.50 & 0.43 & 0.08 & 0.44 \\ 
                             & $\beta_2$         & -0.10 & 0.25 & 0.27 & 0.03 & 0.42 & 0.42 & 0.08 & 0.36 & 0.37 & 0.18 & 0.17 & 0.25 & 0.38 & 0.08 & 0.39 \\ 
  ~~~~~~~~~~~~~~~~balancing weights& $\beta_1$    & -2.31 & 0.11 & 2.31 & -1.66 & 0.12 & 1.67 & -1.06 & 0.12 & 1.07 & -0.49 & 0.12 & 0.51 & 0.05 & 0.12 & 0.14 \\ 
                               & $\beta_2$       & -0.08 & 0.05 & 0.09 & -0.06 & 0.06 & 0.08 & -0.06 & 0.06 & 0.09 & -0.07 & 0.06 & 0.09 & -0.05 & 0.06 & 0.08 \\ 
  $S\{Y(t)\}$  included    &                          &  &  &  &  &  &  &  &  &  &  &  &  &  &  &  \\ 
   ~~~~~~~~correct $Z(t)$  &                     &  &  &  &  &  &  &  &  &  &  &  &  &  &  &  \\ 
  ~~~~~~~~~~~~~~~~MLE weights& $\beta_1$         & -0.11 & 0.07 & 0.13 & -0.14 & 0.06 & 0.16 & -0.16 & 0.08 & 0.18 & -0.09 & 0.12 & 0.15 & 0.07 & 0.15 & 0.16 \\ 
  ~~~~~~~~~~~~~~~~           & $\beta_2$         & -0.00 & 0.07 & 0.07 & 0.02 & 0.07 & 0.07 & 0.05 & 0.08 & 0.09 & 0.14 & 0.10 & 0.17 & 0.29 & 0.13 & 0.31 \\ 
  ~~~~~~~~~~~~~~~~balancing weights& $\beta_1$   & 0.01 & 0.08 & 0.08 & -0.00 & 0.07 & 0.07 & -0.01 & 0.06 & 0.06 & 0.01 & 0.06 & 0.06 & 0.11 & 0.07 & 0.13 \\ 
                                & $\beta_2$       & -0.00 & 0.02 & 0.02 & 0.00 & 0.02 & 0.02 & 0.00 & 0.02 & 0.02 & 0.01 & 0.02 & 0.02 & 0.02 & 0.02 & 0.03 \\ 
 ~~~~~~~~incorrect $Z(t)$  &                     &  &  &  &  &  &  &  &  &  &  &  &  &  &  &  \\ 
  ~~~~~~~~~~~~~~~~MLE weights& $\beta_1$          & -0.56 & 0.09 & 0.56 & -0.92 & 0.11 & 0.93 & -1.43 & 0.27 & 1.46 & -1.43 & 0.47 & 1.50 & -0.81 & 0.32 & 0.87 \\ 
                              & $\beta_2$        & 0.00 & 0.10 & 0.10 & 0.02 & 0.12 & 0.12 & 0.07 & 0.32 & 0.33 & 0.20 & 0.49 & 0.53 & 0.41 & 0.33 & 0.52 \\ 
   ~~~~~~~~~~~~~~~~balancing weights& $\beta_1$  & -0.41 & 0.08 & 0.42 & -0.68 & 0.08 & 0.69 & -1.01 & 0.12 & 1.01 & -1.03 & 0.21 & 1.05 & -0.63 & 0.26 & 0.68 \\ 
                                & $\beta_2$      & -0.01 & 0.04 & 0.04 & -0.02 & 0.04 & 0.04 & -0.06 & 0.06 & 0.08 & -0.09 & 0.10 & 0.14 & -0.09 & 0.11 & 0.14 \\ 
 \hline
\end{tabular}}
\end{sidewaystable}

\begin{sidewaystable}
\caption{\label{ecount200} \footnotesize{Empirical bias, empirical standard deviation (SD), and root mean squared error (RMSE) of the IIWEs with the MLE weights and the balancing weights estimators for $\beta_1$ (group effect) and $\beta_2$ (time effect)  in the marginal regression model of a longitudinal \textbf{count} outcome, when the sample size $\bm{n=200}$, the sensitivity parameter $\phi$ is set at $0, 0.15, 0.30, 0.45, 0.60$ and the visit process is \textbf{\emph{highly}} dependant on time-varying covariates with $\gamma_z=1.25$. The naive analysis without weighting and the analysis based on complete data are also presented. }}
\centering
\scriptsize{
\begin{tabular}{llrrrrrrrrrrrrrrr}
  \hline
                                                &&\multicolumn{3}{c}{$\phi=0$}&\multicolumn{3}{c}{$\phi=0.15$}&\multicolumn{3}{c}{$\phi=0.30$}&\multicolumn{3}{c}{$\phi=0.45$}&\multicolumn{3}{c}{$\phi=0.60$}\\
                                                  && Bias & SD & RMSE & Bias & SD & RMSE & Bias & SD & RMSE & Bias & SD & RMSE& Bias & SD & RMSE \\ 
  \hline
  Complete data & $\beta_1$                         & -0.00 & 0.02 & 0.02 & 0.00 & 0.02 & 0.02 & -0.00 & 0.02 & 0.02 & 0.00 & 0.02 & 0.02 & 0.00 & 0.02 & 0.02 \\ 
               & $\beta_2$                          & -0.00 & 0.01 & 0.01 & -0.00 & 0.01 & 0.01 & 0.00 & 0.01 & 0.01 & 0.00 & 0.01 & 0.01 & -0.00 & 0.01 & 0.01 \\ 
 Naive analysis & $\beta_1$                         & 0.76 & 0.11 & 0.77 & 0.84 & 0.09 & 0.85 & 0.88 & 0.08 & 0.88 & 0.91 & 0.07 & 0.92 & 0.93 & 0.06 & 0.93 \\ 
               & $\beta_2$                          & 0.16 & 0.05 & 0.17 & 0.20 & 0.04 & 0.21 & 0.24 & 0.04 & 0.25 & 0.29 & 0.03 & 0.29 & 0.33 & 0.03 & 0.33 \\ 
  $S\{Y(t)\}$ not included    &                          &  &  &  &  &  &  &  &  &  &  &  &  &  &  &  \\ 
  ~~~~~~~~correct $Z(t)$  &                         &  &  &  &  &  &  &  &  &  &  &  &  &  &  &  \\ 
 ~~~~~~~~~~~~~~~~MLE weights& $\beta_1$             & -0.67 & 0.32 & 0.74 & -0.79 & 0.46 & 0.91 & -0.85 & 0.58 & 1.03 & -0.82 & 0.62 & 1.03 & -0.71 & 0.66 & 0.97 \\ 
 ~~~~~~~~~~~~~~~~           & $\beta_2$             & 0.13 & 0.36 & 0.38 & 0.19 & 0.52 & 0.55 & 0.30 & 0.68 & 0.75 & 0.36 & 0.86 & 0.93 & 0.44 & 1.01 & 1.10 \\ 
 ~~~~~~~~~~~~~~~~balancing weights& $\beta_1$       & -0.04 & 0.25 & 0.25 & -0.02 & 0.24 & 0.24 & -0.02 & 0.24 & 0.24 & 0.03 & 0.28 & 0.28 & 0.01 & 0.25 & 0.25 \\ 
                               & $\beta_2$          & -0.09 & 0.21 & 0.23 & -0.08 & 0.22 & 0.23 & -0.07 & 0.21 & 0.23 & -0.04 & 0.24 & 0.24 & -0.02 & 0.25 & 0.25 \\ 
  ~~~~~~~~incorrect $Z(t)$  &                       &  &  &  &  &  &  &  &  &  &  &  &  &  &  &  \\ 
 ~~~~~~~~~~~~~~~~MLE weights& $\beta_1$             & -0.94 & 0.47 & 1.05 & -0.70 & 0.44 & 0.83 & -0.45 & 0.38 & 0.59 & -0.21 & 0.32 & 0.38 & 0.04 & 0.24 & 0.24 \\ 
                             & $\beta_2$            & 0.17 & 0.60 & 0.63 & 0.19 & 0.46 & 0.50 & 0.23 & 0.38 & 0.44 & 0.27 & 0.32 & 0.42 & 0.35 & 0.23 & 0.42 \\ 
  ~~~~~~~~~~~~~~~~balancing weights& $\beta_1$      & -0.57 & 0.27 & 0.63 & -0.47 & 0.28 & 0.55 & -0.39 & 0.24 & 0.46 & -0.28 & 0.26 & 0.38 & -0.21 & 0.24 & 0.32 \\ 
                               & $\beta_2$          & -0.09 & 0.24 & 0.25 & -0.06 & 0.25 & 0.25 & -0.06 & 0.23 & 0.23 & -0.01 & 0.24 & 0.24 & 0.03 & 0.23 & 0.23 \\ 
  $S\{Y(t)\}$  included    &                             &  &  &  &  &  &  &  &  &  &  &  &  &  &  &  \\ 
   ~~~~~~~~correct $Z(t)$  &                        &  &  &  &  &  &  &  &  &  &  &  &  &  &  &  \\ 
  ~~~~~~~~~~~~~~~~MLE weights& $\beta_1$            & -0.67 & 0.32 & 0.74 & -0.57 & 0.31 & 0.65 & -0.46 & 0.27 & 0.54 & -0.33 & 0.21 & 0.40 & -0.21 & 0.18 & 0.28 \\ 
  ~~~~~~~~~~~~~~~~           & $\beta_2$            & 0.13 & 0.36 & 0.38 & 0.18 & 0.33 & 0.37 & 0.24 & 0.30 & 0.38 & 0.28 & 0.25 & 0.38 & 0.36 & 0.21 & 0.41 \\ 
  ~~~~~~~~~~~~~~~~balancing weights& $\beta_1$      & -0.04 & 0.25 & 0.25 & -0.03 & 0.23 & 0.23 & -0.03 & 0.22 & 0.22 & 0.01 & 0.23 & 0.23 & 0.01 & 0.20 & 0.20 \\ 
                                & $\beta_2$         & -0.09 & 0.21 & 0.23 & -0.07 & 0.21 & 0.22 & -0.06 & 0.20 & 0.21 & -0.01 & 0.20 & 0.20 & 0.01 & 0.20 & 0.20 \\ 
 ~~~~~~~~incorrect $Z(t)$  &                        &  &  &  &  &  &  &  &  &  &  &  &  &  &  &  \\ 
  ~~~~~~~~~~~~~~~~MLE weights& $\beta_1$            & -0.96 & 0.46 & 1.06 & -0.77 & 0.42 & 0.88 & -0.61 & 0.37 & 0.71 & -0.45 & 0.34 & 0.57 & -0.27 & 0.27 & 0.39 \\ 
                              & $\beta_2$           & 0.19 & 0.58 & 0.61 & 0.23 & 0.44 & 0.50 & 0.27 & 0.37 & 0.46 & 0.30 & 0.33 & 0.45 & 0.36 & 0.27 & 0.45 \\ 
   ~~~~~~~~~~~~~~~~balancing weights& $\beta_1$     & -0.59 & 0.26 & 0.64 & -0.51 & 0.27 & 0.58 & -0.45 & 0.24 & 0.51 & -0.35 & 0.25 & 0.43 & -0.29 & 0.22 & 0.36 \\ 
                                & $\beta_2$         & -0.08 & 0.23 & 0.25 & -0.04 & 0.24 & 0.24 & -0.04 & 0.21 & 0.22 & 0.02 & 0.22 & 0.22 & 0.05 & 0.21 & 0.22 \\  
 \hline
 \end{tabular}}
\end{sidewaystable}

\begin{sidewaystable}
\caption{\label{ncount200}\footnotesize{Empirical bias, empirical standard deviation (SD), and root mean squared error (RMSE) of the IIWEs with the MLE weights and the balancing weights estimators for $\beta_1$ (group effect) and $\beta_2$ (time effect)  in the marginal regression model of a longitudinal \textbf{count } outcome, when the sample size $\bm{n=200}$, the sensitivity parameter $\phi$ is set at $0, 0.15, 0.30, 0.45, 0.60$ and the visit process is \textbf{\emph{moderately} }dependant on time-varying covariates with $\gamma_z=0.5$. The naive analysis without weighting and the analysis based on complete data are also presented. }}
\centering
\scriptsize{
\begin{tabular}{llrrrrrrrrrrrrrrr}
  \hline
                                                &&\multicolumn{3}{c}{$\phi=0$}&\multicolumn{3}{c}{$\phi=0.15$}&\multicolumn{3}{c}{$\phi=0.30$}&\multicolumn{3}{c}{$\phi=0.45$}&\multicolumn{3}{c}{$\phi=0.60$}\\
                                                  && Bias & SD & RMSE & Bias & SD & RMSE & Bias & SD & RMSE & Bias & SD & RMSE& Bias & SD & RMSE \\ 
  \hline
   Complete data & $\beta_1$                      & -0.00 & 0.02 & 0.02 & 0.00 & 0.02 & 0.02 & -0.00 & 0.02 & 0.02 & 0.00 & 0.02 & 0.02 & -0.00 & 0.02 & 0.02 \\ 
               & $\beta_2$                       & 0.04 & 0.10 & 0.11 & 0.07 & 0.13 & 0.15 & -0.00 & 0.01 & 0.01 & 0.00 & 0.01 & 0.01 & 0.11 & 0.15 & 0.18 \\ 
 Naive analysis & $\beta_1$                      & -0.33 & 0.15 & 0.36 & -0.12 & 0.12 & 0.17 & 0.08 & 0.10 & 0.12 & 0.23 & 0.08 & 0.25 & 0.38 & 0.07 & 0.38 \\ 
               & $\beta_2$                       & 0.06 & 0.17 & 0.18 & 0.11 & 0.14 & 0.18 & 0.11 & 0.09 & 0.14 & 0.19 & 0.08 & 0.20 & 0.29 & 0.05 & 0.30 \\ 
  $S\{Y(t)\}$ not included    &                       &  &  &  &  &  &  &  &  &  &  &  &  &  &  &  \\ 
  ~~~~~~~~correct $Z(t)$  &                      &  &  &  &  &  &  &  &  &  &  &  &  &  &  &  \\ 
 ~~~~~~~~~~~~~~~~MLE weights& $\beta_1$           & -0.05 & 0.22 & 0.22 & -0.03 & 0.19 & 0.19 & -0.04 & 0.18 & 0.19 & -0.11 & 0.14 & 0.18 & -0.22 & 0.12 & 0.25 \\ 
 ~~~~~~~~~~~~~~~~           & $\beta_2$           & 0.03 & 0.19 & 0.19 & 0.06 & 0.20 & 0.21 & -0.02 & 0.16 & 0.16 & 0.01 & 0.13 & 0.13 & 0.14 & 0.13 & 0.19 \\ 
 ~~~~~~~~~~~~~~~~balancing weights& $\beta_1$    & -0.01 & 0.22 & 0.23 & 0.02 & 0.20 & 0.20 & 0.04 & 0.22 & 0.22 & 0.04 & 0.21 & 0.21 & 0.05 & 0.21 & 0.22 \\ 
                               & $\beta_2$       & 0.01 & 0.19 & 0.19 & 0.06 & 0.20 & 0.21 & -0.01 & 0.17 & 0.17 & 0.01 & 0.16 & 0.16 & 0.15 & 0.19 & 0.24 \\ 
  ~~~~~~~~incorrect $Z(t)$  &                     &  &  &  &  &  &  &  &  &  &  &  &  &  &  &  \\ 
 ~~~~~~~~~~~~~~~~MLE weights& $\beta_1$          & -1.12 & 0.19 & 1.14 & -1.26 & 0.18 & 1.27 & -1.29 & 0.16 & 1.30 & -1.30 & 0.16 & 1.31 & -1.44 & 0.32 & 1.48 \\ 
                             & $\beta_2$          & 0.04 & 0.21 & 0.21 & 0.06 & 0.21 & 0.22 & -0.06 & 0.21 & 0.22 & -0.04 & 0.21 & 0.21 & 0.08 & 0.20 & 0.22 \\ 
  ~~~~~~~~~~~~~~~~balancing weights& $\beta_1$     & -0.79 & 0.17 & 0.81 & -0.72 & 0.14 & 0.74 & -0.62 & 0.12 & 0.64 & -0.54 & 0.12 & 0.55 & -0.49 & 0.11 & 0.50 \\ 
                               & $\beta_2$        & 0.02 & 0.17 & 0.17 & 0.07 & 0.17 & 0.18 & 0.01 & 0.12 & 0.12 & 0.06 & 0.10 & 0.12 & 0.19 & 0.12 & 0.22 \\ 
  $S\{Y(t)\}$  included     &                          &  &  &  &  &  &  &  &  &  &  &  &  &  &  &  \\ 
   ~~~~~~~~correct $Z(t)$  &                      &  &  &  &  &  &  &  &  &  &  &  &  &  &  &  \\ 
  ~~~~~~~~~~~~~~~~MLE weights& $\beta_1$          & -0.05 & 0.22 & 0.22 & -0.04 & 0.16 & 0.16 & -0.04 & 0.13 & 0.14 & -0.05 & 0.10 & 0.11 & -0.04 & 0.08 & 0.09 \\ 
  ~~~~~~~~~~~~~~~~           & $\beta_2$          & 0.03 & 0.19 & 0.19 & 0.07 & 0.18 & 0.20 & 0.02 & 0.13 & 0.13 & 0.06 & 0.10 & 0.12 & 0.19 & 0.11 & 0.22 \\ 
  ~~~~~~~~~~~~~~~~balancing weights& $\beta_1$    & -0.01 & 0.22 & 0.23 & -0.00 & 0.17 & 0.17 & 0.00 & 0.16 & 0.16 & -0.00 & 0.12 & 0.12 & -0.01 & 0.10 & 0.10 \\ 
                                & $\beta_2$        & 0.01 & 0.19 & 0.19 & 0.06 & 0.19 & 0.20 & -0.01 & 0.13 & 0.13 & 0.02 & 0.10 & 0.10 & 0.15 & 0.14 & 0.20 \\ 
 ~~~~~~~~incorrect $Z(t)$  &                      &  &  &  &  &  &  &  &  &  &  &  &  &  &  &  \\ 
  ~~~~~~~~~~~~~~~~MLE weights& $\beta_1$           & -1.08 & 0.18 & 1.09 & -1.10 & 0.17 & 1.11 & -1.02 & 0.13 & 1.03 & -0.93 & 0.12 & 0.94 & -0.93 & 0.18 & 0.95 \\ 
                              & $\beta_2$         & 0.05 & 0.20 & 0.21 & 0.09 & 0.18 & 0.21 & 0.03 & 0.15 & 0.15 & 0.10 & 0.13 & 0.16 & 0.22 & 0.10 & 0.25 \\ 
   ~~~~~~~~~~~~~~~~balancing weights& $\beta_1$   & -0.80 & 0.17 & 0.82 & -0.75 & 0.14 & 0.77 & -0.69 & 0.12 & 0.70 & -0.64 & 0.10 & 0.64 & -0.62 & 0.10 & 0.63 \\ 
                                & $\beta_2$       & 0.02 & 0.17 & 0.17 & 0.07 & 0.17 & 0.18 & 0.01 & 0.11 & 0.11 & 0.05 & 0.09 & 0.10 & 0.18 & 0.12 & 0.22 \\ 
 \hline
\end{tabular}}
\end{sidewaystable}

\begin{sidewaystable}
\caption{\label{ecount500}\footnotesize{Empirical bias, empirical standard deviation (SD), and root mean squared error (RMSE) of the IIWEs with the MLE weights and the balancing weights estimators for $\beta_1$ (group effect) and $\beta_2$ (time effect)  in the marginal regression model of a longitudinal \textbf{count} outcome, when the sample size $\bm{n=500}$, the sensitivity parameter $\phi$ is set at $0, 0.15, 0.30, 0.45, 0.60$ and the visit process is \textbf{\emph{highly}} dependant on time-varying covariates with $\gamma_z=1.25$. The naive analysis without weighting and the analysis based on complete data are also presented. }}
\centering
\scriptsize{
\begin{tabular}{llrrrrrrrrrrrrrrr}
  \hline
                                                &&\multicolumn{3}{c}{$\phi=0$}&\multicolumn{3}{c}{$\phi=0.15$}&\multicolumn{3}{c}{$\phi=0.30$}&\multicolumn{3}{c}{$\phi=0.45$}&\multicolumn{3}{c}{$\phi=0.60$}\\
                                                  && Bias & SD & RMSE & Bias & SD & RMSE & Bias & SD & RMSE & Bias & SD & RMSE& Bias & SD & RMSE \\ 
  \hline
    Complete data & $\beta_1$                     & -0.00 & 0.01 & 0.01 & 0.00 & 0.01 & 0.01 & 0.00 & 0.01 & 0.01 & 0.00 & 0.01 & 0.01 & 0.00 & 0.01 & 0.01 \\ 
                & $\beta_2$                      & 0.04 & 0.10 & 0.11 & 0.07 & 0.13 & 0.15 & 0.11 & 0.15 & 0.18 & 0.07 & 0.13 & 0.15 & 0.08 & 0.13 & 0.15 \\ 
  Naive analysis & $\beta_1$                     & 0.76 & 0.06 & 0.77 & 0.85 & 0.06 & 0.85 & 0.91 & 0.06 & 0.91 & 0.93 & 0.05 & 0.93 & 0.95 & 0.05 & 0.95 \\ 
                & $\beta_2$                      & 0.18 & 0.05 & 0.19 & 0.23 & 0.05 & 0.23 & 0.27 & 0.04 & 0.27 & 0.29 & 0.02 & 0.29 & 0.32 & 0.02 & 0.32 \\ 
   $S\{Y(t)\}$ not included    &                      &  &  &  &  &  &  &  &  &  &  &  &  &  &  &  \\ 
   ~~~~~~~~correct $Z(t)$  &                     &  &  &  &  &  &  &  &  &  &  &  &  &  &  &  \\ 
  ~~~~~~~~~~~~~~~~MLE weights& $\beta_1$          & -0.81 & 0.28 & 0.86 & -1.03 & 0.42 & 1.11 & -1.20 & 0.65 & 1.36 & -1.15 & 0.68 & 1.34 & -1.13 & 2.49 & 2.74 \\ 
  ~~~~~~~~~~~~~~~~           & $\beta_2$          & 0.12 & 0.30 & 0.32 & 0.19 & 0.39 & 0.43 & 0.26 & 0.51 & 0.57 & 0.31 & 0.63 & 0.70 & 0.35 & 0.59 & 0.69 \\ 
  ~~~~~~~~~~~~~~~~balancing weights& $\beta_1$   & -0.01 & 0.17 & 0.17 & 0.01 & 0.16 & 0.16 & 0.04 & 0.17 & 0.17 & 0.05 & 0.17 & 0.18 & 0.07 & 0.18 & 0.20 \\ 
                                & $\beta_2$      & -0.00 & 0.17 & 0.17 & 0.05 & 0.20 & 0.20 & 0.10 & 0.20 & 0.22 & 0.07 & 0.19 & 0.21 & 0.10 & 0.19 & 0.22 \\ 
   ~~~~~~~~incorrect $Z(t)$  &                    &  &  &  &  &  &  &  &  &  &  &  &  &  &  &  \\ 
  ~~~~~~~~~~~~~~~~MLE weights& $\beta_1$         & -1.01 & 0.38 & 1.08 & -0.82 & 0.36 & 0.89 & -0.58 & 0.36 & 0.68 & -0.27 & 0.25 & 0.37 & -0.01 & 0.19 & 0.19 \\ 
                              & $\beta_2$         & 0.12 & 0.42 & 0.44 & 0.19 & 0.34 & 0.39 & 0.23 & 0.25 & 0.34 & 0.26 & 0.18 & 0.32 & 0.32 & 0.14 & 0.35 \\ 
   ~~~~~~~~~~~~~~~~balancing weights& $\beta_1$    & -0.57 & 0.16 & 0.59 & -0.49 & 0.15 & 0.52 & -0.41 & 0.15 & 0.44 & -0.32 & 0.15 & 0.35 & -0.23 & 0.14 & 0.27 \\ 
                                & $\beta_2$       & -0.01 & 0.19 & 0.19 & 0.06 & 0.20 & 0.21 & 0.11 & 0.21 & 0.23 & 0.08 & 0.20 & 0.22 & 0.12 & 0.19 & 0.22 \\ 
   $S\{Y(t)\}$  included    &                          &  &  &  &  &  &  &  &  &  &  &  &  &  &  &  \\ 
    ~~~~~~~~correct $Z(t)$  &                     &  &  &  &  &  &  &  &  &  &  &  &  &  &  &  \\ 
   ~~~~~~~~~~~~~~~~MLE weights& $\beta_1$         & -0.81 & 0.28 & 0.86 & -0.74 & 0.28 & 0.79 & -0.66 & 0.30 & 0.73 & -0.49 & 0.25 & 0.55 & -0.35 & 0.22 & 0.41 \\ 
   ~~~~~~~~~~~~~~~~           & $\beta_2$         & 0.12 & 0.30 & 0.32 & 0.20 & 0.24 & 0.31 & 0.25 & 0.18 & 0.31 & 0.29 & 0.15 & 0.32 & 0.34 & 0.13 & 0.36 \\ 
   ~~~~~~~~~~~~~~~~balancing weights& $\beta_1$   & -0.01 & 0.17 & 0.17 & 0.00 & 0.15 & 0.15 & 0.02 & 0.13 & 0.13 & 0.03 & 0.12 & 0.13 & 0.05 & 0.11 & 0.12 \\ 
                                 & $\beta_2$       & -0.00 & 0.17 & 0.17 & 0.05 & 0.19 & 0.20 & 0.11 & 0.18 & 0.21 & 0.08 & 0.17 & 0.19 & 0.11 & 0.16 & 0.19 \\ 
  ~~~~~~~~incorrect $Z(t)$  &                     &  &  &  &  &  &  &  &  &  &  &  &  &  &  &  \\ 
   ~~~~~~~~~~~~~~~~MLE weights& $\beta_1$          & -1.02 & 0.37 & 1.09 & -0.89 & 0.35 & 0.95 & -0.73 & 0.35 & 0.81 & -0.53 & 0.26 & 0.59 & -0.35 & 0.21 & 0.41 \\ 
                               & $\beta_2$        & 0.14 & 0.41 & 0.43 & 0.23 & 0.33 & 0.40 & 0.26 & 0.25 & 0.36 & 0.29 & 0.20 & 0.35 & 0.33 & 0.17 & 0.37 \\ 
    ~~~~~~~~~~~~~~~~balancing weights& $\beta_1$  & -0.59 & 0.15 & 0.61 & -0.53 & 0.15 & 0.55 & -0.47 & 0.15 & 0.49 & -0.39 & 0.14 & 0.42 & -0.32 & 0.13 & 0.34 \\ 
                                 & $\beta_2$      & -0.00 & 0.19 & 0.19 & 0.06 & 0.20 & 0.21 & 0.11 & 0.19 & 0.22 & 0.09 & 0.18 & 0.21 & 0.13 & 0.17 & 0.21 \\ 
 \hline
\end{tabular}}
\end{sidewaystable}

\begin{sidewaystable}
\caption{\label{ncount500}\footnotesize{Empirical bias, empirical standard deviation (SD), and root mean squared error (RMSE) of the IIWEs with the MLE weights and the balancing weights estimators for $\beta_1$ (group effect) and $\beta_2$ (time effect)  in the marginal regression model of a longitudinal \textbf{count }outcome, when the sample size $\bm{n=500}$, the sensitivity parameter $\phi$ is set at $0, 0.15, 0.30, 0.45, 0.60$ and the visit process is \textbf{\emph{moderately}} dependant on time-varying covariates with $\gamma_z=0.5$. The naive analysis without weighting and the analysis based on complete data are also presented. }}
\centering
\scriptsize{
\begin{tabular}{llrrrrrrrrrrrrrrr}
  \hline
                                                &&\multicolumn{3}{c}{$\phi=0$}&\multicolumn{3}{c}{$\phi=0.15$}&\multicolumn{3}{c}{$\phi=0.30$}&\multicolumn{3}{c}{$\phi=0.45$}&\multicolumn{3}{c}{$\phi=0.60$}\\
                                                  && Bias & SD & RMSE & Bias & SD & RMSE & Bias & SD & RMSE & Bias & SD & RMSE& Bias & SD & RMSE \\ 
  \hline
  Complete data & $\beta_1$                     & 0.00 & 0.01 & 0.01 & -0.00 & 0.01 & 0.01 & -0.00 & 0.01 & 0.01 & -0.00 & 0.01 & 0.01 & 0.00 & 0.01 & 0.01 \\ 
               & $\beta_2$                      & 0.04 & 0.10 & 0.11 & 0.00 & 0.00 & 0.00 & -0.00 & 0.00 & 0.00 & 0.00 & 0.00 & 0.00 & -0.00 & 0.00 & 0.00 \\ 
 Naive analysis & $\beta_1$                     & -0.32 & 0.09 & 0.33 & -0.11 & 0.08 & 0.14 & 0.08 & 0.06 & 0.10 & 0.24 & 0.05 & 0.25 & 0.36 & 0.05 & 0.36 \\ 
               & $\beta_2$                      & 0.06 & 0.13 & 0.14 & 0.06 & 0.07 & 0.09 & 0.11 & 0.06 & 0.12 & 0.19 & 0.05 & 0.19 & 0.29 & 0.04 & 0.29 \\ 
  $S\{Y(t)\}$ not included    &                      &  &  &  &  &  &  &  &  &  &  &  &  &  &  &  \\ 
  ~~~~~~~~correct $Z(t)$  &                     &  &  &  &  &  &  &  &  &  &  &  &  &  &  &  \\ 
 ~~~~~~~~~~~~~~~~MLE weights& $\beta_1$          & -0.05 & 0.14 & 0.14 & -0.03 & 0.12 & 0.13 & -0.05 & 0.12 & 0.13 & -0.12 & 0.10 & 0.16 & -0.25 & 0.09 & 0.26 \\ 
 ~~~~~~~~~~~~~~~~           & $\beta_2$          & 0.03 & 0.16 & 0.16 & -0.01 & 0.11 & 0.11 & -0.01 & 0.10 & 0.11 & 0.01 & 0.09 & 0.09 & 0.07 & 0.08 & 0.11 \\ 
 ~~~~~~~~~~~~~~~~balancing weights& $\beta_1$   & -0.00 & 0.14 & 0.14 & 0.02 & 0.14 & 0.14 & 0.04 & 0.16 & 0.16 & 0.05 & 0.16 & 0.17 & 0.05 & 0.19 & 0.19 \\ 
                               & $\beta_2$      & 0.02 & 0.15 & 0.15 & -0.01 & 0.12 & 0.12 & 0.00 & 0.12 & 0.12 & 0.02 & 0.13 & 0.13 & 0.05 & 0.15 & 0.16 \\ 
  ~~~~~~~~incorrect $Z(t)$  &                    &  &  &  &  &  &  &  &  &  &  &  &  &  &  &  \\ 
 ~~~~~~~~~~~~~~~~MLE weights& $\beta_1$         & -1.12 & 0.12 & 1.13 & -1.21 & 0.10 & 1.22 & -1.29 & 0.10 & 1.29 & -1.30 & 0.09 & 1.30 & -1.23 & 0.09 & 1.24 \\ 
                             & $\beta_2$         & 0.03 & 0.15 & 0.15 & -0.02 & 0.12 & 0.12 & -0.05 & 0.12 & 0.13 & -0.06 & 0.13 & 0.14 & -0.01 & 0.14 & 0.14 \\ 
  ~~~~~~~~~~~~~~~~balancing weights& $\beta_1$    & -0.78 & 0.10 & 0.79 & -0.70 & 0.09 & 0.71 & -0.61 & 0.08 & 0.62 & -0.53 & 0.07 & 0.53 & -0.45 & 0.06 & 0.45 \\ 
                               & $\beta_2$       & 0.02 & 0.13 & 0.13 & -0.00 & 0.07 & 0.07 & 0.02 & 0.07 & 0.07 & 0.06 & 0.07 & 0.09 & 0.12 & 0.06 & 0.14 \\ 
  $S\{Y(t)\}$  included    &                          &  &  &  &  &  &  &  &  &  &  &  &  &  &  &  \\ 
   ~~~~~~~~correct $Z(t)$  &                     &  &  &  &  &  &  &  &  &  &  &  &  &  &  &  \\ 
  ~~~~~~~~~~~~~~~~MLE weights& $\beta_1$         & -0.05 & 0.14 & 0.14 & -0.04 & 0.10 & 0.11 & -0.04 & 0.08 & 0.10 & -0.04 & 0.06 & 0.08 & -0.04 & 0.05 & 0.07 \\ 
  ~~~~~~~~~~~~~~~~           & $\beta_2$         & 0.03 & 0.16 & 0.16 & 0.01 & 0.10 & 0.10 & 0.03 & 0.08 & 0.08 & 0.06 & 0.06 & 0.09 & 0.12 & 0.06 & 0.14 \\ 
  ~~~~~~~~~~~~~~~~balancing weights& $\beta_1$   & -0.00 & 0.14 & 0.14 & 0.00 & 0.11 & 0.11 & 0.01 & 0.10 & 0.10 & 0.00 & 0.08 & 0.08 & -0.00 & 0.06 & 0.06 \\ 
                                & $\beta_2$       & 0.02 & 0.15 & 0.15 & -0.01 & 0.10 & 0.10 & 0.01 & 0.09 & 0.09 & 0.02 & 0.07 & 0.08 & 0.05 & 0.06 & 0.08 \\ 
 ~~~~~~~~incorrect $Z(t)$  &                     &  &  &  &  &  &  &  &  &  &  &  &  &  &  &  \\ 
  ~~~~~~~~~~~~~~~~MLE weights& $\beta_1$          & -1.07 & 0.12 & 1.08 & -1.06 & 0.10 & 1.06 & -1.01 & 0.08 & 1.02 & -0.93 & 0.07 & 0.93 & -0.82 & 0.06 & 0.82 \\ 
                              & $\beta_2$        & 0.04 & 0.15 & 0.15 & 0.02 & 0.10 & 0.10 & 0.05 & 0.08 & 0.10 & 0.10 & 0.08 & 0.12 & 0.18 & 0.07 & 0.19 \\ 
   ~~~~~~~~~~~~~~~~balancing weights& $\beta_1$  & -0.80 & 0.10 & 0.80 & -0.74 & 0.08 & 0.74 & -0.68 & 0.07 & 0.68 & -0.62 & 0.06 & 0.63 & -0.57 & 0.05 & 0.57 \\ 
                                & $\beta_2$      & 0.02 & 0.13 & 0.13 & -0.00 & 0.07 & 0.07 & 0.02 & 0.06 & 0.07 & 0.06 & 0.06 & 0.08 & 0.11 & 0.05 & 0.12 \\  
 \hline
\end{tabular}}
\end{sidewaystable}

\begin{sidewaystable}
\caption{\label{ecount1000}\footnotesize{Empirical bias, empirical standard deviation (SD), and root mean squared error (RMSE) of the IIWEs with the MLE weights and the balancing weights estimators for $\beta_1$ (group effect) and $\beta_2$ (time effect)  in the marginal regression model of a longitudinal \textbf{count }outcome, when the sample size $\bm{n=1000}$, the sensitivity parameter $\phi$ is set at $0, 0.15, 0.30, 0.45, 0.60$ and the visit process is \textbf{\emph{highly} }dependant on time-varying covariates with $\gamma_z=1.25$. The naive analysis without weighting and the analysis based on complete data are also presented. }}
\centering
\scriptsize{
\begin{tabular}{llrrrrrrrrrrrrrrr}
  \hline
                                                &&\multicolumn{3}{c}{$\phi=0$}&\multicolumn{3}{c}{$\phi=0.15$}&\multicolumn{3}{c}{$\phi=0.30$}&\multicolumn{3}{c}{$\phi=0.45$}&\multicolumn{3}{c}{$\phi=0.60$}\\
                                                  && Bias & SD & RMSE & Bias & SD & RMSE & Bias & SD & RMSE & Bias & SD & RMSE& Bias & SD & RMSE \\ 
  \hline
  Complete data & $\beta_1$                     & -0.00 & 0.01 & 0.01 & 0.00 & 0.01 & 0.01 & -0.00 & 0.01 & 0.01 & -0.00 & 0.01 & 0.01 & -0.00 & 0.05 & 0.05 \\ 
               & $\beta_2$                      & -0.00 & 0.00 & 0.00 & 0.00 & 0.00 & 0.00 & 0.00 & 0.00 & 0.00 & 0.00 & 0.00 & 0.00 & 0.00 & 0.01 & 0.01 \\ 
 Naive analysis & $\beta_1$                     & 0.76 & 0.05 & 0.76 & 0.84 & 0.04 & 0.84 & 0.89 & 0.03 & 0.89 & 0.92 & 0.03 & 0.92 & 0.93 & 0.03 & 0.93 \\ 
               & $\beta_2$                      & 0.16 & 0.02 & 0.16 & 0.20 & 0.02 & 0.20 & 0.24 & 0.02 & 0.24 & 0.29 & 0.02 & 0.29 & 0.33 & 0.02 & 0.33 \\ 
  $S\{Y(t)\}$ not included    &                      &  &  &  &  &  &  &  &  &  &  &  &  &  &  &  \\ 
  ~~~~~~~~correct $Z(t)$  &                     &  &  &  &  &  &  &  &  &  &  &  &  &  &  &  \\ 
 ~~~~~~~~~~~~~~~~MLE weights& $\beta_1$          & -0.84 & 0.23 & 0.87 & -0.99 & 0.30 & 1.03 & -1.07 & 0.41 & 1.15 & -1.05 & 0.49 & 1.16 & -0.96 & 0.53 & 1.10 \\ 
 ~~~~~~~~~~~~~~~~           & $\beta_2$          & 0.11 & 0.25 & 0.27 & 0.15 & 0.39 & 0.42 & 0.21 & 0.54 & 0.58 & 0.28 & 0.56 & 0.62 & 0.42 & 0.70 & 0.82 \\ 
 ~~~~~~~~~~~~~~~~balancing weights& $\beta_1$   & 0.01 & 0.13 & 0.13 & 0.02 & 0.11 & 0.12 & 0.05 & 0.13 & 0.14 & 0.07 & 0.13 & 0.15 & 0.09 & 0.16 & 0.18 \\ 
                               & $\beta_2$      & -0.02 & 0.11 & 0.11 & -0.02 & 0.11 & 0.11 & -0.00 & 0.11 & 0.12 & 0.02 & 0.13 & 0.13 & 0.06 & 0.14 & 0.15 \\ 
  ~~~~~~~~incorrect $Z(t)$  &                    &  &  &  &  &  &  &  &  &  &  &  &  &  &  &  \\ 
 ~~~~~~~~~~~~~~~~MLE weights& $\beta_1$         & -0.97 & 0.30 & 1.01 & -0.73 & 0.23 & 0.77 & -0.46 & 0.19 & 0.50 & -0.19 & 0.15 & 0.24 & 0.04 & 0.12 & 0.13 \\ 
                             & $\beta_2$         & 0.11 & 0.34 & 0.36 & 0.14 & 0.28 & 0.31 & 0.17 & 0.21 & 0.27 & 0.25 & 0.15 & 0.29 & 0.34 & 0.11 & 0.35 \\ 
  ~~~~~~~~~~~~~~~~balancing weights& $\beta_1$    & -0.55 & 0.11 & 0.56 & -0.46 & 0.11 & 0.47 & -0.37 & 0.11 & 0.39 & -0.28 & 0.11 & 0.30 & -0.19 & 0.11 & 0.22 \\ 
                               & $\beta_2$       & -0.04 & 0.12 & 0.13 & -0.02 & 0.13 & 0.13 & 0.01 & 0.14 & 0.14 & 0.04 & 0.14 & 0.15 & 0.07 & 0.15 & 0.17 \\ 
  $S\{Y(t)\}$  included    &                          &  &  &  &  &  &  &  &  &  &  &  &  &  &  &  \\ 
   ~~~~~~~~correct $Z(t)$  &                     &  &  &  &  &  &  &  &  &  &  &  &  &  &  &  \\ 
  ~~~~~~~~~~~~~~~~MLE weights& $\beta_1$         & -0.84 & 0.23 & 0.87 & -0.71 & 0.18 & 0.73 & -0.56 & 0.15 & 0.58 & -0.40 & 0.12 & 0.42 & -0.27 & 0.09 & 0.28 \\ 
  ~~~~~~~~~~~~~~~~           & $\beta_2$         & 0.11 & 0.25 & 0.27 & 0.15 & 0.21 & 0.26 & 0.20 & 0.17 & 0.26 & 0.27 & 0.13 & 0.30 & 0.34 & 0.10 & 0.36 \\ 
  ~~~~~~~~~~~~~~~~balancing weights& $\beta_1$   & 0.01 & 0.13 & 0.13 & 0.01 & 0.10 & 0.10 & 0.03 & 0.10 & 0.10 & 0.04 & 0.09 & 0.10 & 0.06 & 0.09 & 0.11 \\ 
                                & $\beta_2$       & -0.02 & 0.11 & 0.11 & -0.01 & 0.10 & 0.10 & 0.00 & 0.10 & 0.10 & 0.03 & 0.09 & 0.10 & 0.06 & 0.09 & 0.11 \\ 
 ~~~~~~~~incorrect $Z(t)$  &                     &  &  &  &  &  &  &  &  &  &  &  &  &  &  &  \\ 
  ~~~~~~~~~~~~~~~~MLE weights& $\beta_1$          & -0.99 & 0.29 & 1.03 & -0.82 & 0.23 & 0.85 & -0.64 & 0.19 & 0.67 & -0.46 & 0.16 & 0.49 & -0.30 & 0.14 & 0.33 \\ 
                              & $\beta_2$        & 0.13 & 0.33 & 0.35 & 0.18 & 0.28 & 0.33 & 0.21 & 0.21 & 0.29 & 0.27 & 0.17 & 0.32 & 0.34 & 0.14 & 0.37 \\ 
   ~~~~~~~~~~~~~~~~balancing weights& $\beta_1$  & -0.57 & 0.11 & 0.58 & -0.50 & 0.10 & 0.51 & -0.43 & 0.10 & 0.45 & -0.36 & 0.09 & 0.37 & -0.28 & 0.10 & 0.30 \\ 
                                & $\beta_2$      & -0.03 & 0.12 & 0.12 & -0.01 & 0.12 & 0.12 & 0.02 & 0.12 & 0.12 & 0.05 & 0.12 & 0.13 & 0.08 & 0.12 & 0.15 \\ 
 \hline
\end{tabular}}
\end{sidewaystable}

\begin{sidewaystable}
\caption{\label{ncount1000}\footnotesize{Empirical bias, empirical standard deviation (SD), and root mean squared error (RMSE) of the IIWEs with the MLE weights and the balancing weights estimators for $\beta_1$ (group effect) and $\beta_2$ (time effect)  in the marginal regression model of a longitudinal \textbf{count }outcome, when the sample size $\bm{n=1000}$, the sensitivity parameter $\phi$ is set at $0, 0.15, 0.30, 0.45, 0.60$ and the visit process is \textbf{\emph{moderately} }dependant on time-varying covariates with $\gamma_z=0.5$. The naive analysis without weighting and the analysis based on complete data are also presented. }}
\centering
\scriptsize{
\begin{tabular}{llrrrrrrrrrrrrrrr}
  \hline
                                                &&\multicolumn{3}{c}{$\phi=0$}&\multicolumn{3}{c}{$\phi=0.15$}&\multicolumn{3}{c}{$\phi=0.30$}&\multicolumn{3}{c}{$\phi=0.45$}&\multicolumn{3}{c}{$\phi=0.60$}\\
                                                  && Bias & SD & RMSE & Bias & SD & RMSE & Bias & SD & RMSE & Bias & SD & RMSE& Bias & SD & RMSE \\ 
  \hline
  Complete data & $\beta_1$                     & 0.00 & 0.01 & 0.01 & -0.00 & 0.01 & 0.01 & 0.00 & 0.01 & 0.01 & 0.00 & 0.01 & 0.01 & -0.00 & 0.01 & 0.01 \\ 
               & $\beta_2$                      & -0.00 & 0.00 & 0.00 & 0.00 & 0.00 & 0.00 & 0.00 & 0.00 & 0.00 & 0.00 & 0.00 & 0.00 & 0.00 & 0.00 & 0.00 \\ 
 Naive analysis & $\beta_1$                     & -0.32 & 0.07 & 0.32 & -0.11 & 0.06 & 0.12 & 0.08 & 0.04 & 0.10 & 0.24 & 0.04 & 0.25 & 0.36 & 0.03 & 0.36 \\ 
               & $\beta_2$                      & 0.04 & 0.07 & 0.08 & 0.06 & 0.05 & 0.08 & 0.10 & 0.04 & 0.11 & 0.19 & 0.03 & 0.19 & 0.29 & 0.03 & 0.29 \\ 
  $S\{Y(t)\}$ not included    &                      &  &  &  &  &  &  &  &  &  &  &  &  &  &  &  \\ 
  ~~~~~~~~correct $Z(t)$  &                     &  &  &  &  &  &  &  &  &  &  &  &  &  &  &  \\ 
 ~~~~~~~~~~~~~~~~MLE weights& $\beta_1$          & -0.04 & 0.11 & 0.12 & -0.03 & 0.11 & 0.11 & -0.06 & 0.09 & 0.11 & -0.13 & 0.09 & 0.16 & -0.25 & 0.07 & 0.26 \\ 
 ~~~~~~~~~~~~~~~~           & $\beta_2$          & 0.00 & 0.12 & 0.12 & -0.01 & 0.09 & 0.09 & -0.02 & 0.08 & 0.08 & 0.01 & 0.07 & 0.07 & 0.07 & 0.07 & 0.10 \\ 
 ~~~~~~~~~~~~~~~~balancing weights& $\beta_1$   & 0.00 & 0.12 & 0.12 & 0.02 & 0.13 & 0.13 & 0.03 & 0.13 & 0.13 & 0.04 & 0.15 & 0.16 & 0.06 & 0.19 & 0.20 \\ 
                               & $\beta_2$      & -0.00 & 0.10 & 0.11 & -0.00 & 0.10 & 0.10 & 0.01 & 0.10 & 0.10 & 0.03 & 0.12 & 0.12 & 0.07 & 0.14 & 0.16 \\ 
  ~~~~~~~~incorrect $Z(t)$  &                    &  &  &  &  &  &  &  &  &  &  &  &  &  &  &  \\ 
 ~~~~~~~~~~~~~~~~MLE weights& $\beta_1$         & -1.10 & 0.08 & 1.11 & -1.21 & 0.08 & 1.21 & -1.29 & 0.07 & 1.29 & -1.30 & 0.07 & 1.30 & -1.23 & 0.07 & 1.24 \\ 
                             & $\beta_2$         & 0.01 & 0.09 & 0.09 & -0.03 & 0.08 & 0.09 & -0.06 & 0.09 & 0.11 & -0.06 & 0.09 & 0.11 & -0.01 & 0.10 & 0.10 \\ 
  ~~~~~~~~~~~~~~~~balancing weights& $\beta_1$    & -0.78 & 0.07 & 0.78 & -0.70 & 0.07 & 0.70 & -0.61 & 0.05 & 0.61 & -0.53 & 0.06 & 0.53 & -0.44 & 0.05 & 0.45 \\ 
                               & $\beta_2$       & -0.01 & 0.06 & 0.06 & -0.00 & 0.05 & 0.05 & 0.02 & 0.05 & 0.05 & 0.07 & 0.05 & 0.08 & 0.13 & 0.05 & 0.14 \\ 
  $S\{Y(t)\}$  included    &                          &  &  &  &  &  &  &  &  &  &  &  &  &  &  &  \\ 
   ~~~~~~~~correct $Z(t)$  &                     &  &  &  &  &  &  &  &  &  &  &  &  &  &  &  \\ 
  ~~~~~~~~~~~~~~~~MLE weights& $\beta_1$         & -0.04 & 0.11 & 0.12 & -0.04 & 0.08 & 0.10 & -0.05 & 0.06 & 0.08 & -0.05 & 0.05 & 0.07 & -0.04 & 0.04 & 0.06 \\ 
  ~~~~~~~~~~~~~~~~           & $\beta_2$         & 0.00 & 0.12 & 0.12 & 0.01 & 0.08 & 0.08 & 0.02 & 0.06 & 0.06 & 0.06 & 0.05 & 0.08 & 0.13 & 0.04 & 0.13 \\ 
  ~~~~~~~~~~~~~~~~balancing weights& $\beta_1$   & 0.00 & 0.12 & 0.12 & 0.00 & 0.10 & 0.10 & 0.00 & 0.08 & 0.08 & 0.00 & 0.06 & 0.06 & -0.00 & 0.06 & 0.06 \\ 
                                & $\beta_2$       & -0.00 & 0.10 & 0.11 & -0.00 & 0.08 & 0.08 & 0.01 & 0.06 & 0.06 & 0.02 & 0.06 & 0.06 & 0.06 & 0.05 & 0.08 \\ 
 ~~~~~~~~incorrect $Z(t)$  &                     &  &  &  &  &  &  &  &  &  &  &  &  &  &  &  \\ 
  ~~~~~~~~~~~~~~~~MLE weights& $\beta_1$          & -1.06 & 0.08 & 1.06 & -1.06 & 0.07 & 1.06 & -1.02 & 0.06 & 1.02 & -0.93 & 0.05 & 0.93 & -0.82 & 0.04 & 0.82 \\ 
                              & $\beta_2$        & 0.02 & 0.08 & 0.09 & 0.02 & 0.07 & 0.07 & 0.04 & 0.06 & 0.07 & 0.10 & 0.05 & 0.11 & 0.18 & 0.05 & 0.19 \\ 
   ~~~~~~~~~~~~~~~~balancing weights& $\beta_1$  & -0.79 & 0.07 & 0.79 & -0.73 & 0.06 & 0.74 & -0.67 & 0.05 & 0.68 & -0.62 & 0.04 & 0.62 & -0.57 & 0.04 & 0.57 \\ 
                                & $\beta_2$      & -0.01 & 0.06 & 0.06 & -0.00 & 0.05 & 0.05 & 0.02 & 0.04 & 0.05 & 0.06 & 0.04 & 0.07 & 0.11 & 0.04 & 0.12 \\ 
  \hline
\end{tabular}}
\end{sidewaystable}

\newpage
\section{Additional details and results of the PsA clinic sub-cohort data analysis}

\subsection{Specification of the marginal regression model for the longitudinal active joint count }
\tb{We assumed that the active joint count follows a negative binomial model with mean $\mu\{t,\bm{X}(t)\}$ and a dispersion parameter $\theta$. Specifically, 
and}
\tb{
\begin{eqnarray}
    \log\{\mu\{t,\bm{X}(t)\}\}&=&
    \beta_0\cdot {\mbox{I}}(\mbox{Year [1980, 2008)})+\beta_1\cdot\mbox{I}(\mbox{Year [2008, 2012)}) \nonumber \\
      &+&\beta_2\cdot\mbox{I}(\mbox{Year [2012, 2015)})+
   \beta_3 \cdot {\mbox{I}}(\mbox{1-2 years after  baseline})\nonumber \\
     &+& \beta_4 \cdot {\mbox{I}}(\mbox{2-3 years after  baseline})
    + \beta_5 \cdot {\mbox{I}}(\mbox{3-4 years after  baseline})\nonumber \\
    &+& \beta_6 \cdot {\mbox{I}}(\mbox{4-5 years after  baseline})
    + \beta_7 \cdot {\mbox{I}}(\mbox{$\ge$ years after  baseline})\nonumber \\
     &+& \beta_8 \cdot \mbox{I}(\mbox{Male}) +\beta_{9} \cdot \mbox{Age at baseline} \nonumber \\
     &+&\beta_{10} \cdot \mbox{PsA disease duration at baseline}+ \beta_{11}\cdot \mbox{baseline ESR}\nonumber \\ 
     &+&\beta_{12} \cdot   \mbox{baseline active joint count}+\beta_{13} \cdot \mbox{baseline damaged joint count}\nonumber \\
      &+&\beta_{14} \cdot  \mbox{I} (\mbox{baseline use of NSAIDs})+\beta_{15} \cdot \mbox{I} (\mbox{baseline use of DMARDs})\nonumber \\
       &+&  \beta_{16} \cdot {\mbox{I}}(\mbox{0-1 years after  baseline})\cdot{\mbox{I}}(\mbox{baseline biologics use})\nonumber \\
      &+&  \beta_{17} \cdot {\mbox{I}}(\mbox{1-2 years after  baseline})\cdot{\mbox{I}}(\mbox{baseline biologics use})\nonumber \\
     &+& \beta_{18}\cdot {\mbox{I}}(\mbox{2-3 years after  baseline})\cdot{\mbox{I}}(\mbox{baseline biologics use})\nonumber \\
     &+& \beta_{19} \cdot {\mbox{I}}(\mbox{3-4 years after  baseline})\cdot{\mbox{I}}(\mbox{baseline biologics use})\nonumber \\
    &+& \beta_{20} \cdot {\mbox{I}}(\mbox{4-5 years after  baseline})\cdot{\mbox{I}}(\mbox{baseline biologics use})\nonumber \\
     &+& \beta_{21} \cdot {\mbox{I}}(\mbox{$\ge$ years after  baseline})\cdot{\mbox{I}}(\mbox{baseline biologics use}).\nonumber 
\end{eqnarray}
Baseline active joint count and baseline damaged joint count were transformed by taking $\log(x+1)$. All non-binary variables were standardised to have mean zero and standard deviation 0.5. }

\subsection{More details of calibrating the sensitivity parameter}
\begin{figure}[!p]
\centering\includegraphics[scale=0.5]{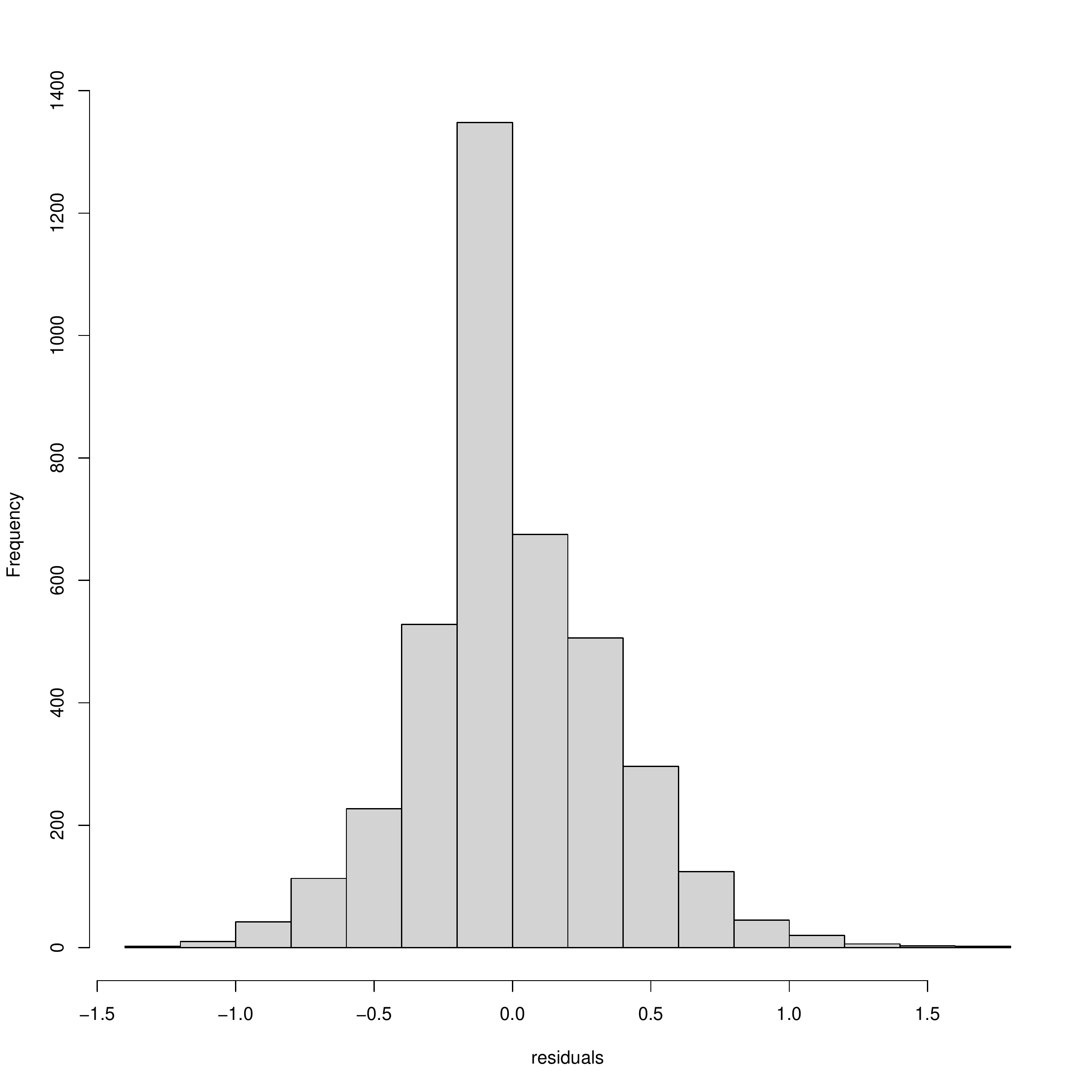}
\caption{Histogram of the residuals from the linear regression model for the log-transformed observed active joint count given the same observed history variables as in the Cox model for the visit process and time (modelled as natural cubic splines with five degrees of freedom).    }
 \label{Fig13}
\end{figure}

\tb{Figure~\ref{Fig13} presented the histogram of the residuals from the linear  regression model for the log-transformed observed active joint count given the same observed history variables as in the Cox model for the visit process and time (modelled as natural cubic splines with five degrees of freedom). The distribution of these residuals was approximately symmetric after the $\log(x+1)$ transformation, which suggested that the normality assumption for the log-transformed observed active joint count $\log\{Y(t)+1\} \mid dN^*(t)=1$ was plausible.}

\subsection{Additional results}

Table~\ref{Table2} presented the results of the fitted Cox model for the visit process in the PsA clinic cohort, assuming the visiting at random assumption was satisfied (i.e., $\phi=0$). Results in Table~\ref{Table2} indicated that patients with more active joints at the last visit were more likely to visit, but those with more damaged joints at the last visit were less likely to visit. Patients with longer median previous inter-visit gap times were less likely to visit. \tb{However, patients were more likely to visit as the time since the last visit increased.}  Finally, visits occurred more frequently as time elapsed since baseline.  
\begin{table}[!p]
\caption{Estimated hazard ratios with 95\% confidence intervals and Wald test p-values from the fitted Cox model for the visit process in the PsA clinic cohort, assuming the visiting at random assumption was satisfied (i.e., $\phi=0$).   \label{Table2}}

\centering
\footnotesize{
\begin{tabular}{lcrrr}
  \hline
  \hline
 & Hazard ratio &\multicolumn{2}{c}{95 \% CI} & p-value \\ 
  \hline
    Gender (male)                     & 1.028 & 0.960 & 1.100 & 0.433 \\  
    PsA disease duration at baseline  & 2.697 & 1.946 & 3.737 & $<$0.001 \\  
  Damaged joints at baseline          & 1.157 & 0.984 & 1.361 & 0.078 \\ 
  Active joints at baseline           & 0.933 & 0.853 & 1.021 & 0.131 \\ 
  ESR at baseline                     & 1.013 & 0.929 & 1.103 & 0.775 \\ 
  &&&&\\
  1-2 years (vs. 0-1 year) since baseline       & 2.630 & 2.198 & 3.146 & $<$0.001 \\  
  2-3 years  (vs. 0-1 year) since baseline      & 2.700 & 2.235 & 3.262 & $<$0.001 \\  
  3-4 years  (vs. 0-1 year) since baseline      & 3.193 & 2.615 & 3.899 & $<$0.001 \\  
  4-5 years  (vs. 0-1 year) since baseline      & 3.428 & 2.763 & 4.253 & $<$0.001 \\  
  $\ge 5$ years (vs. 0-1 year) since baseline   & 4.725 & 3.835 & 5.821 & $<$0.001 \\  
   &&&&\\
  0-1 years since baseline $*$ biologics use    & 1.087 & 0.898 & 1.316 & 0.393 \\ 
  1-2 years since baseline $*$ biologics use    & 1.206 & 1.008 & 1.443 & 0.040 \\ 
  2-3 years since baseline $*$ biologics use    & 1.233 & 1.020 & 1.489 & 0.030 \\ 
  3-4 years since baseline $*$ biologics use    & 1.225 & 1.008 & 1.490 & 0.042 \\ 
  4-5 years since baseline $*$ biologics use    & 1.342 & 1.084 & 1.662 & 0.007 \\ 
  $\ge5$ years since baseline $*$ biologics use & 1.150 & 1.022 & 1.296 & 0.021 \\ 
   &&&&\\
    PsA disease duration at $t$ & 0.335 & 0.232 & 0.484 & $<$0.001 \\ 
  Age at $t$                    & 0.992 & 0.920 & 1.069 & 0.832 \\ 

 Year $[1980, 2008)$ $*$ latest $\mbox{ESR}$ prior to $t$            & 1.022 & 0.889 & 1.174 & 0.762 \\ 
 Year  $[2008, 2012)$ $*$ latest $\mbox{ESR}$ prior to $t$          & 0.997 & 0.882 & 1.127 & 0.967 \\ 
 Year  $[2012, 2015)$ $*$ latest $\mbox{ESR}$ prior to $t$          & 0.971 & 0.854 & 1.104 & 0.653 \\ 
 Year  $[1980, 2008)$ $*$ latest active joint count prior to $t$    & 1.098 & 1.004 & 1.201 & 0.041 \\ 
 Year  $[2008, 2012)$ $*$ latest active joint count prior to $t$    & 1.104 & 1.012 & 1.204 & 0.025 \\ 
 Year $[2012, 2015)$ $*$ latest active joint count prior to $t$     & 1.073 & 0.991 & 1.163 & 0.084 \\ 
 Year $[1980, 2008)$ $*$ latest damaged joint count prior to $t$    & 0.907 & 0.781 & 1.054 & 0.203 \\ 
 Year $[2008, 2012)$ $*$ latest damaged joint count prior to $t$    & 0.862 & 0.733 & 1.013 & 0.071 \\ 
 Year $[2012, 2015)$ $*$ latest damaged joint  count prior to $t$   & 0.931 & 0.809 & 1.073 & 0.324 \\ 
 Year $[1980, 2008)$ $*$ latest NSAID use  prior to $t$             & 1.113 & 0.974 & 1.273 & 0.117 \\ 
 Year $[2008, 2012)$ $*$ latest NSAID use prior to $t$              & 1.005 & 0.902 & 1.120 & 0.927 \\ 
 Year $[2012, 2015)$ $*$ latest NSAID use prior to $t$              & 1.000 & 0.883 & 1.131 & 0.994 \\ 
 Year $[1980, 2008)$ $*$ latest DMARD use prior to $t$              & 1.046 & 0.918 & 1.192 & 0.497 \\ 
 Year $[2008, 2012)$ $*$ latest DMARD use prior to $t$              & 1.086 & 0.971 & 1.215 & 0.146 \\ 
 Year $[2012, 2015)$ $*$ latest DMARD  use prior to $t$             & 1.065 & 0.938 & 1.209 & 0.332 \\ 
 Year $[1980, 2008)$ $*$ biologics use                              & 0.736 & 0.642 & 0.843 & $<$0.001 \\ 
 Year $[2008, 2012)$ $*$ biologics use                              & 0.797 & 0.704 & 0.903 & $<$0.001 \\ 
 Year $[2012, 2015)$ $*$ biologics use                              & 0.809 & 0.704 & 0.928 & 0.003 \\ 
 Year $[1980, 2008)$ $*$ median inter-visit time by $t$             & 0.742 & 0.635 & 0.868 & $<$0.001 \\ 
 Year $[2008, 2012)$ $*$ median inter-visit  time by $t$            & 0.598 & 0.469 & 0.763 & $<$0.001 \\ 
 Year  $[2012, 2015)$ $*$ median inter-visit  time by $t$           & 0.652 & 0.484 & 0.878 & 0.005 \\
 Time since last visit                 & 1.001         & 1.001 & 1.002 & 0.001 \\ 
\hline
\end{tabular}}
\end{table}

\begin{sidewaysfigure}[!p]
\centering\includegraphics[scale=0.5]{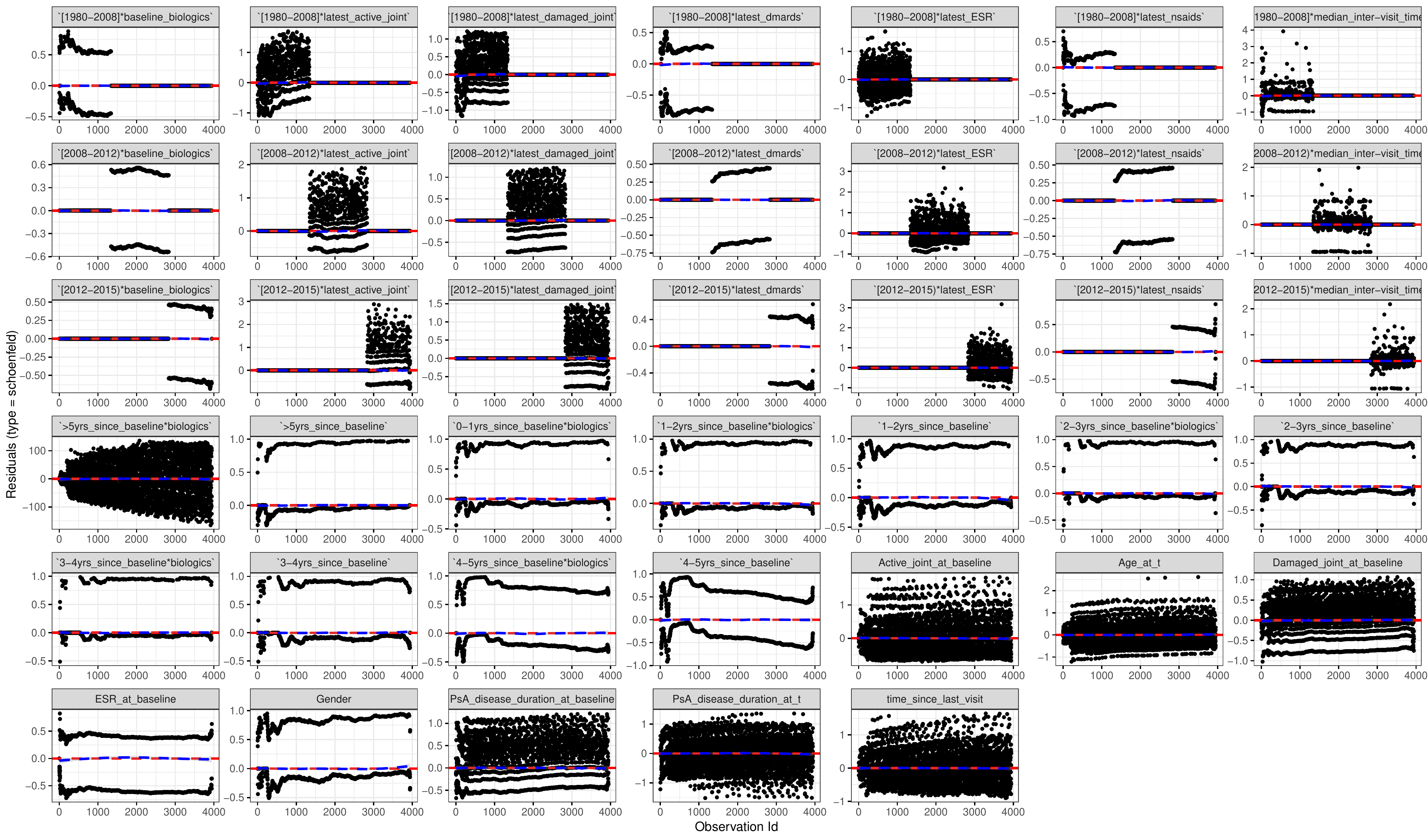}
\caption{Schoenfeld residuals from the fitted Cox model for the visit process in the PsA clinic cohort, assuming the visiting at random assumption was satisfied (i.e., $\phi=0$).  }
 \label{schoenfeld}
\end{sidewaysfigure}

\tb{Figure~\ref{schoenfeld} shows the Schoenfeld residuals from the fitted Cox model for the visit process in the PsA clinic cohort, assuming the visiting at random assumption was satisfied (i.e., $\phi=0$). These residuals and their smoothing line fits (blue lines)  did not show obvious patterns with time, which suggested that the proportionality assumption made in the Cox model was not violated. } 

\begin{figure}[!p]
\centering\includegraphics[scale=0.5]{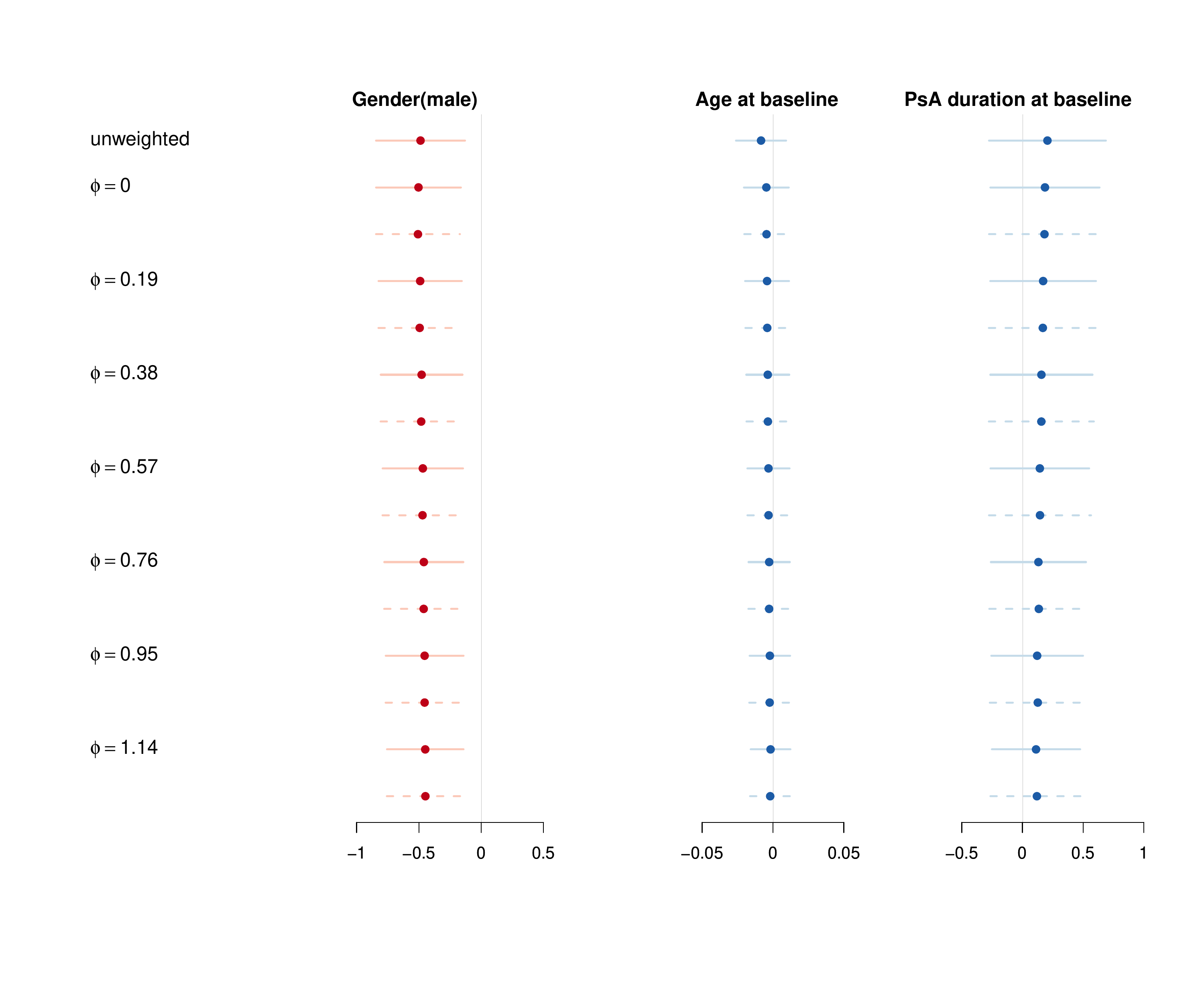}
\caption{Estimates and 95\% confidence intervals (based on jackknife standard errors) of the \textbf{{demographics variable}} effects on active joint counts in the PsA clinic sub-cohort.  Solid lines (\protect\solidblueline, \protect\solidredline): 95\% confidence intervals when using the  IIWEs with the MLE weights or the naive unweighted estimator; dashed lines (\protect\dashblueline,\protect\dashredline): 95\% confidence intervals when using the balancing weights estimators.  The
estimated effects with 95\% confidence intervals covering zero and not covering zero are in light blue (\protect\solidblueline,\protect\dashblueline)  and pink  (\protect\solidredline,\protect\dashredline), respectively. }
 \label{demongraphics}
\end{figure}

\begin{figure}[!p]
\centering\includegraphics[scale=0.5]{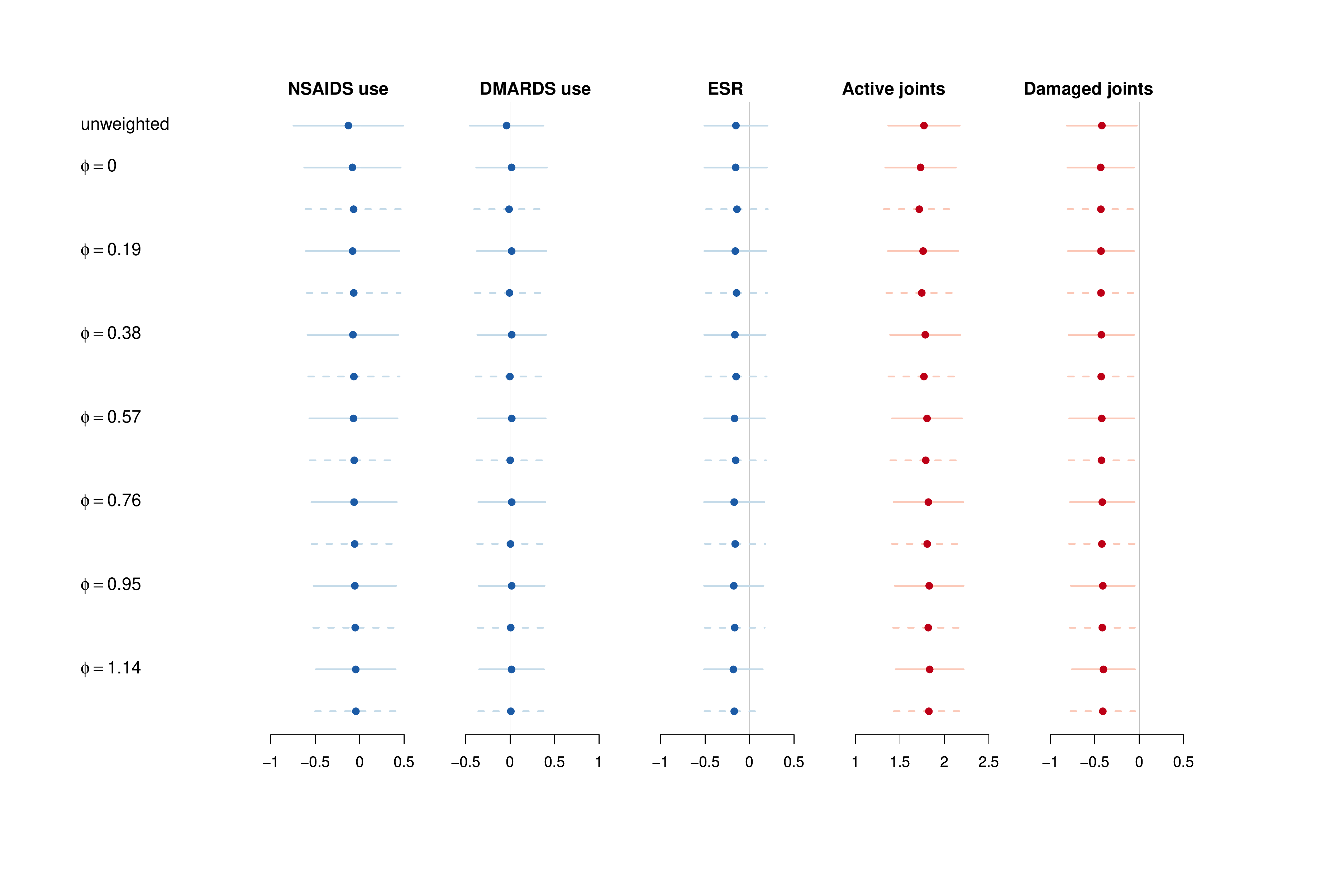}
\caption{Estimates and 95\% confidence intervals (based on jackknife standard errors) of the \textbf{{baseline clinical variable}} effects on active joint counts in the PsA clinic sub-cohort.  Solid lines (\protect\solidblueline, \protect\solidredline): 95\% confidence intervals when using the  IIWEs with the MLE weights or the naive unweighted estimator; dashed lines (\protect\dashblueline,\protect\dashredline): 95\% confidence intervals when using the balancing weights estimators.  The
estimated effects with 95\% confidence intervals covering zero and not covering zero are in light blue (\protect\solidblueline,\protect\dashblueline)  and pink  (\protect\solidredline,\protect\dashredline), respectively. }
 \label{clinical}
\end{figure}

\begin{figure}[!p]
\centering\includegraphics[scale=0.5]{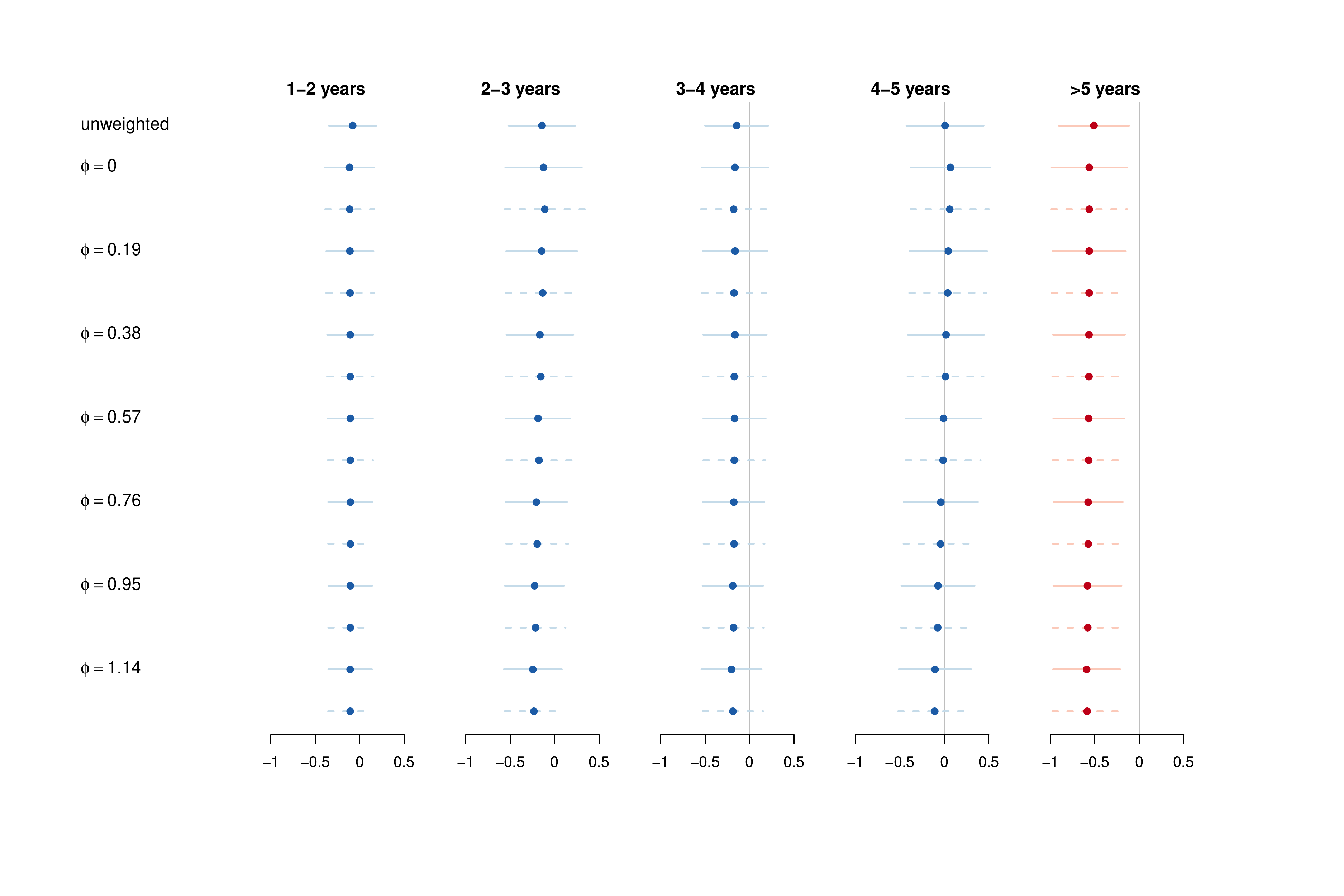}
\caption{Estimates and 95\% confidence intervals (based on jackknife standard errors) of the \textbf{{time (since baseline)}} effects (reference: 0-1 years since baseline) on active joint counts in the PsA clinic sub-cohort.  Solid lines (\protect\solidblueline, \protect\solidredline): 95\% confidence intervals when using the  IIWEs with the MLE weights or the naive unweighted estimator; dashed lines (\protect\dashblueline,\protect\dashredline): 95\% confidence intervals when using the balancing weights estimators.  The
estimated effects with 95\% confidence intervals covering zero and not covering zero are in light blue (\protect\solidblueline,\protect\dashblueline)  and pink  (\protect\solidredline,\protect\dashredline), respectively. }
 \label{time}
\end{figure}

\tb{Figures~\ref{demongraphics}--\ref{time} presented the results for the effects of the demographics variables,  baseline clinical variables and the time since baseline. These effects were not as sensitive as the calendar time effects and the effects of biologics use over time presented in Figures 1 and 2 of  the main text. }

\bibliographystyle{rss}
\bibliography{iiwref}